\documentclass[natbib]{svjour3}                     
\smartqed  
\usepackage{graphicx}
\pdfoutput=1
\usepackage{times}
\usepackage{amssymb}
\usepackage[usenames]{color}
\usepackage{subfigure}
\usepackage{longtable}
\newcommand{\mvm}{mVm$^{-1}$}
\newcommand{\kms}{kms$^{-1}$}
\newcommand{\cmc}{cm$^{-3}$}
\begin{document}

\title{The dynamic quasiperpendicular shock: Cluster discoveries}

\titlerunning{Dynamic quasiperpendicular shock}        

\author{V. Krasnoselskikh \and
             M. Balikhin \and
             S. N. Walker \and
             S. Schwartz \and
             D. Sundkvist \and 
             V. Lobzin \and
             M. Gedalin \and
             S. D. Bale \and
             F. Mozer \and
             J. Soucek \and
             Y. Hobara \and
             H. Comisel
}

\authorrunning{Krasnoselskikh {\it et al.}} 

\institute{V. Krasnoselskikh \at
              LPC2E, CNRS-University of Orleans, Orleans, France \\
              Tel.: +33 02.38.25.52.75\\
              Fax: +\\
              \email{vkrasnos@cnrs-orleans.fr}           
           \and
             M. Balikhin, S. N. Walker \at
             ACSE, University of Sheffield, Sheffield S1 3JD, UK.
             \and
             S. J. Schwartz \at
             Blackett Laboratory, Imperial College London, London SW7 2AZ, UK.
             \and
             D. Sundkvist, S. D. Bale, F. Mozer \at
             Space Sciences Laboratory, University of California, Berkeley, California, USA.
             \and 
             V. Lobzin \at
             School of Physics, University of Sydney, NSW, Austrailia.
             \and
             M. Gedalin \at
             Department of Physics, Ben-Gurion University, Beer-Sheva, Israel
              \and
             J. Soucek \at
             Institute of Atmospheric Physics, Academy of Sciences of the Czech Republic, Prague, Czech Republic.
             \and
             Y. Hobara \at
             Research Center of Space Physics and Radio Engineering, University of Electro-Communications, Tokyo, Japan.
              \and
             H. Comisel \at
             Institute for Space Sciences, Bucharest, Romania.
}

\date{Received: date / Accepted: date}

\maketitle

\begin{abstract}
The physics of collisionless shocks is a very broad topic which has been studied for more than five decades. However, there are a number of important issues which remain unresolved. The energy repartition amongst particle populations in quasiperpendicular shocks is a multi-scale process related to the spatial and temporal structure of the electromagnetic fields within the shock layer. The most important processes take place in the close vicinity of the major magnetic transition or ramp region. The distribution of electromagnetic fields in this region determines the characteristics of ion reflection and thus defines the conditions for ion heating and energy dissipation for supercritical shocks and also the region where an important part of electron heating takes place. In other words, the ramp region determines the main characteristics of energy repartition. All of these processes are crucially dependent upon the characteristic spatial scales of the ramp and foot region provided that the shock is stationary. The process of shock formation consists of the  steepening of a large amplitude nonlinear wave. At some point in its evolution the steepening is arrested by processes occurring within the shock transition. From the earliest studies of collisionless shocks these processes were identified as nonlinearity, dissipation, and dispersion. Their relative role determines the scales of electric and magnetic fields, and so control the characteristics of processes such as of ion reflection, electron heating and particle acceleration. The determination of the scales of the electric and magnetic field is one of the key issues in the physics of collisionless shocks. Moreover, it is well known that under certain conditions shocks manifest a nonstationary dynamic behaviour called reformation. It was suggested that the transition from stationary to nonstationary quasiperiodic dynamics is related to gradients, e.g. scales of the ramp region and its associated  whistler waves that form a precursor wave train. This implies that the ramp region should be considered as the source of these waves. All these questions have been studied making use observations from the Cluster satellites. The Cluster project continues to provide a unique viewpoint from which to study the scales of shocks. During is lifetime the inter-satellite distance between the Cluster satellites has varied from 100 km to 10000 km allowing scientists to use the data best adapted for the given scientific objective. 

The purpose of this review is to address a subset of unresolved problems in collisionless shock physics from experimental point of view making use multi-point observations onboard Cluster satellites. The problems we address are determination of scales of fields and of a scale of electron heating, identification of energy source of precursor wave train, an estimate of the role of anomalous resistivity in energy dissipation process by means of measuring short scale wave fields, and direct observation of reformation process during one single shock front crossing.

\end{abstract}

\tableofcontents


\section{Introduction}

\begin{figure}
\centering
\includegraphics[width=4in]{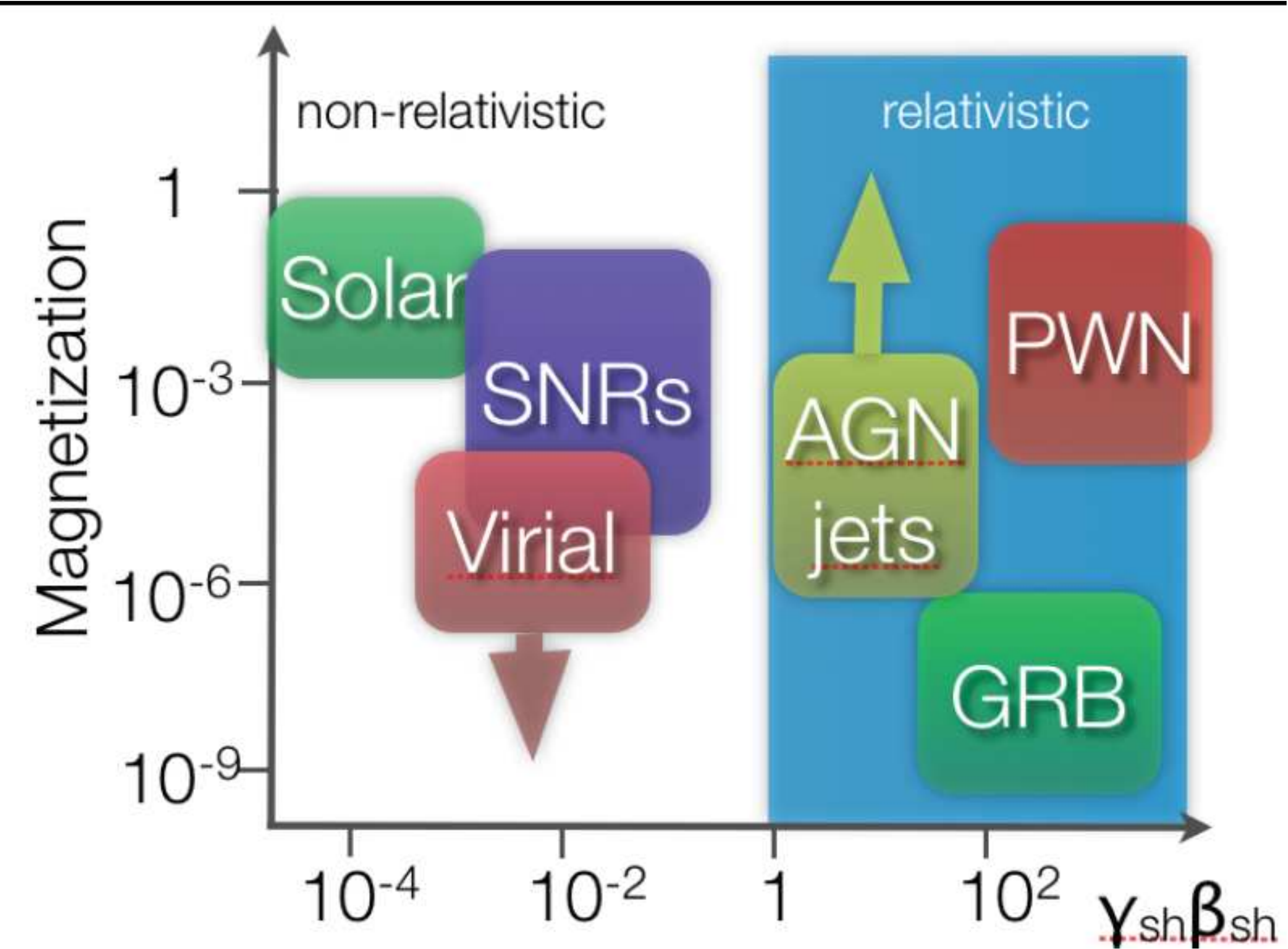}
\caption{(courtesy of A. Spitkovsky). Parametric range of observations of
collisionless shocks associated with different astrophysical objects.\label{fig:range_shocks}}
\end{figure}

Collisionless shocks (CS) are ubiquitous in the universe. They play
an important role in the interaction of the solar wind  with the planets \citep{russell85:_planet, russell77, greenstadt79:_shock, ness74:_magnet_field_obser_venus, ness81:_magnet_voyag_prelim_satur}, they also are
supposed to have vital role in fundamental astrophysical
problems such as cosmic ray acceleration \citep{krymskii77, axford77, bell78:_accel_cr_i, bell78:_accel_cr_ii, blandford78:_partic}. CS's are of crucial importance for
understanding physical processes in the vicinity of such astrophysical
objects as supernova remnants \citep{koyama95:_eviden_sn100, bamba03:_small_scale_struc_sn_shock_chand_obser}, plasma jets \citep{piran05:_magnet_field_gamma_ray_burst}, binary systems and ordinary stars. In spite of this great variety of CS in the Universe only those shocks in the Solar system can be probed using in-situ observations. Moreover, comprehensive in-situ data exist only for interplanetary shocks and planetary bow shocks, however, it is worth noting that some astrophysical shocks are similar to those in
the solar system. As was noted by \citet{kennel85:_quart_centur} 'The density,
temperature and magnetic field in the hot interstellar medium are similar to
those in the solar wind, and the Mach numbers of supernova shocks at the
phase when they accelerate the most cosmic rays are similar to those of
solar wind shocks'. Astrophysical shocks associated with different objects exhibit large differences in the parameters that characterise them. Figure~\ref{fig:range_shocks} shows the variation of astrophysical shocks as a function of magnetisation (Y-axis),  determined as $1/M_{A}$ where $M_{A}$ is the Alfv\'{e}n Mach number (the ratio of the upstream flow velocity to the characteristic velocity of propagation of magnetic perturbations in a plasma or Alfv\'{e}n velocity) and the characteristic plasma pressure to magnetic pressure ratio (X-axis) where $\gamma_{sh}$ is the ratio of the upstream flow velocity to the velocity of light and $\beta_{sh}$ the ratio of total plasma particle thermal pressure to the magnetic field pressure in the reference frame of the upstream flow. Collisionless shocks associated with different astrophysical objects such as Supernovae Remnants (SNR), Active Galactic Nuclei
jets (AGN), Pulsar Wind Nebulae (PWN), and Gamma Ray Bursts (GBR) are indicated.

As can be seen from this figure, different 'families' of shocks occupy different regions of parameter space.
One can also see that the parameters of SNR shocks are quite similar to those of solar system shocks. It allows one to suggest that the studies of solar system shocks, and in particular Earth's bow shock, represents an interest for wider scientific community than only for geophysics. The majority of astrophysical and Solar system shocks are developed in magnetised plasmas.

Collisional shocks have been studied for many decades, beginning with the earliest observations of Mach \citep{mach86:_einig_versuc_reflex_disper, mach87:_photog_fixir_projec_luft_vorgaen}. A shock occurs when an obstacle finds itself immersed in a supersonic gas flow. Before reaching the surface of the obstacle the flow should be
decelerated to velocities lower than the velocity of sound so that it may flow around the body. This process of
flow deceleration and the redistribution of its directed energy occurs over distances of the order of the collisional mean free path of gas particles and the energy difference. The energy  taken from the flow during deceleration is mainly
transformed into the thermal energy of the gas as it is heated. As a result
the sound velocity in the gas increases and, after the shock transition,
becomes larger than the remaining directed velocity of the flow so that the
motion downstream of the shock is subsonic. Thus the shock represents the
transition from supersonic directed motion to subsonic in the reference
frame of the obstacle immersed into the flow.

The notion of the collisionless shock was introduced by several authors in the
late 50's \citep{adlam58, davis58:_struc_hydrom_shock_waves, sagdeyev60}. The modern form of the description was presented in an almost complete form in the famous review paper by \citet{sagdeev66:_cooper_phenom_shock_waves_collis_plasm}. 
The first problem to be overcome is related to the the existence of a shock. For collisional shocks (as mentioned above) the shock thickness is related to the mean free path of the gas particles. However, in space plasmas the mean free path can be as large as 5AU (i.e. similar to the distance of Jupiter from the Sun)! So, how can a shock exist whose width is much smaller than the mean free path? Historically, a very similar problem first appeared in  laboratory devices and only later in space plasmas. However the crucial issue in both cases is exactly the same \citep{paul65:_exper_obser_struc_collis_shock, kurtmullaev65:_excit, kurtmullaev67:_shock_waves_partial_ioniz_plasm, ascoli-bartoli66:_format_carid_cn, goldenbaum65:_exper_study_shock_wave_format}.

The solution proposed initially relied on the processes of anomalous dissipation, namely, anomalous resistivity. Thin shock transition contains quite strong variations of the magnetic field components perpendicular to its normal. This implies  that there is a very intense current inside the shock transition layer. The current carriers, charged particles move with respect to another. The plasma state supporting these intense currents is, in general, unstable. The instabilities in the plasma result in the generation of intense waves. Wave generation opens new channel of impulse and energy exchange between the different populations of plasma particles. For instance, current carrying electrons can emit/generate the waves, and these waves can be absorbed by ions. This generation-absorption exchange using waves as a transmission media between plasma components leads to an exchange of energy and pulse between them. Typically, the characteristic scale of energy exchange between different particle populations can be much smaller than the mean free path of particles. As a result the characteristic scale of the dissipation process can be determined by this anomalous dissipation. Thus the principal difference between collisional and colisionless shocks is the change of the dissipation scale that is determined by additional process involved, but the nature of the transition and its characteristics remain very similar. In both cases the shock redistributes the directed bulk plasma flow energy to plasma thermal energy. However, the dissipation rate and characteristic scales of collisionless shocks are determined by the anomalous process of energy dissipation. The notion of anomalous resistivity was already well known and widely used in plasma physics. The theory of anomalous resistivity based on current instabilities and the generation of ion-sound waves was directly applied to the theory of collisionless shocks by \citet{galeev76:_collis}. Later this idea was further developed in series of papers by \citet{papadopoulos85:_microin, papadopoulos85:_microin_anomal_trans_collis_shock}, who noticed that in the case of currents perpendicular to the magnetic field the ion-sound instability is less efficient than the instability of lower-hybrid waves that propagate almost perpendicular to the magnetic field and the theory of anomalous resistivity in this case should account for these rather than ion-sound waves.

From the earliest experimental studies of shocks in space and laboratory plasmas it was found that the characteristics of the shocks observed can be quite different even in the range of parameters that correspond to solar system shocks and those in laboratory plasmas. There were observations of quite small scales for the ramp with much longer precursor wave train, there were shocks consisting of a long transition region with large amplitude structures in the magnetic field filling a very large area in space. These early observations gave rise to attempts to classify shocks. 

The first systematic classification was proposed by \citet{formisano85:_collis}. He noticed that there are three basic parameters of the upstream flow that are important for the classification. These are the angle between the magnetic field and shock front normal, $\theta _{Bn}$, the plasma beta $\beta $ i.e. the ratio of total particle thermal  pressure to the magnetic field pressure $\beta =8\pi nT/B^{2}$, where $T$ is the total plasma temperature, $n$ is the plasma density, and $B$ the magnitude of the magnetic field; and the magnetosonic Mach number $M_{Ms}$ = $(V_{up}/V_{Ms})$, where $V_{up}$ is the normal component of the velocity to the shock, $V_{Ms}$ is the velocity of the magnetosonic wave propagating in the same direction as the shock. Later this classification was slightly modified and is used in the form proposed by \citet{kennel85:_quart_centur}. This paper divides shocks in to two broad classes that are related to the ion dynamics, namely, quasiparallel and quasiperpendicular. The characteristic feature of the first group is that the ions that are reflected from the shock front can freely propagate to the upstream region. These shocks correspond to the range of angles between the magnetic field and the normal vector to the shock front $\theta_{Bn}<45^{\circ}$. In the second group, quasiperpendicular, part of the ion population is reflected. After reflection they turn back to the upstream region and can gain extra energy due to the acceleration by inductive electric field tangential to the shock surface and perpendicular to the magnetic field. Then they can cross the shock front \citep{woods69:_alfven_mach, woods71, sckopke83:_evolut_specul}. This process can occur when $\theta_{Bn}>45^{\circ}$.

Low Mach number shocks could dissipate the necessary energy entirely through some anomalous resistivity within the current-carrying shock layer.  The right-hand fast magnetosonic/whistler waves have phase and group velocities that increase with decreasing wavelength beyond the fluid regime. Thus, steepened fast mode shocks are expected to radiate short wavelength waves, and hence energy, into the unshocked oncoming flow. The shortest wavelength capable of standing in the flow then forms a ``precursor wavetrain'' that has been observed at these sub-critical shocks \citep{mellott85:_subcr} and as we shall show later this process occurs in supercritical shocks also. 
However, above a critical Mach number, anomalous resistivity within the layer carrying the limited shock current is unable to convert the required amount of energy from directed bulk flow into thermal energy. At quasi-perpendicular shocks, where the magnetic field in the unshocked region makes an angle $\theta_{Bn} >45^\circ$ with the shock propagation direction (the shock ``normal'' $\hat{\vec{n}}$), a fraction of the incident ions are reflected by the steep shock ramp as described above. They gyrate around the magnetic field and gain energy due to acceleration by the transverse motional electric field ($-\vec{V}\times\vec{B}$). Returning to the shock layer they have sufficient energy to pass through into the downstream shocked region \citep{woods69:_alfven_mach, woods71, sckopke83:_evolut_specul}. The separation of ions onto two groups, crossing front directly and after the reflection  results in the  dispersal of particles in velocity space. Group of reflected particles  is separated from the bulk ion population due to an increase in peculiar velocity relative to the bulk motion.  This process corresponds to the kinetic ``heating'' required by the shock jump conditions and it ensures the major part of energy dissipation necessary for directed energy transfer to thermal energy of plasma ion population. The process of ion thermalization takes place on rather large scale downstream of the shock front. The spatial length of the transition to ion thermal equilibrium can be treated in a similar fashion to that of the shock front thickness in collisional shocks. Detailed measurements of ion distributions onboard ISEE mission resulted in establishing all major characteristics of this process \citep{ sckopke83:_evolut_specul}. This result is probably one of the most important obtained in this outstanding program. In theory this critical Mach number corresponds to the multi-fluid hydrodynamic limit in which the resistivity and viscosity can nolonger provide sufficient dissipation \citep{coroniti70:_dissip}. The reflection occurs on sufficiently smaller scales than thermalization due to a combination of magnetic forces and an electrostatic cross-shock potential. The main potential, which corresponds to the frame-invariant $\vec{E}\cdot\vec{B}$ electric field, is known as the deHoffmann-Teller potential \citep{hoffmann50:_magnet_hydrod_shock, goodrich84}. It results directly from the leading electron pressure gradient term in the Generalised Ohm's Law \citep{scudder86:_2}. However in more detailed two-fluid descriptions, the quasiperpendicular shock has fine structure that depends upon the characteristics of the nonlinear shock  profile \citep{galeev88:_fine, galeev89:_mach, gedalin97:_ion_heat, krasnoselskikh02:_nonst}. 
In this paper we shall present the results of the studies of quasiperpendicular shocks only. Quasiparallel shocks will be discussed in a paper by Burgess and Scholer (this review) .

The basic ideas of the shock formation can be understood by considering the
propagation and evolution of large amplitude wave. In gas dynamics the wave corresponds to a sound wave whose evolution in terms of gas dynamics leads to
the formation of discontinuities. In reality, however, narrow transition regions are formed in which the dynamics are dominated by dissipative processes. In plasmas, the characteristics of the main shock transition are determined not only by an interplay between nonlinearity and dissipation, but also by another important physical effect, the wave dispersion.
It is well known that a subcritical shock has a nonlinear whistler wave train upstream of its front \citep{sagdeev66:_cooper_phenom_shock_waves_collis_plasm, mellott85:_subcr}. The presence of whistler/fast magnetosonic precursor wave trains in supercritical shocks as well, was experimentally established in \citet{krasnoselskikh91, balikhin97:_nonst_low_frequen_turbul_quasi_shock_front, oka06:_whist_mach}. These whistler waves have rather large amplitudes and their role in energy transformation and redistribution between different particle populations and in the formation of the structure of the shock front is still an open question. The major transition of such a dispersive shock, the ramp, may behave in a similar fashion to either the largest peak of the whistler precursor wave packet \citep{karpman73:_elect, sagdeev66:_cooper_phenom_shock_waves_collis_plasm, kennel85:_quart_centur, galeev88:_fine, galeev89:_mach, krasnoselskikh02:_nonst} or the dissipative shock region in which the major dissipation due to current driven instabilities occurs.
The nonlinear steepening process can be described as the transfer of energy to smaller scales. The steepening can be terminated either by collisionless dissipation, as described above, or by wave dispersion. Typically the dissipative scale $L_{d}$ exceeds the dispersive one $L_{disp}$, the former is reached first and further steepening can be prevented by the dissipation that takes away energy. When steepening is balanced by dissipation, a dissipative subcritical shock forms. Most subcritical collisionless shocks observed in situ are supposed to be dissipative even though dispersive processes play a role in forming a dispersive precursor wave train. Such a shock is characterized by a monotonic transition in the magnetic field (magnetic ramp) of the width $L_{d}$. The dissipative length is determined by the most important anomalous dissipative process. Its major features are the generation of intense short-scale waves and their dissipation. This form of the evolution of a nonlinear wave takes place at low Mach numbers. 
However, if the nonlinearity is strong enough (as determined by the velocity and density of the incoming flow), dissipation is not capable of stopping the steepening, and the gradients continue to grow then energy transfer to smaller scales continues and the characteristic scale of the transition can become smaller. The next process that comes into play to counterbalance the steepening is dispersion. Dispersion becomes important when the gradients become comparable with the dispersive scale $L_{disp}$. In this case the shock front structure becomes multi-scale. The steepening is prevented by short-scale dispersive waves which are able to propagate away from the evolving shock front. These waves effectively remove some part of the energy and, most importantly, restrain further growth of the gradients. For perpendicular shocks the phase velocity of the dispersive waves decreases with decreasing scale and a wave train is formed downstream of the
magnetic ramp. For shocks with a more oblique geometry (quasiperpendicular shocks) the phase
velocity increases with decreasing spatial scale and an upstream
wave train is formed. The upstream wave precursor is approximately phase
standing in the upstream flow. Its amplitude decreases with the distance
from the shock ramp due to dissipation processes as was discussed in the 
early theoretical papers describing subcritical shocks \citep{sagdeev65:_asymp, sagdeev66:_cooper_phenom_shock_waves_collis_plasm}.

The transition to reflection shock takes place when downstream bulk velocity
reaches the downstream ion-sound speed. Supercritical reflection shocks have
a more complex structure in comparison to subcritical shocks. 
In quasi-perpendicular shocks the upstream magnetic field does not allow reflected ions to travel far upstream before turning them back to the shock front. The upstream region in which the beam of reflected ions perform part of their Larmor orbit before being turned back to the shock is called foot. The foot region is characterised by a 15-20\% increase in the magnitude of the magnetic field. The consideration of a Larmor orbit of a reflected ion gives relatively accurate estimate of the spatial size of the foot $L_{f}=0.68R_{Li}\sin{\theta_{Bn}}$ where $R_{Li}$ is the gyroradius of ions moving with the velocity equal to normal component of the velocity of upstream flow \citep{woods69:_alfven_mach, livesey84}. The coefficient $0.68$ corresponds the case of to specular reflection. For non-specular reflection this relation is slightly modified \citep{gedalin96:_ion}. Downstream of the quasi-perpendicular shock's main
transition the joint gyration of the bulk plasma ions and beam of reflected ions
leads to an overshoot-undershoot structure. Again, the size of this overshoot/undershoot can be estimated in a straightforward manner in terms of the ion gyroradius.
However, the main transition layer lies between the foot and the overshoot. This is the region 
where the most dramatic changes in the plasma parameters occur. In a
supercritical, quasi-perpendicular shock this layer is characterised by the steepest
increase of the magnetic field referred to as the ramp. The change of the
electrostatic potential, reflection of ions, and electron thermalisation take
place within the ramp and its spatial scale determines the major physical processes within the shock and the mechanisms for the interaction of the shock front with the incoming electrons and ions. For instance, 
several theoretical models suggest that in the ramp of high Mach number shocks very small scale electric fields can be present \citep{krasnoselskikh85:_nonlin, galeev88:_fine, galeev89:_mach}. 
There are several critical issues regarding supercritical quasiperpendicular shock physics for which alternative explanations for the observational features of the shock front have been proposed. Theoretical considerations \citep{galeev88:_fine, galeev89:_mach, krasnoselskikh02:_nonst}, that treat the supercritical shock front as being similar to a nonlinear dispersive wave, predict that the ramp scales (gradients) should decrease with increasing Mach number, eventually reaching characteristic values as small as several electron inertial lengths $L_e=c/{\omega_p}$. Moreover, after some critical Mach number corresponding to nonlinear whistler critical Mach number whose value is approximately 1.4 times the linear whistler critical Mach number, the shock should become nonstationary. These critical Mach numbers determine the characteristic flow velocities when they become larger than the maximum velocity of a linear or nonlinear whistler wave propagating upstream along the shock front normal.  
Many computer simulations \citep{scholer04:_nonst, matsukiyo06:_mach} come to the conclusion that the thickness of the ramp is determined by the dissipative process due to either the modified Buneman instability (MBI) or the modified two stream instability (MTSI).   
Both theoretical studies and computer simulations have associated pitfalls. Theoretical models can not accurately take into consideration the presence of reflected ions whilst simulations are carried out with an unrealistic ratio of the plasma frequency to gyrofrequency that strongly changes the ratio of electric to magnetic wave fields and often with an unrealistic ion to electron mass ratio. Both theoretical models and simulations predict the transition to nonstationary dynamics. However, they strongly differ in the determination of the scales of the electric and magnetic fields in the ramp region, in the energy sources for the upstream whistler waves that form the precursor wave train, and in the characteristics of the shock dynamics when it becomes nonstationary. For these reasons experimental studies of these questions are crucial for our understanding of the physical processes in quasiperpendicular collisionless shocks.
Our Review aims to report the studies of all these questions making use mainly of Cluster measurements (adding some other data where it is necessary, in particular THEMIS data in studies of magnetic field scales of shocks).

The first critical issue we shall address is magnetic ramp width and spatial scale. The main motivation for the study of the magnetic ramp width $L_r$ is that it is this scale that determines the nature of the shock, i.e., the dominant physical processes that counteract nonlinear steepening. The shock width can be determined either by the solitary structure of nonlinear whistler slightly modified by the presence of reflected ions or by characteristic anomalous resistivity scale associated with one of instabilities mentioned above. For instance if it is indeed the case that the ramp width increases with increasing Mach number as concluded by \citet{bale03:_densit_trans_scale_quasip_collis_shock}, then the evolution of nonlinear whistler waves must be excluded from the processes that are involved in the formation of supercritical shocks. The characteristics of the major transition within the shock in which the flow deceleration and the magnetic field and electrostatic potential variations take place are determined by the interplay between nonlinearity, dispersion and dissipation. The presence of a population of reflected ions makes it difficult to construct a reliable
theoretical model based on an analytical or semi-analytical description. However, the establishment of the scales of this transition allows one to determine
the characteristics of the dominant physical processes in play. 
The ramp thickness is also crucial for a redistribution of energy between electrons and ions. An important characteristic involved in this process is the gradient scale of the transition. Two reasons cause the need in introducing this ramp gradient spatial scale. The first is the interaction between the incoming electrons and the electromagnetic field at the shock front. As was shown by \citet{balikhin93:_new, balikhin94:_kinem, gedalin95:_demag_e, gedalin95:_demag}, and \citet{balikhin98:_e_b}, an important effect of this interaction is the possible violation of adiabaticity even in the case when the width of the magnetic ramp considerably exceeds the formally calculated electron gyroradius. Two very different scenario of electron heating can occur depending upon if the conditions for adiabaticity are satisfied or violated. This effect is crucially dependent upon the ramp spatial scale. The change between adiabatic/non-adiabatic regimes  is related to the ability of the nonuniform electric field within the ramp to rectify the electron motion and increase their effective gyration radius \citep{balikhin93:_new, balikhin94:_kinem, gedalin95:_demag_e, gedalin95:_demag, balikhin98:_e_b}. The parameter that determines the transition from adiabatic to nonadiabatic motion of the electrons is the inhomogeneity of the magnetic field. The characteristic spatial scale of such a layer may be defined as the product of the change in magnetic field $\Delta B$ normalised to the upstream field $B_0$ and the spatial distance $dx$ over which this change occurs i.e. $l_{gr}=L_{Br}(B_{0}/\Delta B)$. To illustrate the effect of this parameter, one can consider two cases. For a weak shock for which $(B_{d}/B_{0})\approx 1.2$, here $B_{d}$ is the magnetic field magnitude after the shock transition (downstream), and whose ramp width is of the order 5-6 $R_{Le}$ ( electron Larmor radii) the electron motion will be adiabatic. However, in a stronger shock of similar magnetic ramp width for which the maximum magnetic field observed in the overshoot exceeds that of the upstream field by a factor 5-6 the electron behaviour becomes non-adiabatic. 
This makes it necessary to carry out the statistical study of the magnetic ramp spatial scale in addition to the ramp width (size).
The ramp width and its gradient scale are also important for the problem of stability of the ramp region of the shock front. According to \citet{krasnoselskikh85:_nonlin}, \citet{galeev88:_fine, galeev89:_mach} and \citet{krasnoselskikh02:_nonst} the nonlinear whistler wave structure becomes unstable when the characteristic gradient exceeds some critical value. It was suggested by \citet{krasnoselskikh85:_nonlin} that it takes place when the Mach number becomes equal to nonlinear critical whistler Mach number. When this happens dispersive process can no longer counterbalance the nonlinearity and the shock front overturns. Thus, the characteristic gradient scale provides a rather universal characteristic of the degree of steepness of the shock front. Thus its determination completed a comprehensive statistical study of the magnetic ramp spatial gradient scale in addition to the ramp width (size).
Many papers were devoted to the magnetic field structure of collisionless shocks \cite[e.g.][]{russell83:_multip, krasnoselskikh91,farris91, newbury96:_obser, hobara10:_statis, mazelle10:_self_refor_quasi_perpen_shock}. In particular, the spatial scales of its various regions have been comprehensively investigated \citep{balikhin95,
 farris93:_magnet, hobara10:_statis, mazelle10:_self_refor_quasi_perpen_shock}. ISEE and AMPTE measurements of the magnetic field profiles of the shock front structure led to evaluation of the scale sizes of the foot and overshoot regions that were supposed to be of the order of ion inertial length $c/{\omega}_{pi}$ and $3c/{\omega}_{pi}$ respectively, here ${\omega}_{pi}$ is the ion plasma frequency. The ramp scale has been estimated to be less than an ion inertial length with reports of one or two very particular shocks whose ramp scale was sufficiently smaller, of the order $0.1c/\omega_{pi}$ \citep{newbury96:_obser, walker99:_obser_very_thin_shock}. We report here the statistical studies of scales based on papers by \citet{hobara10:_statis} and \citet{mazelle10:_self_refor_quasi_perpen_shock}.
Another critical issue we shall address hereafter is the electric field distribution inside the ramp region. The energy transfer to smaller scales due to steepening can achieve the scales comparable to electron inertial scale where the whistler waves become quasi-electrostatic. In nonlinear-dispersive scenario of the shock front description the field can have multiple short scale electric field spikes. The experimental study can answer the question do they exist or not and to determine their characteristics. 
Studies of the magnetic field profile across the terrestrial bow shock significantly outnumber those based on electric field measurements. Despite the fundamental effect that the electric field has on the plasma dynamics across collisionless shocks, the complexity of the interpretation of electric field data has impeded studies of the electric field structure within the shock front. It is worth noting that only a handful of studies are dedicated to the electric field structure within the shock front \citep {heppner78:_early_isee, formisano82:_measur, scudder86:_1, wygant87:_elect, balikhin02:_obser, walker04:_elect, balikhin05:_ion, bale07:_measur, hobara08:_clust, dimmock12:_mach_clust, bale08:_direc}
In contrast, there have been very few reports regarding the scale lengths of features observed in the electric field at quasi-perpendicular shocks. The scale size over which the potential varies at the front of a quasi-perpendicular bow shock is an issue that requires resolving in order to gain a full understanding of the physical processes that are occurring in the front. Several different points of view have been published on the relationship between the scale size of the magnetic ramp and that over which the change in potential is observed. Some studies \citep{eselevich71:_isomag, balikhin93:_new, formisano82:_ion, formisano82:_measur, formisano85:_collis, balikhin02:_obser, krasnoselskikh85:_nonlin, leroy82, liewer91:_numer, scholer03:_quasi} have proposed that the spatial scale of electrostatic potential is of the same order or smaller than that of the magnetic ramp under certain conditions. Such shocks have been observed in numerous experimental and numerical studies of quasi-perpendicular supercritical shocks. On the other hand \citet{scudder95:_ae_review_physic_elect_heatin_collis} claimed that the potential scale length is larger than that of the magnetic scale length.
    Actual measurements of the electric field variations within the bow shock are very sparse. The main reason for this is due to the difficulties encountered when making electric field measurements. Only a small number of space-based measurements of the electric field during the shock front crossing have been reported. \citet{heppner78:_early_isee} reported observations of a short lived spikes in the electric field making use of ISEE measurements. However, being short duration, these features were not observed at every shock crossing. Subsequent investigations by \citet{wygant87:_elect} have shown the existence of spike-like features in the electric field both at the shock ramp and in the region just upstream. From the study of spin averaged ISEE-1 data, \cite{formisano82:_measur} determined that the increase in the observed electric field E intensity began just upstream of the magnetic ramp and lasted longer than the ramp crossing itself. Whilst the E-field intensity in the regions upstream and downstream of the shock could be interpreted as due to the $V \times B$ motion of the plasma the enhancement observed during the shock crossing must be due to the processes occurring within the shock front itself.
    In laboratory experiments where the conditions are not exactly the same as in space plasmas \citet{eselevich71:_isomag} reported that the major change in potential  across the shock occurs within the magnetic ramp region. 

Using data of numerical simulations, \citet{lembege99:_spatial_sizes_elect_magnet_field} analysed the scale size of both the magnetic ramp region $L_{Br}$ and the scale of the major change in potential $L_{\phi}$ inside and around the ramp. They concluded that the scale lengths were of the same order, i.e. $L_{Br} \approx L_{\phi}$. This view is supported by the simulations of \cite{scholer03:_quasi}. The latter authors also show that during the shock reformation process, the main potential drop occurs over several ion inertial scales in the foot region and they noticed that the steepened magnetic ramp region also contributes a significant fraction of the change in total potential over much smaller scales, typically 5-10 Debye lengths. Despite the simulation shocks parameters are still rather far from observations (see Section~\ref{sec:electricscales} for more details) the tendencies in majority of simulations are well pronounced and are similar to those in laboratory experiments.  
Hereafter we report the observations of electric field spikes observed onboard Cluster satellites first reported by \citet{walker04:_elect} and the statistical study of their characteristics.
The third important problem of quasiperpendicular shock physics addressed in this Review is the problem of electron heating. By contrast to ion heating problem well advanced due to detailed studies onboard ISEE mission, the electron heating problem has remained controversial. The action of shock quasistatic electric and magnetic fields on the electron population (which can have thermal speeds far in excess of the shock speed) is to inflate and open up a hole in the phase space distribution by accelerating (decelerating) incoming (escaping) electrons \citep{scudder86:_3, feldman82:_plasm_isee}. This inflation in itself is reversible thus it is not dissipation or heating if other processes would not be involved. Irreversibility may be imposed if additional scattering would take place infilling the hole. If adiabatic invariant of electrons is conserved while electrons cross the shock front it can not happen. One should conclude that some non-adiabatic process should occur inside the shock front. One of the possibilities can be related to Debye-scale electric fields \citep{bale98:_bipol}. Another possibility is to suggest that the phase space inflation is indeed accompanied by instabilities which could scatter the electrons. Demagnetisation of the electrons due to the strong gradients in the electric field as it was mentioned above \citep{balikhin94:_kinem} or nonlinear wave phenomena \cite{krasnoselskikh02:_nonst} combined with wave particle interaction can offer alternative scattering processes.
Thus the partition of energy between ions and electrons is a complex, self-consistent multi-scale interplay between electron heating, magnetic/electric field profile, shock potential, and ion reflection. This interplay remains poorly understood despite 40 years of research. That research has included detailed case studies \citep{scudder86:_2}, statistics of the inferred potential and electric field structures \citep{schwartz88:_elect, walker04:_elect}, theoretical studies \citep{galeev88:_fine, gedalin97:_ion_heat, krasnoselskikh02:_nonst} and increasingly sophisticated numerical simulations \citep{lembege04:_selec_probl_collis_shock_physic, scholer06:_trans}.
Direct measurements of the thickness of the shock transition layer combined with the rapid simultaneous measurements of the electron distribution function can allow solving this long standing opened problem in shock physics. If the electron heating can be attributed to kinetic instabilities, the shock thickness will be measured in ion inertial lengths ($c/\omega_{pi}$) \citep{papadopoulos85:_microin, matsukiyo06:_mach}. If such instabilities prove ineffective, above a second critical Mach number the shock steepening is expected to be limited by whistler dispersion and/or be unstable to shock reformation \citep{krasnoselskikh02:_nonst}. Recent studies of the shock thickness \citep{hobara10:_statis, mazelle10:_self_refor_quasi_perpen_shock} do show scales comparable to whistler wavelengths. These contrasted an earlier study \citep{bale03:_densit_trans_scale_quasip_collis_shock} reporting scalings that matched the gyro-scales of reflected ions.
To date, studies have relied on the high temporal cadence available from magnetic or electric field experiments. However, field profiles provide only indirect evidence of the shock dissipation scales. A recent study \citep{lefebvre07:_elect} used sub-populations of electrons to determine the electrostatic potential profile at one shock, suggesting that it rose in concert with the magnetic field. In the work reported here, first published in \citep{schwartz11:_elect_temper_gradien_scale_collis_shock}, the electron distribution function major characteristics are measured at sufficient cadence to reveal directly for the first time the scale of the electron temperature profile. Many shock crossings by Cluster satellites take place on the flanks of the magnetosphere that creates quite favourable conditions for the studies of the relatively narrow shock transitions allowing one to have many measurements on small spatial scale. Hereafter we present unprecedently rapid measurements of electron distribution moments that allow to shed new light on electron heating problem and its scales.

The fourth problem presented in the Review, is closely related to the problem of magnetic and electric field scales, is determination of the source of waves forming upstream precursor wave train. The presence of whistler/fast magnetosonic precursor wave trains in supercritical shocks was experimentally established in \citet{balikhin97:_exper, krasnoselskikh91, oka06:_whist_mach}. These whistler waves have rather large amplitudes and their role in energy transformation and redistribution between different particle populations and in the formation of the shock front structure is still an open question. The energy source responsible for the generation of these waves is the subject of active debate in shock physics \citep[see][]{galeev88:_fine,galeev89:_mach, krasnoselskikh02:_nonst, matsukiyo06:_mach, comisel11:_non}. Often the precursor waves are almost phase-standing in the shock frame. However, if they are generated by the ramp region as the dispersive precursor their group velocity can still be greater than zero in the shock reference frame, which would allow energy flow in the form of Poynting flux to be emitted towards the upstream of the shock transition. On the other hand, if the waves are generated by instabilities related to reflected ions their energy flux will be directed from the upstream region towards the shock ramp. The goal of Section~\ref{sec:upstr_waves} is to address this problem, to present the direct measurement of the Poynting flux of the upstream whistler waves aiming to establish the direction of the Poynting flux. There are two different points of view on this subject also.
It has been suggested that the shock front structure of quasi-perpendicular supercritical shocks is formed in a way similar to that of subcritical shocks \citep{galeev88:_fine, galeev89:_mach, krasnoselskikh02:_nonst}. In such scenario the precursor wave train is a part of the shock front structure emitted by the ramp region upstream due to positive dispersion of whistler waves.
The observed dynamic features of shocks have also been studied extensively using computer PIC- or hybrid simulations, often with focus on the precursor wave activity and reflected ions \citep{hellinger97:_upstr, hellinger07:_emiss, matsukiyo06:_mach}. 
From a kinetic viewpoint, however, it may be argued that the shock-reflected ions change the physical picture and that the principal scales, temporal and spatial, could be determined by the characteristics of the reflected ion population \citep{bale03:_densit_trans_scale_quasip_collis_shock}. Upstream waves can then be generated due to counterstreaming ions and electrons in the shock front region, forming unstable particle distributions with respect to some wave modes \citep{papadopoulos85:_microin, hellinger07:_emiss, scholer04:_nonst, matsukiyo06:_mach}. 
While this is probably the case for some higher frequency waves, we present here an analysis that leads to the conclusion that the source of the upstream low frequency whistler waves is indeed related to the presence of the nonlinear ramp transition, emitting smaller scale dispersive waves towards the upstream flow. 
The existence of phase-standing upstream whistler waves depends on the value of the upstream flow speed Mach number relative to the phase velocity. If the Mach number of the shock does not exceed the nonlinear whistler critical Mach number $M_{w} = V_{w,max}/V_A = 1/2\sqrt{m_i/m_e}\cos \theta_{Bn}$, where $V_{w,max}$ represents the highest possible velocity of nonlinear whistler wave, then phase-standing (nonlinear) whistler wave trains can exist upstream of the shock \citep{galeev88:_fine,galeev89:_mach, krasnoselskikh02:_nonst}. The results we report here were first published in \citep{sundkvist12:_disper_natur_high_mach_number}. Similar results were reported making use the Time Domain Sampling instrument (TDS) onboard Wind satellite \citep{wilson12:_obser} for three of four crossings of interplanetary shocks. In one case the polarization of waves was found to be different from whistler wave. Unfortunately, one satellite measurements do not allow to establish unambiguously the reason of this anomaly, it could associated with some particular perturbation in the solar wind.  

The fifth problem intimately related to previous is the problem of nonstationary dynamics of high Mach number shocks. 
Shock waves are usually considered to be nonlinear waves that cause irreversible changes of state of the media and from macroscopic point of view they are stationary (for a review, see, e.g., \cite{tidman71:_shock}. However, in the very beginning of the collisionless shock physics \citet{paul67:_measur_elect_temper_collis_shock} hypothesized that high-Mach-number shocks can be nonstationary, and the first unambiguous evidence of the nonstationarity was obtained by  \citet{morse72:_nonst_behav_collis_shock} in  laboratory experiments. New evidence of shock front nonstationarity was found in the 1980s. In particular, \citet{vaisberg84:_relax} reported low frequency oscillations of the ion flux in the Earth's bow shock. Later \citet{begenal87:_uranian_voyag_mach}  observed a similar phenomenon in the Uranian bow shock. In the very beginning of computer simulations of the collisionless shocks \citet{biskamp72:_ion_heatin_high_mach_number} have observed the process of shock dynamic behaviour. The inflowing ions formed vortices in the phase space and dynamics of the front was definitely nonstationary.  Later,  numerical simulations performed by \citet{leroy82} using 1-D hybrid code showed that the front structure of perpendicular shocks varies with time, for instance, the maximum value of the magnetic field exhibits temporal variations with a characteristic time of the order of ion gyroperiod, the magnitude of these variations being about of 20\% if the parameters are typical for the Earth bow shock ($M_A = 8$ and ${\beta}_{e,i} = 0.6$, where $M_A$ is the Alfv\'{e}n Mach number, $\beta_{e,i}$ is the ratio of the thermal electron/ion and magnetic pressures). They also found that for $M_A = 10$ and $\beta_{e,i} = 0.1$ the ion reflection was bursty, oscillating between 0 and 70-75\%. Hybrid simulation of perpendicular shocks with very high Mach numbers carried out for the first time by \citet{quest86:_simul_mach} have shown that the ion reflection in the shocks can be periodic, the stages with 100\% ion reflection alternating with the stages of 100\% ion transmission. As a result, instead of a stationary structure, he observed a periodic wave breaking and shock front reformation. Later \citet{hellinger02:_refor}  reexamined the properties of perpendicular shocks with the use of the 1-D hybrid code and observed the front reformation for a wide range of parameters if upstream protons are cold and/or Mach number is high. \citet{scholer03:_quasi} and  \citet{scholer04:_nonst, matsukiyo06:_mach} in their 1-D full-particle simulations with the physical ion to electron mass ratio reproduced the reformation of exactly and approximately perpendicular high-Mach-number shocks in plasmas with $\beta_i = 0.4$ and demonstrated an importance of modified two-stream instability for the reformation process.
\citet{krasnoselskikh85:_nonlin} and \citet{galeev88:_fine, galeev89:_mach} proposed models describing the shock front instability due to domination of nonlinearity over dispersion and dissipation. This instability results in a gradient catastrophe within a finite time interval. Several aspects of the model, including the role of nonlinear whistler oscillations and existence of a critical Mach number above which a nonstationarity appears, were developed in further detail and more rigorously by \citet{krasnoselskikh02:_nonst}  and complemented by numerical simulations with the use of the 1-D full particle electromagnetic code with a relatively small ratio of electron and ion masses, ${m_e}/{m_i} = 0.005$. It was also shown that the transition to nonstationarity is always accompanied by disappearance and re-appearance of the phase-standing whistler wave train upstream of the shock front. Moreover, for large Mach numbers the nonstationarity manifests itself as a periodic ramp reformation, which influences considerably the ion reflection, in particular, the reflection becomes bursty and sometimes the ions are reflected from both old and new ramps simultaneously.
The four-spacecraft Cluster mission gave much more new opportunities for experimental studies of the shocks. The first examples of some aspects of shock nonstationarity were presented by \citet{horbury01:_clust}. These authors analyzed magnetic field data for two quasiperpendicular shocks with moderate and high Alfv\'{e}n Mach number. While for moderate $M_A$ the shock profiles measured by different spacecraft were approximately the same, with the exception of a small-amplitude wave activity in the foot, for high $M_A$ the amplitude of the fluctuations attains ~10 nT, making profiles considerably different for different spacecraft. However, the authors argued that these fluctuations stop before the ramp and do not appear to disrupt the shock structure; on the other hand, they didn't reject an opportunity that the fluctuations observed may represent the signatures of the unsteady shock reformation. Hereafter we report the first direct observation that clearly evidence the shock front reformation observed onboard Cluster mission on 24th of January 2001. This material was first published in \citet{lobzin07:_nonst_mach}.

The sixth problem we discuss in this Review is important for the definition of the relative role of dissipative and dispersive effects, namely the problem of anomalous resistivity. The problem of electron heating mentioned above is considered for many years to be 'solved' for subcritical shocks and conventional solution proposed and widely accepted is formulated in terms of magic words 'anomalous resistivity'. This notion introduced first by \citet{sagdeev65:_landau} and then analyzed in more detail by \citet{galeev76:_collis}, who made estimates of the characteristic scale of the shock transition relying on ion-sound instability. \citet{papadopoulos85:_microin_anomal_trans_collis_shock} has noticed that in case of quasiperpendicular shocks the most important instabilities should be related to lower hybrid waves and has revised the model taking these effects into account. However there were no any measurements that might be used to confirm or reject theoretical models of the dissipation due to anomalous resistivity. It is worth noting that this problem is very important for the determination of energy redistribution between electrons and ions, especially for the electron heating and electron acceleration. We can not present here theoretical studies of anomalous resistivity, interested reader can find general ideas in the review papers by \citet{galeev89:_basic} and \citet{papadopoulos85:_microin_anomal_trans_collis_shock}. 
The second important problem where the short scale length waves are important is an energization of electrons within a collisionless shock. It requires the transfer of a portion of the energy associated with the incoming upstream plasma flow to the electron population. In order for this energy transfer to occur, there has to be some media that can channel energy from the incoming ion population to the electrons. One mechanism that has commonly been proposed, both for solar systems and particularly for astrophysical applications is based on excitation of lower-hybrid waves \citep{papadopoulos81:_elect, papadopoulos81:_elect, laming01:_accel, krasnoselskikh85:_fast_ii}.
The increased level of electric field fluctuations in the frequency
range $10^2-10^3$Hz observed in the vicinity of a quasiperpendicular 
shock front is usually attributed to either ion-sound or whistler waves. One of the most comprehensive studies of the plasma waves in this frequency range was conducted by \citet{gurnett85:_plasm}. It's main conclusion was that waves observed above local electron cyclotron frequency are Doppler shifted ion-sound waves whilst those below are whistler mode waves.

The main reason of the lack of measurements of ion sound and lower hybrid waves is related to technical difficulties of small scale electric field measurements. Recently two papers were published that represent first attempts to create the experimental basis for the anomalous resistivity studies in collisionless shocks. We present in this Section a short summary of the results obtained following to \citet{balikhin05:_ion,walker08:_lower}.   
The rapid changes that are observed in the plasma at the front of a supercritical, quasi-perpendicular shock and described in previous Sectons lead to the creation of multiple free energy sources for various plasma instabilities. Twin satellite missions, such as ISEE or AMPTE, have provided data for a number of comprehensive surveys of the waves observed in the frequency range
($10^{-2}-10^{1}$Hz) of the plasma turbulence encountered at the shock
front. The use of multisatellite data for wave identification and
turbulence studies is limited to the analysis of those waves whose
coherence lengths are of the same order of magnitude as the satellite
separation distance. Plasma wave modes such as the ion-sound or
Lower-Hybrid, that are supposed to play an important role at the shock front, possess coherence lengths that are very short in comparison with any realistic satellite separation distance \citep{smirnov87:_eviden}. For
majority of these waves the coherence length is either comparable to or a few
times greater than their wavelength. In such cases the waves observed by
different satellites in a multisatellite mission will be uncorrelated.
This will make it impossible to apply wave identification methods based on
intersatellite phase delays \citep{balikhin97:_exper, balikhin03:_minim} or k-filtering \citep{pincon08:_multi_spacec_method_wave_field_charac}. Nevertheless the identification of waves with short wavelengths and study of their dynamics remains very important because of their potential role in the transfer of energy associated with the upstream directed motion into other degrees of freedom. In the classical model of a quasiperpendicular low $\beta$ shock anomalous resistivity occurs due to ion-sound turbulence in the shock front \citep{galeev76:_collis}. Lower hybrid waves also may play an important role at the shock front since they also can be involved in resonance interactions both with electrons and ions and so may be extremely efficient at channelling the energy exchange between the two spices. In order to assess the importance of ion-sound, lower hybrid and other waves with relatively short wavelengths within the plasma dynamics of the shock front the mode of the observed waves should first be identified. The strong Doppler shift that results from the large values of wavevector $|k|$ precludes the reliable use of the observed frequency for correct identification as was done in many previous studies. Here we show that the data from a single spacecraft can be used to determine propagation modes of waves observed in the frequency range $10^2-10^4$ Hz at the front of the terrestrial bow shock. A similar approach has been used by \citet{tjulin03:_lower} in a study of lower-hybrid waves in the inner magnetosphere.

The lower-hybrid wave is an electrostatic plasma wave mode whose plasma frame frequency is in the vicinity of the lower-hybrid resonance frequency 
$\omega_{lh}\sim\sqrt{\Omega_{ci}\Omega_{ce}}$ where $\Omega_{ci}$ and 
$\Omega_{ce}$ are the ion and electron gyrofrequencies respectively. The wave 
has linear polarisation and propagates almost perpendicular to the magnetic field ($\cos (\theta_{kB})\sim \sqrt{m_{e}/m_{i}}\sim 89^{\circ}$). The 
maximum growth rate $\gamma_{MAX}$ occurs when $k_{||}/k \sim 
\omega_{pi}/\omega_{pe}$, where $k_{||}$ is parallel component of the wavevector. 
Since the waves are propagating in a plasma that is moving with respect to the satellite, their frequencies will be Doppler shifted in the spacecraft frame. The magnitude of this shift can be estimated using the resonance condition of the Modified Two Stream Instability (MTSI) $2V_{A}M_{A}k=\omega_{lh}$. This gives a maximum estimate for the correction in observed wave frequency due to the Doppler shift $kV_{sw}\sim\omega_{lh}/2$, here $V_{sw}$ is the velocity of the solar wind supposed to be equal to normal component of the upstream flow velocity. 

Current models of wave turbulence that determines anomalous resistivity in the front of quasiperpendicular shocks are based on the occurrence of 
lower hybrid waves at a shock front being generated due to counter-streaming populations of ions and relative motion of reflected ions \citep{leroy82} and bulk electrons and ions at the front via the modulational two-stream instability (MTSI) \citep{papadopoulos85:_microin, matsukiyo06:_mach} or modified Buneman instability. These models are often used to explain the electron acceleration observed at various astrophysical shocks such as supernova remnants \citep{laming01:_accel}. However, there is currently no substantial experimental evidence that these waves do indeed exist in the fronts of supercritical, quasiperpendicular, collisionless shocks. The results of data analysis from the Intershock electric field experiment, in which wave activity was observed at frequencies of a few Hertz, has been used to argue for the existence of lower hybrid waves \citep{vaisberg83:_elect}.
An alternative explanation, however, has been proposed in which Intershock may have simply observed the electric field component of whistler wave packets propagating in the foot region \citep{krasnoselskikh91, balikhin97:_exper, walker99:_ramp}. Electric field observations of Comet Halley also showed evidence for waves observed in the vicinity of the lower hybrid frequency \citep{klimov86:_extrem_halley}. However, their exact wave mode was not determined.
The natural way to differentiate these modes is to examine their polarisations. Whistler mode waves are elliptically polarised whilst lower hybrid waves, as mentioned above, are linearly polarised. We present here the summary of direct observations of the ion-sound \citep{balikhin05:_ion} and lower-hybrid waves \citep{walker08:_lower} that we complet by estimates of characteristic effective collision frequencies.
The paper is organized as follows.
In Section~\ref{sec:magneticscales} we present the statistical studies of quasiperpendicular shock ramp widths. Section~\ref{sec:electricscales} is dedicated to electric field scales of the ramp of quasiperpendicular shocks. In Section~\ref{sec:upstr_waves} the results of evaluation of the Poynting flux of oblique whistler waves upstream of the shock front are outlined. Recent results on electron heating scale at High Mach number quasiperpendicular shocks are presented in Section~\ref{sec:elec_heat}. In Section~\ref{sec:anom_res} we use the data of measurements of lower hybrid and ion sound waves intensities to evaluate the characteristics of anomalous resistivity aiming to determine its role in the shock front formation. In Section~\ref{sec:nonstationarity} we present results of direct observations onboard Cluster satellites of nonstationarity and reformation of high-Mach number quasiperpendicular shock. In Section~\ref{sec:conclusions} we resume the results of experimental observations and discuss the conclusions that follow from Cluster and THEMIS observations. In Appendix~\ref{sec:simulations} we present some short comments concerning comparison of computer simulations with the observations. Appendix~\ref{sec:notation} contains a notation table defining the parameters used in this paper.

\section{Statistical studies of quasi-perpendicular shocks ramp widths}
\label{sec:magneticscales}

\begin{figure}
\includegraphics[width=0.6\textwidth]{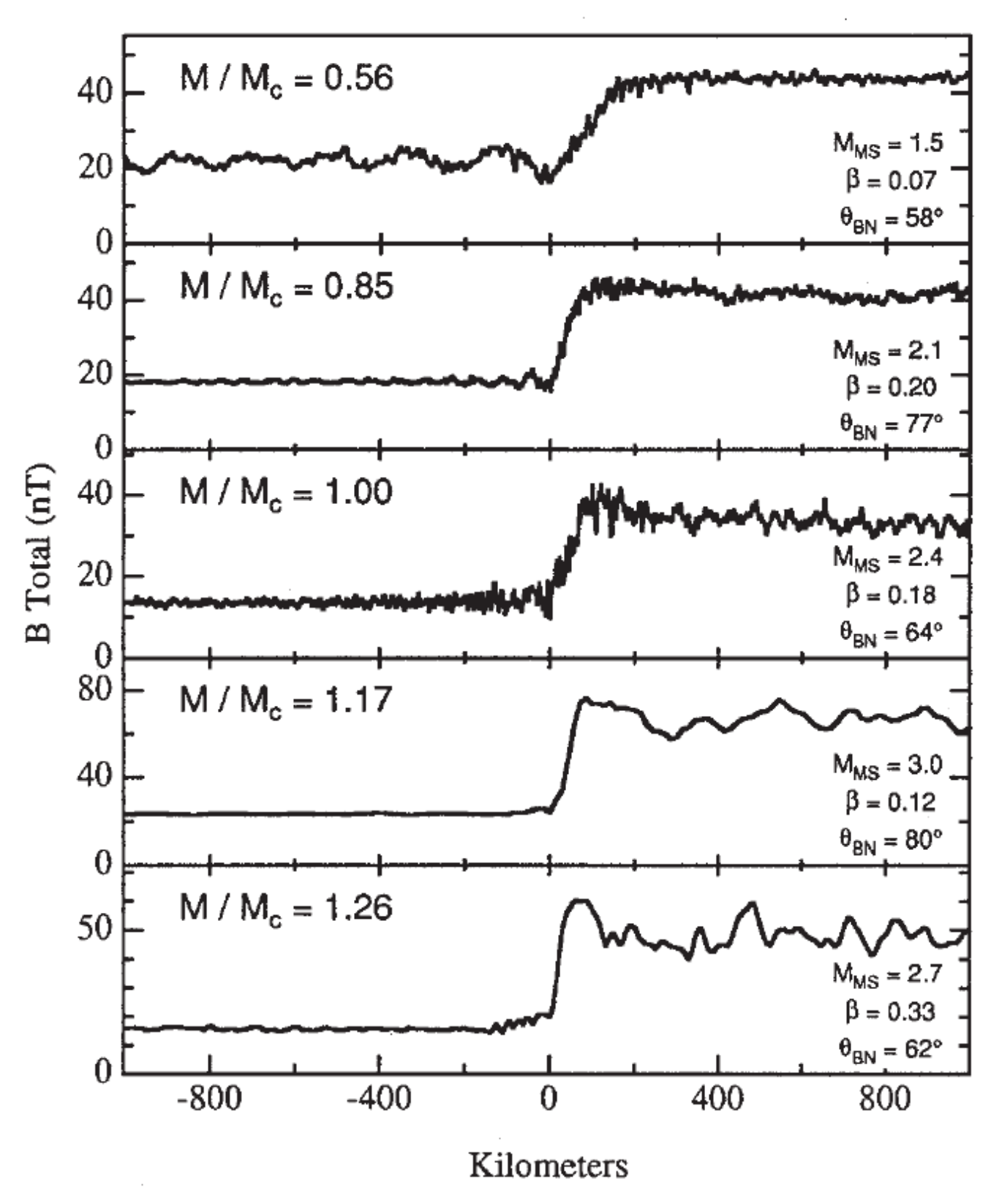}
\caption{The total magnetic field strength as a function of distance
through five low beta, quasi-perpendicular shocks in order of increasing ratio of criticality. The Mach number, $\beta$, and $\theta_{BN}$ for each shock are displayed. The data are shown at the highest temporal resolution available. The sampling rate for the first three shocks is 16 Hz and 4Hz for the last two \citep[Adapted from][]{farris93:_magnet}.}
\label{fig:MS1}       
\end{figure}

As was discussed in the Introduction the characteristics of the major transition within the shock in which the flow deceleration and the magnetic field and
electrostatic potential variations take place are determined by the
interplay between nonlinearity, dispersion and dissipation. The presence of
a population of reflected ions makes it difficult to construct a reliable
theoretical model based on an analytical or semi-analytical description. However, the establishment of the scales of this transition allows one to determine the characteristics of the dominant physical processes in play. Single
satellite missions provide very poor possibilities for the reliable
identification of the shock width and evaluation of the characteristic scales
of structures within it. In such cases the spatial size of the foot or overshoot have been used \citep{balikhin95} to evaluate the thickness of the ramp region. 
The first shock crossings by two satellites were studied in the frame of ISEE and AMPTE projects. These missions provided the first insight into the thickness of the shock transition. 
The decrease of the shock width and consequent increase of gradients
as a function of increasing Mach number was clearly demonstrated by \citet{farris93:_magnet}. In their study the Mach number was normalized to the critical Mach number that determines the transition from sub-critical to supercritical shocks. Figure~\ref{fig:MS1}  \citep[From][]{farris93:_magnet} shows magnetic field magnitude measurements made by ISEE 1 for five low beta quasiperpendicular shocks ordered by the ratio of their Mach number to critical Mach number increasing from top to bottom. The increase in the gradient of the
shock transition is clearly independent of the differences in the angles between the normal to the shock front and the magnetic field and of the value of beta.

\begin{figure}
\includegraphics[width=0.6\textwidth]{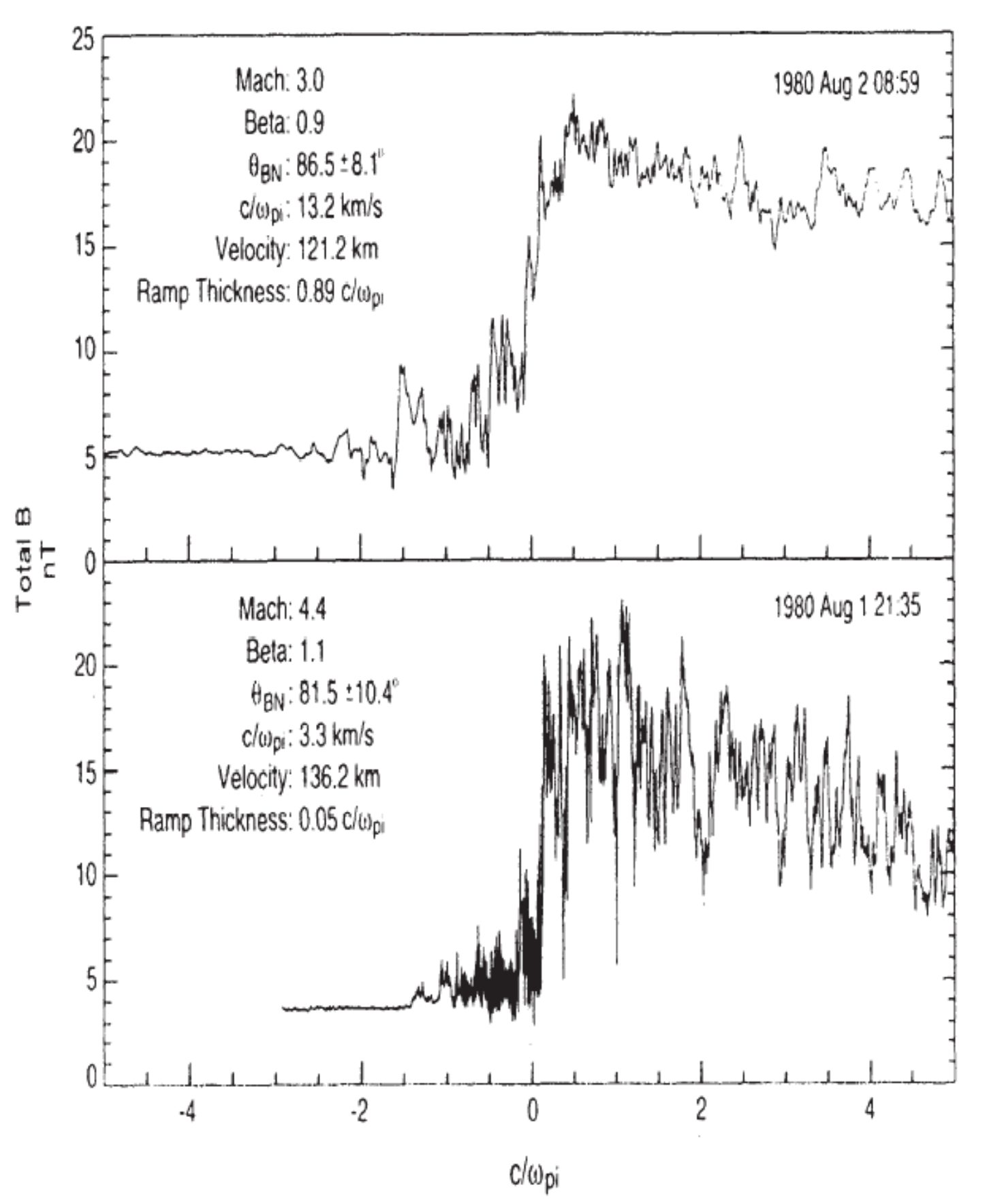}
\caption{These shocks formed under similar solar wind conditions, but there is great disparity between their ramp widths \citep[Adapted from][]{newbury96:_obser}.}
\label{fig:MS2}       
\end{figure}

Several studies have been dedicated to the investigation of structural elements of the shock front making use of ISEE 1,2 magnetometer data. Scudder and co-authors \citep{scudder86:_1, scudder86:_2, scudder86:_3} carried out the detailed study of the shock crossing on 7th November 1977.
 This is presumably the most detailed study of a single event, in which
all the elements of the structure were put together and compared with detailed measurements of the particle distribution functions. These authors succeeded
in relating the evolution of the ion distribution function to the
characteristic features of the magnetic field structure and in the determination
of the major macro-features of structure of the shock front. This study concluded that the size of the magnetic field transition was determined by the dissipative process related to reflection of ions. 
Twin satellite measurements by ISEE provided the first indications that some shocks have quite narrow fronts 
 \citep{newbury96:_obser, newbury98:_mach}. \citet{newbury96:_obser}  reported the observation of an extremely thin,  quasiperpendicular shock whose ramp width was determined to be $0.05L_i$ (where $L_i$ is ion inertial length), i.e. of the order of electron inertial length.  Figure~\ref{fig:MS2} shows this particular shock crossing (bottom panel) together with a second shock, observed  
under similar solar wind conditions, whose ramp width was $0.89L_i$.

Cluster and THEMIS have provided new opportunities for a comprehensive study of the shock ramp scales.
Recently, there have been two papers dedicated to studies of the ramp
scales of the magnetic field transition \citep{hobara10:_statis, mazelle10:_self_refor_quasi_perpen_shock}. They have used slightly different definition of the ramp thickness
and scale and applied different methodology, however they came to very
similar conclusions about statistical characteristics of the scales of the
ramp transition. Hereafter we present the summary of results reported in these papers.

The  article by \citet{hobara10:_statis} is devoted to statistical studies of the
magnetic field spatial scales in the ramp region of the shock front based on
Cluster and THEMIS observations. Due to their highly inclined orbit, the Cluster satellites enable the observation of shock crossings away from ecliptic plane. These shocks typically exhibit Mach numbers that are in the lower range of the whole space of terrestrial Mach numbers, since the shock normal deviates from the sunward direction. To increase the range of available Mach numbers, THEMIS shock crossings were added to the set of Cluster observations. Magnetic ramps cannot be always treated as uniform. Nonlinear substructures have been observed and reported within the ramp in several cases \citep[e.g.][]{balikhin02:_obser, walker04:_elect}. The study of spatial-temporal characteristics of such
substructures requires at least two point measurements separated by a
distance that is sufficiently smaller than the inter satellite distances of both
THEMIS and Cluster missions, thus the authors restricted themselves with the study of the ramp spatial scales only.

\subsection{Criteria for Choosing Shocks and Definition of Notions
\textquotedblleft Size\textquotedblright\ and \textquotedblleft
Scale\textquotedblright\ }

\citet{hobara10:_statis} have used the data from Cluster and THEMIS for a statistical study of the spatial size of the ramp. Both these missions assembled a huge stockpile of shock crossings. These data sets complement each other because of the difference in the orbits of Cluster and THEMIS satellites.
The THEMIS orbit is close to equatorial plane providing an opportunity to
sample the terrestrial bow shock in the vicinity of the subsolar point.
Cluster crossings of the terrestrial bow shock occur mainly on the flanks.
The solar wind flow in the vicinity of the terrestrial orbit is almost along Sun-Earth line, so that the Mach numbers of flank shocks are relatively low due to the greater deviation of the local shock normal from the sunward direction. Therefore, the combination of THEMIS and Cluster crossings allowed to cover a greater dynamical range of Mach numbers available for the analysis than each of these missions provides separately. Cluster crossings for two time intervals, February--April 2001 and February--March 2002, were used in the study. THEMIS shock crossings included in the study took place from the beginning of July 2007 to the end of August 2007. The magnetic field data used in the present paper came from Cluster and THEMIS fluxgate magnetometers (FGM) \citep{balogh97:_clust_magnet_field_inves, auster08:_themis_fluxg_magnet}. Another reason to use the THEMIS data from the initial phase of the mission was that THEMIS C and D spacecraft separation was not very large.

The set of shocks that have been used by \citet{hobara10:_statis} for the study of statistical properties of the ramp width and gradient scale in the paper included 77 individual crossings of the terrestrial bow shock (30 by THEMIS and 47 by Cluster).  
In order to determine the spatial scale of the shock ramp by means of transformation from temporal to spatial variables an estimate of the relative shock spacecraft velocity along the shock front normal was used. These estimates are very sensitive to shock normal definition. That’s why to perform reliable identification of the local normal to the shock front one of them \citep{hobara10:_statis} compared normals found making use of four different methods, using the model shape of the terrestrial bow shock similar to \citet{farris91}, using timing differences methods between the 4 Cluster spacecraft shock crossings, minimum variance and coplanarity theorem. In order to validate the results the evolution of the magnetic field component along the normal direction $B_{n}$ was used. The reliability of the shock normal identification served as the only shock selection criteria. Those shock crossings, for which the calculated normal could not be considered reliable (because of the $B_{n}$ evolution or due to large discrepancy in the shock normal directions found by different methods), have been excluded from consideration. The relative shock spacecraft velocity $V_{ss}$ has been calculated using the shock normal direction, satellite separation vectors and time delay between two subsequent shock crossings.

The second recent  study dedicated to the same problem \citep{mazelle10:_self_refor_quasi_perpen_shock}  was also based on Cluster magnetic field measurements during spring seasons of 2001 and 2002 corresponding to small interspacecraft separation (100 to 600 km typically). The shock parameters (angle between upstream magnetic field and local shock normal $\theta _{Bn}$, Alfv\'{e}nic Mach number $ M_{A}$ and ion beta $\beta _{i}$) were computed from the data of Cluster and ACE.

\citet{mazelle10:_self_refor_quasi_perpen_shock} selected the shocks for the statistical study according to following criteria. First, in order to restrict the study by almost perpendicular shocks the shocks with $\theta_{Bn}$ as close as possible to $90^{\circ}$ were chosen. The other criteria were the existence of well-defined upstream and downstream intervals for the 4 s/c, the stability of the upstream averaged field from one s/c to another, the validity of the normal determination by checking the weak variability of the
normal field component $B_{n}$ around the ramp and low value of $B_{n}$ upstream for $\theta _{Bn}$ to be close to $90^{\circ}$. Only 24 from 455 crossings of Cluster satelite quartet in 2001 and 2002 were left with all criteria validated. This means that 96 individual shock crossings were analyzed. Selected $\theta _{Bn}$ values were chosen to be in the range from nearly $90^{\circ}$ to $75^{\circ}$ but about 80\% of the shocks selected were above 84$^{\circ}$. Mach numbers $M_{A}$ were found to be equally distributed from 2 to 6.5 and corresponding $\beta _{i}$ between very low values to 0.6 but with 67ø of values less than 0.1.

The major difference between two works consists in using different methods of the determination of the shock ramp thickness. If the beginning of the ramp region was defined quite similarly as the beginning of the monotonous increase of the magnetic field, the end of ramp or exit from the ramp region was determined differently.
\citet{hobara10:_statis} defined the ramp crossing duration as a time interval between the upstream edge of the ramp and the maximum of overshoot.
The spatial size (width) of the magnetic ramp has been estimated as a
product of $V_{ss}$ and the duration $D_{t}$ of the ramp crossing. 
\citet{mazelle10:_self_refor_quasi_perpen_shock}  determined at first a stationary asymptotic level of the magnetic field that might be considered as satisfying to Rankine-Hugoniot conditions. To this end a downstream interval where the magnetic field magnitude is quite steady was used. It is then considered as giving an approximate estimate of the value of the magnetic field corresponding to exit from the ramp/entry in the overshoot. From the initial values of the entry in and exit from the ramp a linear fit of the data points inside the estimate time interval is made and this later one is allowed to vary. The choice of a linear fit allows excluding any pollution of the ramp region by a part of the extended foot. The same analysis was repeated for all four satellites for each shock crossing. The steeper slope found for the ramp defines the 'reference satellite'. The times of the middle of the four samples of ramps for one shock crossing are then used to compute both the shock normal and velocity in the GSE frame by the `timing-method' described in \citet{horbury02:_four}.
This makes it possible to derive the 'apparent' width (along each satellite
trajectory) and compare between the 4 s/c. Then, the velocity vector of the
shock in each s/c frame is computed. Its angle with the shock normal allows
reconstructing a local profile along the normal and determining the real local spatial width of each shock sub-structure.
Readers interested in more technical details can find them in \citet{mazelle10:_self_refor_quasi_perpen_shock} and \citet{hobara10:_statis}.

\begin{figure}
\includegraphics[width=\textwidth]{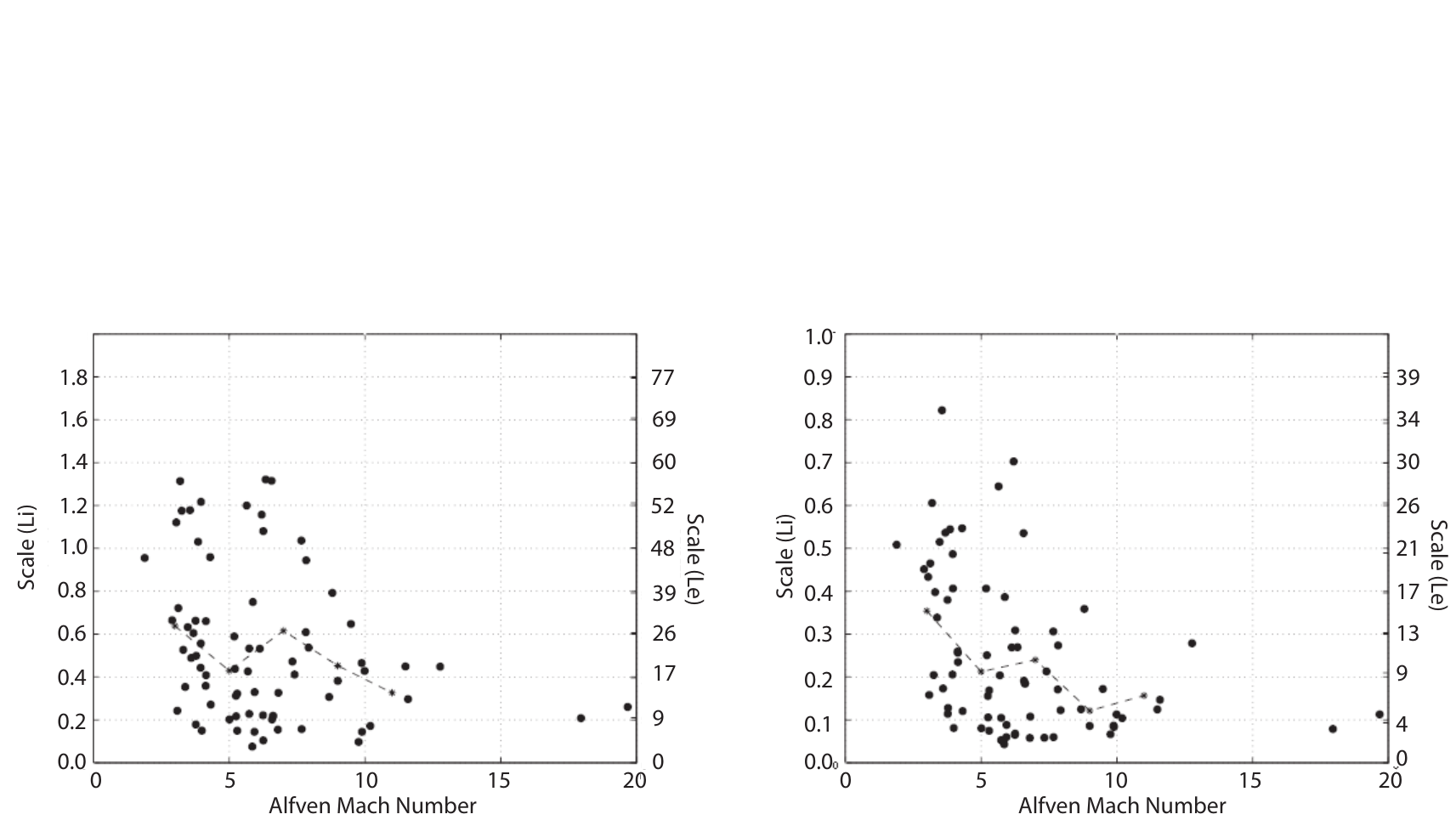}
\caption{Scatterplot of experimentally derived shock size (left panel) and shock spatial scale (right panel) normalized to the ion inertial scale length $c/\omega _{pi}$ (left axis) and electron inertial scale $c/\omega _{pe}$(right axis) as a function of Alfv\'{e}n Mach number. The dashed line represents the averaged values of shock width averaged over shocks with Alfv\'{e}n Mach number in the ranges 2--4, 4--6, 6--8, 8--10, 10--12. The vertical lines represent the statistical error bars for each of these Mach number ranges. \citep[Adapted from ][]{hobara10:_statis}} 
\label{fig:MS4}       
\end{figure}

A scatterplot of the spatial sizes of the ramp as a function of Alfv\'{e}n Mach
number is shown in the left hand panel of Figure~\ref{fig:MS4}. The lefthand panel shows the scale sizes in terms of the ion inertial length whilst the right hand panel shows the width in terms of the electron inertial length. The width of the magnetic ramp varies by about an order of magnitude from $L_{r} = 1.4L_{i}$ ($\approx 60L_{e}$) to $0.1L_{i}$ ($\approx 4L_{e}$). The general trend in Figure~\ref{fig:MS4} indicates that the magnetic ramp becomes thinner with increasing of Alfv\'{e}nic Mach number. This trend is evident even without taking into account two shock crossings with peculiarly high Mach numbers M in the range 17--20 that correspond to the two markers in the bottom right corner of the scatterplot.  The figure also shows a distinct decrease in the maximum width of the shock front with increasing Mach number. To make these tendencies more clear
the characteristic width of the magnetic ramp averaged for the shocks with
Alfv\'{e}nic Mach numbers in ranges 2--4, 4--6, 6--8, 8--10 and 10--12 (dashed
curve) are presented.

The vertical lines on Figure~\ref{fig:MS4} (left panel) represent the statistical error bars for
each range of Mach numbers 2--4, 4--6, 6--8, 8--10 and 10--12. The decrease
of statistical errors with the Mach number is in complete agreement with the
significant decrease of the maximum shock width, while the minimum shock
width undergoes much smaller changes.

The right hand panel of Figure~\ref{fig:MS4} shows the scatterplot of the spatial gradient scale of the magnetic ramp. It clearly demonstrates the same characteristic features as were evidenced in the left hand panel for the ramp width, namely a quite wide range of values, especially for low Mach number shocks and the trend toward shorter scales with the increase of the Mach number as well as the decrease of the maximum gradient scale while Mach number increases. As the change of the magnetic field for all chosen shocks exceeds upstream magnetic field $B_{0}$ (for many of them quite significantly) the values of the gradient scale are smaller then the width of the corresponding shocks. The ramp gradient spatial scale varies in the range $0.05-0.82L_i$ ($2-35L_{e}$).




\subsection{Statistical Analysis \citep{mazelle10:_self_refor_quasi_perpen_shock}}

\begin{figure}
\begin{minipage}{0.45\linewidth}
\centering
\includegraphics[width=2in]{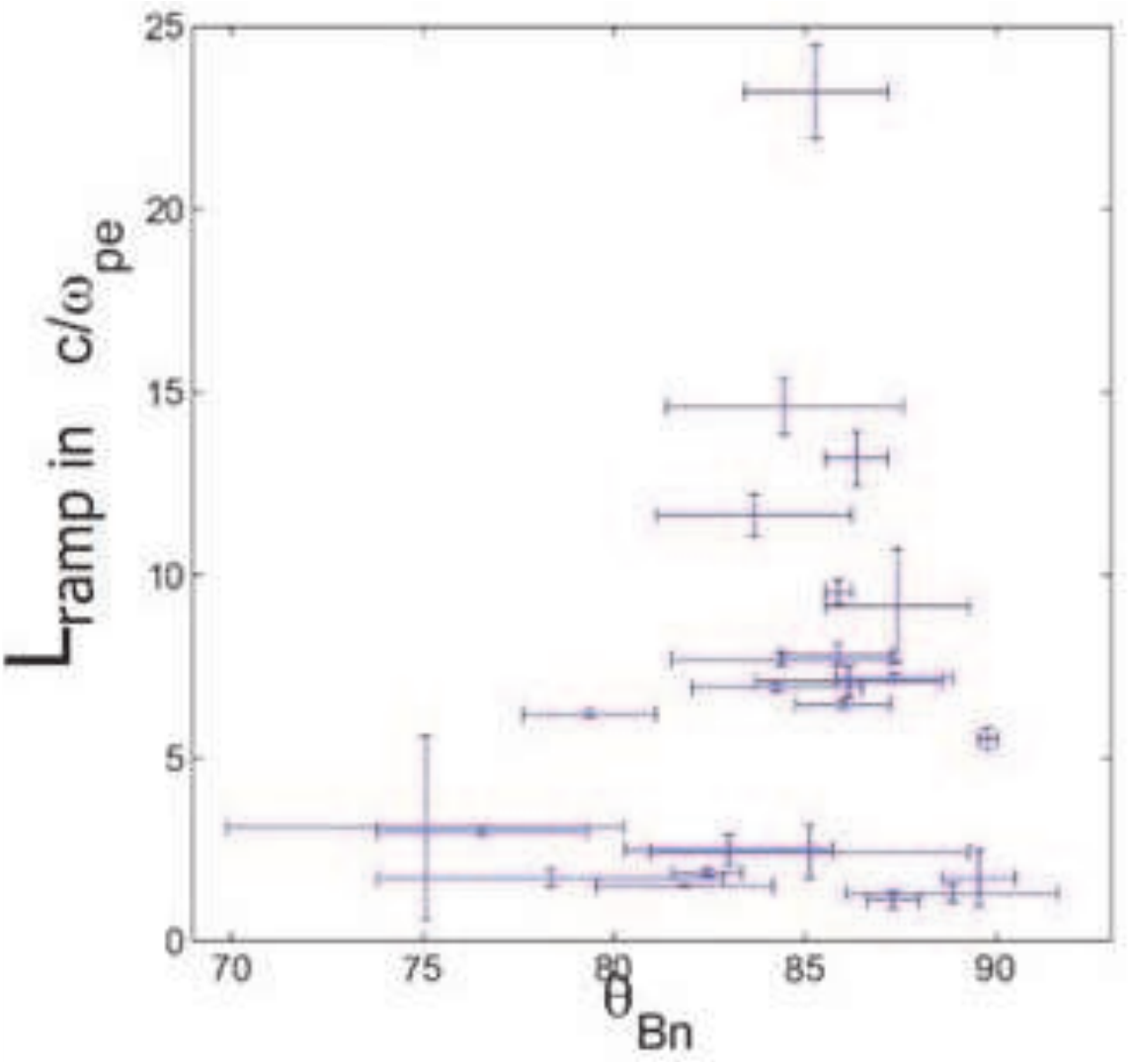}
\caption{Thinnest ramps observed versus shock $\theta _{Bn}$. Reprinted with permission from \citep{mazelle10:_self_refor_quasi_perpen_shock}. Copyright 2010 , American Institute of Physics.}
\label{fig:MS8}
\end{minipage}%
\quad
\begin{minipage}{0.45\linewidth}
\centering
\includegraphics[width=2in]{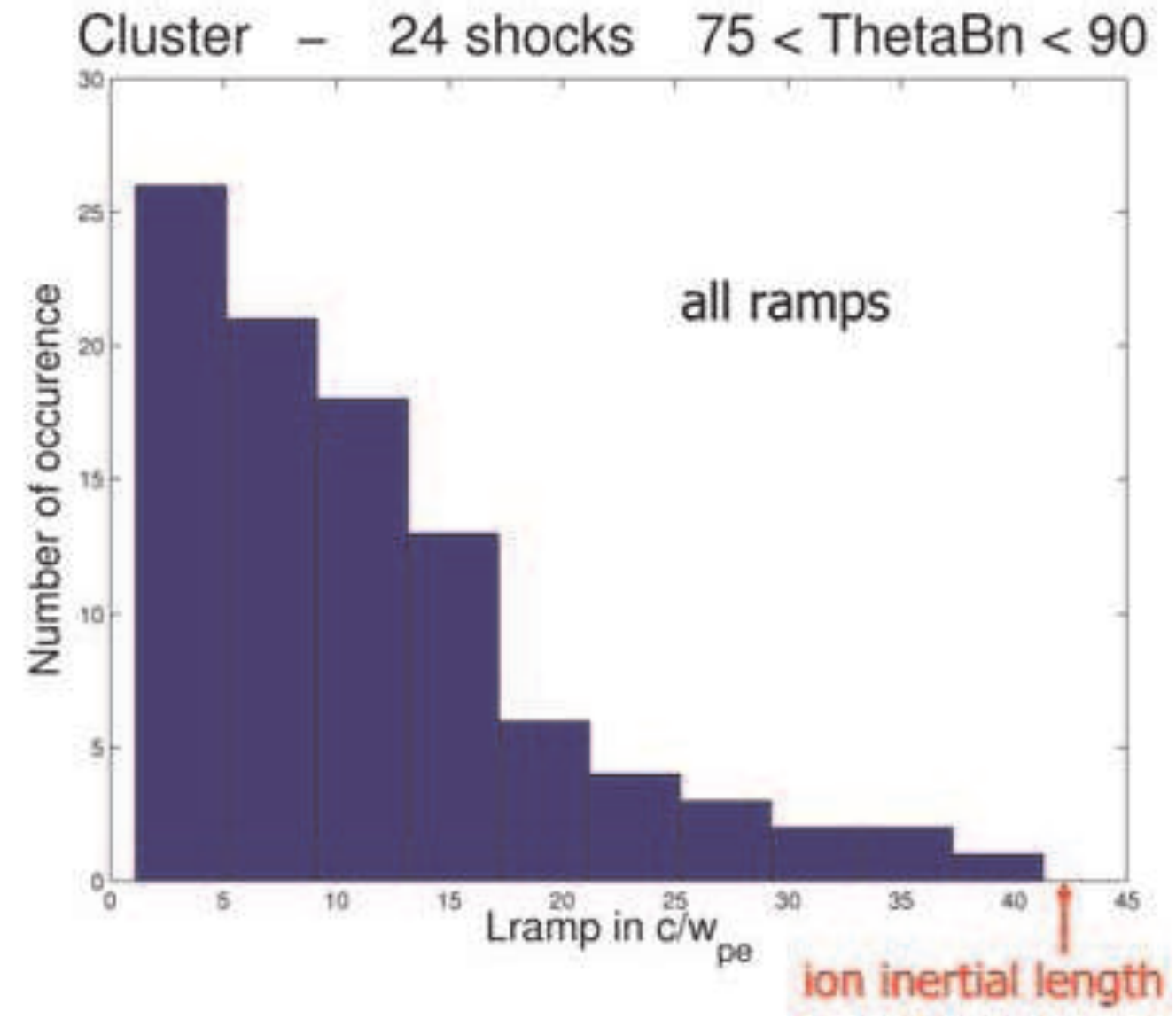}
\caption{Histogram of the 96 shock ramp thicknesses. Reprinted with permission from \citep{mazelle10:_self_refor_quasi_perpen_shock}. Copyright 2010 , American Institute of Physics.}
\label{fig:MS9}
\end{minipage}
\end{figure}

Figure~\ref{fig:MS8} displays the thinnest ramp found among each quartet of
crossings for each individual shock versus $\theta _{Bn}$. There were no
simple relation found. However, it is unambiguously established that many observed thinnest ramps are less than 5 $c/\omega _{pe}$ thick and there was found an apparent trend for lower values as $\theta _{Bn} \rightarrow 90^{\circ}$.


The histogram of all ramp thicknesses in Figure~\ref{fig:MS9} reveals the predominance of narrow ramp width with a Gaussian-like regular decrease towards an asymptotic limit that is still less than $c/\omega _{pi}$.


The authors came to conclusion that their analysis confirms statistically
that the magnetic field ramp of the supercritical quasi-perpendicular shock often reaches a few $c/\omega _{pe}$. 

So, the results of two independent studies by two different groups come to
the same conclusion, the ramp width for quasiperpendicular high Mach number
shocks as seen in magnetic field is of the order of several $c/\omega _{pe}$
and is in a perfect agreement with  estimates  determined from the dispersive scale of a nonlinear whistler wave, modified by the presence of reflected ions \citep{galeev88:_fine,galeev89:_mach, krasnoselskikh02:_nonst}.

\section{Electric field scales of the ramp of quasiperpendicular shocks}
\label{sec:electricscales}

As mentioned in the introduction, there have only been a few reports regarding the scales and structure of the electric field transition at quasi-perpendicular shocks. Based on laboratory experiments, in which the conditions are not exactly the same as in space plasmas, \citet{eselevich71:_isomag} reported that the major change in potential  across the shock occurs within the magnetic ramp region. 

\begin{figure}
\centering
\includegraphics[scale=0.55]{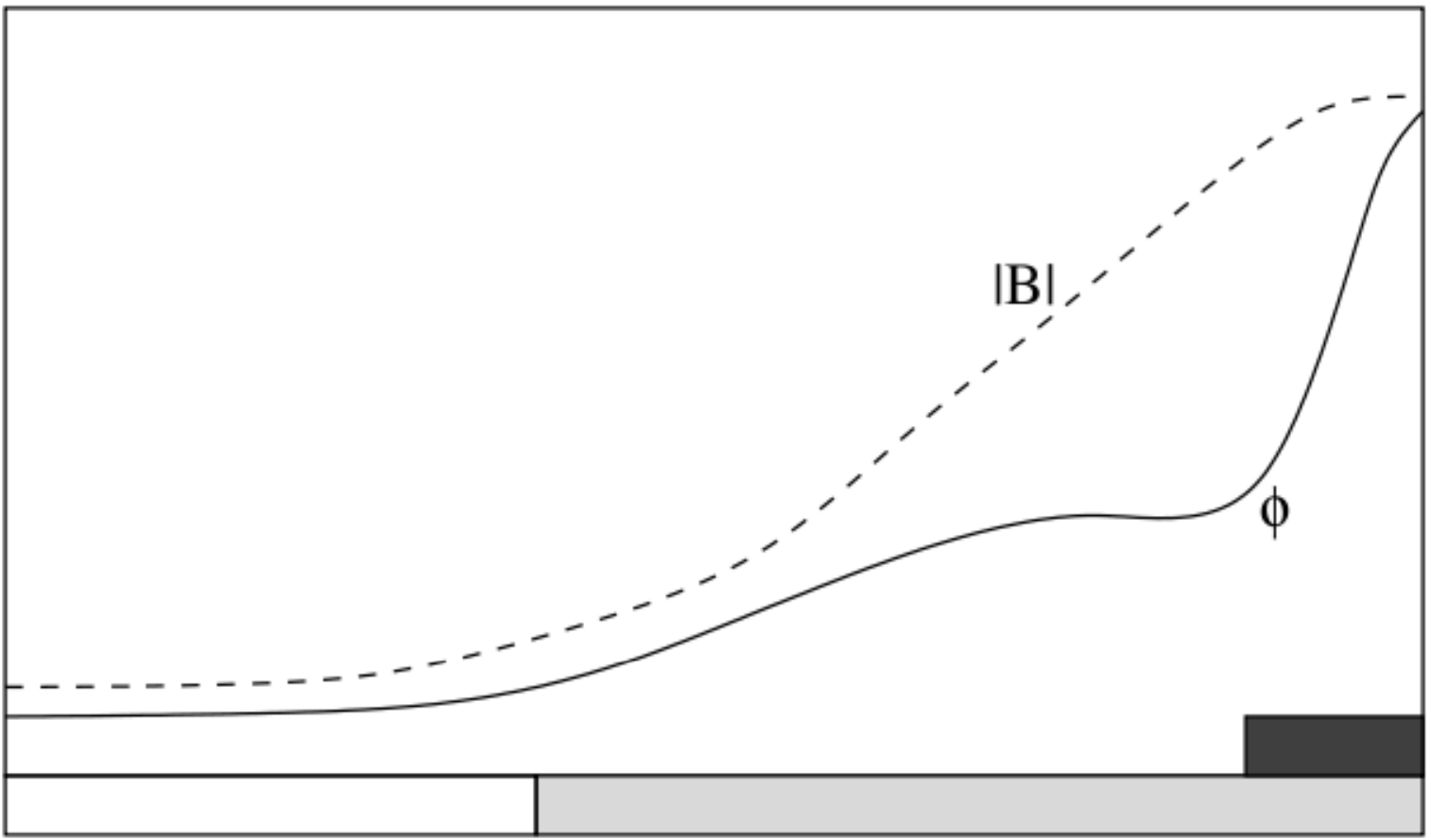}
\caption{Sketch of the changes observed in the magnetic field and electrostatic potential during the crossing of a quasi-perpendicular shock (based upon the experimental results of \citep{eselevich71:_isomag}) \label{f_shkpot}}
\end{figure}
    
Figure~\ref{f_shkpot} is a sketch (based on the results of \citet{eselevich71:_isomag}) of the change in the magnitude of the magnetic field and the accompanying change in the electrostatic potential. These authors interpreted it as a viscous subshock similar to isomagnetic jump. From Figure~\ref{f_shkpot} it can be seen that there are two different length scales that may be associated with the change in the electrostatic potential as the shock is crossed. The first, indicated by the lightly shaded bar at the foot of the plot, shows that overall the potential changes on scales similar to that of the magnetic ramp region in agreement with the results of  \citet{lembege99:_spatial_sizes_elect_magnet_field}. This corresponds to an enhancement of the electric field observed as the shock is crossed. The second scale, indicated by the darkly shaded bar, corresponds to a region within the shock front in which a large increase in the potential is observed over a small spatial scale. Such changes in the potential result from large amplitude spike like features in the electric field.

The results reported here present a study of the large amplitude, short duration features in the electric field observed by the Cluster satellites during a number of crossings of the quasi-perpendicular bow shock published in \citet{walker04:_elect}. These features contribute significantly to the overall change in potential observed at a shock crossing but their short duration implies that they are very localised. The aim of the study was to determine their scale size and amplitudes. These parameters were studied in relation to the upstream shock parameters. 

\begin{figure}
\includegraphics[width=0.9\textwidth]{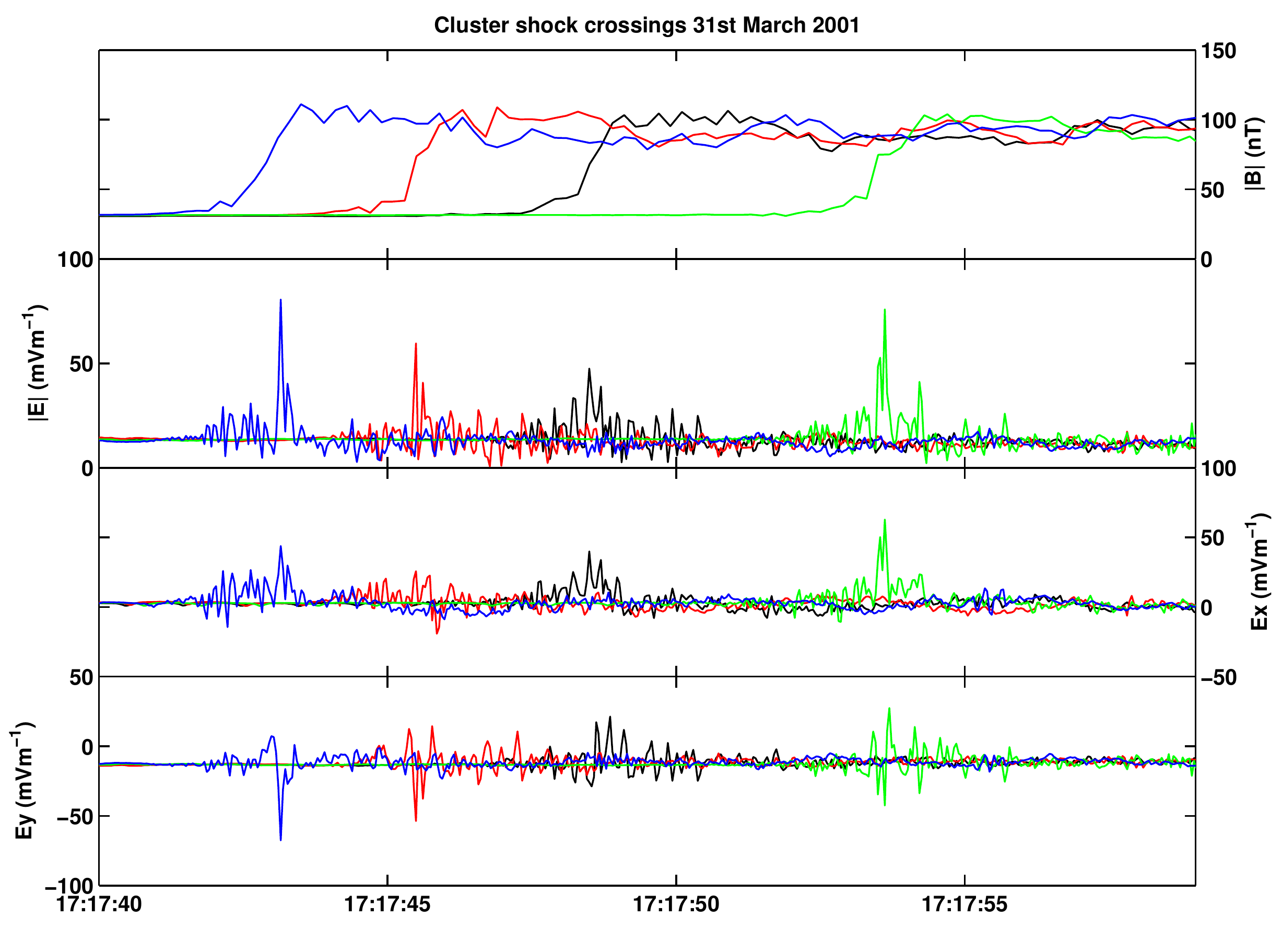}
\caption{Overview of the shock crossing on March $31^{\rm st}$, 2001 at 1718 UT. The top panel shows the magnitude of the magnetic field measured by FGM. The second panel shows the magnitude of the electric field measured in the satellites spin plane. The lower two panels show the spin plane components $E_x$ and $E_y$. \citep[Adapted from][]{walker04:_elect}}
\label{fig:ov2_010331_1718}       
\end{figure}

A total of 54 shock crossings, occurring on 11 separate days were investigated but not all could be analysed fully for various reasons such as unavailability of certain data sets, or the accuracy of the shock normal determination.
In this section we present a case study of the electric field within the shock front, namely the crossing that occurred on March $31^{\rm st}$, 2001 at around 1718 UT. On this day the conditions in the solar wind were some what abnormal due to the passage of a CME. Measurements in the solar wind by Cluster indicated that the magnitude of the magnetic field was of the order of 30nT, the normal for this shock (based upon FGM crossing times) is $n_{B}=(0.94, -0.17, 0.293$) (in the GSE frame), and the shock velocity was determined to be 48.92\kms. Other relevant parameters are $\theta_{Bn}\approx 87^{\circ}$- and a density $n\sim 19$\cmc. The high value of the field resulted in an unusually small ${\beta}\sim 0.07$. The Alfv\'{e}n Mach number for this shock ($M_{A}\sim 3.6$) lies close to the First Critical Mach number and to the whistler critical Mach number so the  conditions of the solar wind are quite favourable for the formation of quasi-electrostatic sub-shocks at the shock front \citep{balikhin02:_obser}.
Figure~\ref{fig:ov2_010331_1718} (adopted from \citet{walker04:_elect}) shows an overview of the magnetic and electric field measurements made by FGM and EFW respectively during this shock crossing. The top panel shows the magnitude of the magnetic field measured by FGM. Initially, all four Cluster spacecraft are in the solar wind just upstream of the outward moving bow shock which subsequently swept over the satellites in the order C4 (17:17:43.5), C2 (17:17:45.5), C1 (17:17:48.5), and finally C3 (17:17:53.5). The magnetic field profiles show a set of clean shock crossings that possess clearly discernible foot, ramp and overshoot regions.  The second panel shows the magnitude of the electric field measured by EFW in the spin plane of each satellite ($|E|^2=E_x^2+E_y^2$). In the solar wind, the typical magnitude of the electric field is around $14$\mvm in the satellite spin plane. It is possible to estimate the $E_z$ component of the upstream electric field assuming that $\bf E \cdot \bf B=0$. This assumption is valid for estimating the field upstream and downstream of the shock but not within the shock region itself. Upstream of the shock, $Ez\approx 5$\mvm. This value is higher than the measured $E_x$ component ($\sim 2.5 $\mvm) and less than the $E_y$ component (-$13$\mvm). Comparing the top two panels it can be seen that the disturbances measured in the electric field begin in the foot region of the shock and continue until the satellites are downstream of the overshoot/undershoot. These general disturbances have amplitudes generally in the range $5-30$\mvm. During their crossings, each of the satellites recorded a number of large amplitude, short duration features in the electric field. The largest of these spikes have maximum amplitudes of approximately 30, 40, 60, and $65$\mvm for satellites 1, 2, 3, and 4 respectively above the field measured in the solar wind just upstream of the shock front. These values represent lower limits of the strength of the electric field since the component perpendicular to the spin plane is not considered. They are seen to occur within the ramp region but there is no strong feature within the FGM data with which they correlate. It is also observed that the largest electric field peaks observed on each satellite appear to occur in pairs which may suggest field rotation. The two lower panels show the components of the electric field measured in the satellite spin planes. Both panels show that the components of the field exhibit a twin peaked structure, similar to that observed in the field magnitude and that the direction of rotation is the same for both peaks. Thus the overall structure is not due to a single rotation of the field. Our goal is a statistical study of the widths of these short living, large amplitude features.
   
\begin{figure}
\centering
\includegraphics[width=0.7\textwidth]{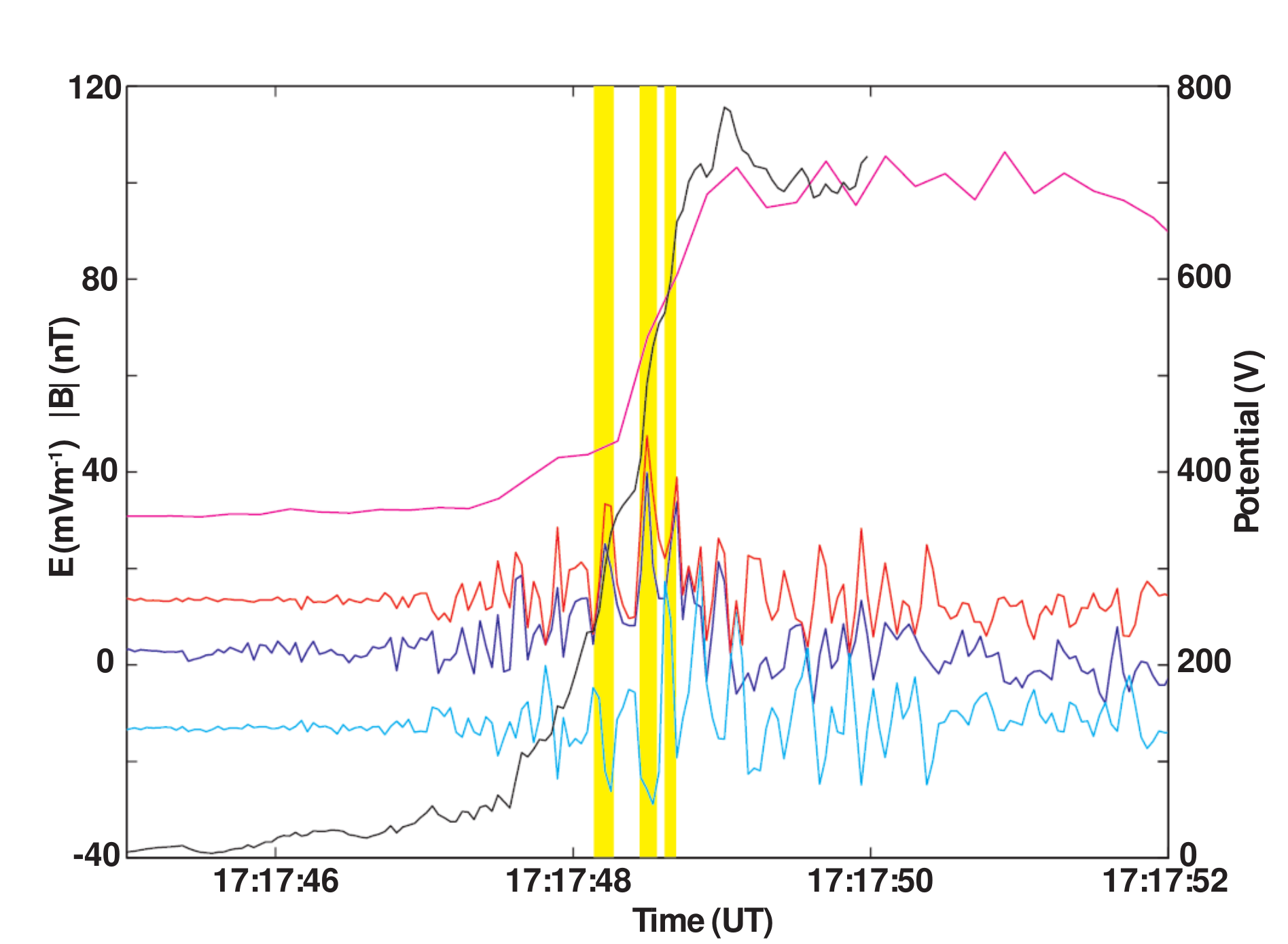}
\caption{The FGM magnetic and EFW electric fields measured by Cluster 1 on March $31^{\rm st}$, 2001 around 17:18 UT. The magnetic field magnitude is shown by the magenta line. The spin plane electric field magnitude, and $E_x$ and $E_y$ components are shown in red, blue and cyan respectively. The yellow regions highlight the periods when large amplitude short duration spikes in the electric field are observed. The black line (Y scale of RHS) represents the change in potential within the shock. \citep[Adapted from][]{walker04:_elect}}
\label{fig:f_c1_010331_1718}       
\end{figure}

   Using the four point measurements one can determine the occurrence time of these peaks in the electric field and hence compute a normal. Examining the Ex component, the time differences between the observations of the first peak in the electric field are $\Delta t_{12}=-3.01$s, $\Delta t_{13}=5.03$s, and $\Delta t_{14}=-5.35$s. When coupled with the respective positions of the satellites this yields a normal direction $n_{E}=(0.946, -0.155, 0.283)$ and a velocity of $\sim 50$\kms. The difference between this normal $n_{E}$ and that determined from the magnetic field ($n_{B}$) is less than a degree. Thus it appears that the electric field spikes correspond to layers within the overall shock structure.
    Figure~\ref{fig:f_c1_010331_1718} shows the results from Cluster 1 in greater detail. The magenta line shows the magnitude of the magnetic field. The foot region was entered around 17:17:47.3 UT whilst the ramp was crossed between 17:17:48.3 and 17:17:48.9 UT. Several large spikes in the electric field are observed in the region of the foot and shock ramp. The three largest occur around 17:17:48.2 (20\mvm), 17:17:48.5 (30\mvm), and 17:17:48.6 (15\mvm). Their short duration implies that their scale size is of the order $3-5c/{\omega}_{pe}$. The black line in Figure~\ref{fig:f_c1_010331_1718} represents an estimation of the electrostatic potential measured in the normal direction. This was calculated by removing an average of the field measured in the region just upstream of the shock from the field measured within the shock region and then integrating the projection of this electric field along the normal direction. Whilst the actual potential cannot be calculated due to the incomplete vector measurements, it can be estimated by assuming that the field perpendicular to the spin plane $E_{z}=0$. This assumption is valid because for this particular shock, the normal lies very close to the spin plane. This calculation can be used to show that the largest jumps in the potential coincide with the spikes observed in the electric field and that these occurrences contribute a significant fraction of the total potential change observed at the shock. During this period, the electric field enhancements contribute around 40\% of the total change.
    
\subsection{Scale size}
\label{sec: scale_size}
The preceding sections have presented evidence for localized increases in the electric field strength measured as the satellite traverses a quasi-perpendicular bow shock. All shocks analyzed show evidence for an enhancement in the background electric field. In most cases, the region in which this field enhancement occurs lasts longer than the crossing of the magnetic ramp. The field typically increases of the order 1-3\mvm above that measured in the solar wind. However, as has been noted above, the turbulence in this region is dominated by spike-like fluctuations lasting a few milliseconds and with magnitudes of typically 4-20\mvm with a maximum magnitude of the order of 70\mvm. This existence of large gradients in the electric field has repercussions for processes involved in the heating of electrons. In the presence of strong electric field gradients the electron gyration frequency can deviate from its classically calculated value \citep{cole76:_effec, balikhin98:_e_b}, leading to an increase in its Larmor radius and the possibility of a breakdown in adiabaticity \citep{cole76:_effec, balikhin98:_e_b}.
    Having shown that the spikes observed in the electric field at the front of a quasiperpendicular shock appear to be physical structures that form a layer within the shock front as opposed to being the result of noise in the data or motion of the shock a statistical study of these features was performed to investigate their relationship to the properties of the shock front. Now we shall present statistical study of the data collected from a number of such spike-like features.

\begin{figure}
\centering
\includegraphics[width=0.8\textwidth]{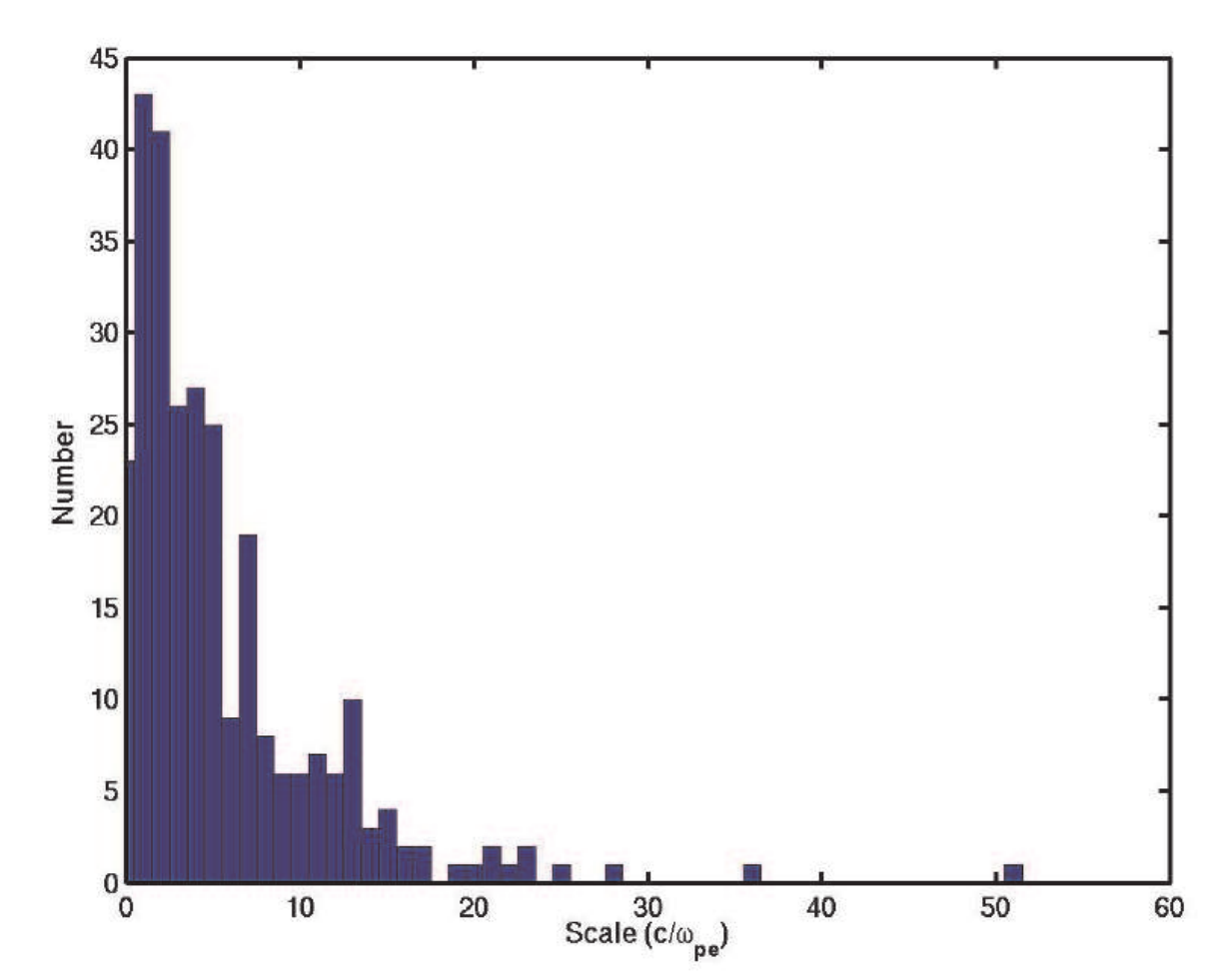}
\caption{Histogram of the scale sizes for the spike-like enhancements observed during a number of crossings of the quasi-perpendicular bow shock. \citep[Adapted from][]{walker04:_elect}}
\label{fig:f_scale_hist}       
\end{figure}

    Figure~\ref{fig:f_scale_hist} shows the distribution of the scale sizes determined from the event duration and the shock velocity evaluation of these features in terms of the electron inertial length. The scale size of these events will be unaffected by the incomplete vector measurements of the electric field. The vast majority of crossings have scale sizes of the order of $1-5c/\omega_{pe}$. The data that form tail of the distribution at longer scale sizes typically comprise events that have a multi-peak structure. This type of event represents an upper limit to the scale size of these short-lived events. In comparison, the typical scale of the magnetic ramp is characterized by the ion inertial length \citep{newbury96:_obser} although these authors also report one particular shock as having a ramp scale as small as $0.05c/{\omega}_{pi}$ or $2c/{\omega}_{pe}$.
    
\begin{figure}
\centering
\includegraphics[width=0.8\textwidth]{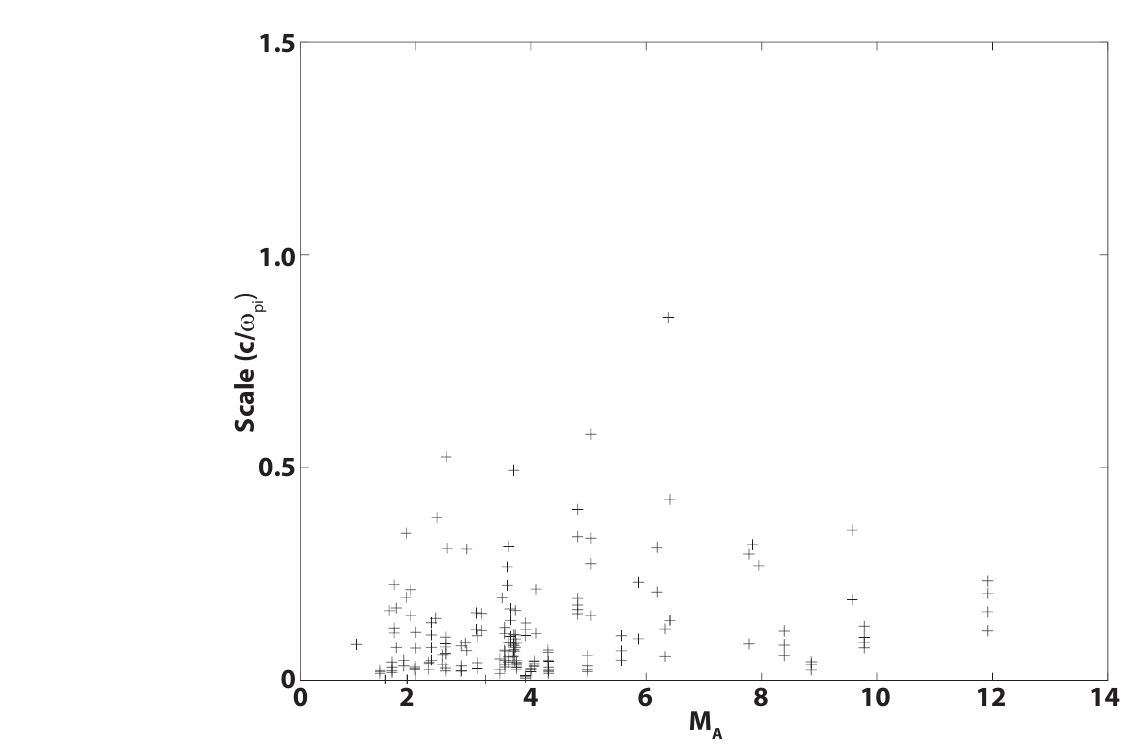}
\caption{Dependence of scale size on upstream Mach number. \citep[From][]{walker04:_elect}}
\label{fig:f_ma}       
\end{figure}

    Figure~\ref{fig:f_ma} shows the relationship between the Mach number and scale size of the spikes observed in the electric field. From the figure, it is clear that the scale size has a lower limit that  decreases as the Mach number  increases. One should notice that these results represent the tendency rather than the proof, the number of points at high Mach numbers is not sufficient for valuable statistical study. 

\begin{figure}[t]
\centering
\includegraphics[width=0.8\textwidth]{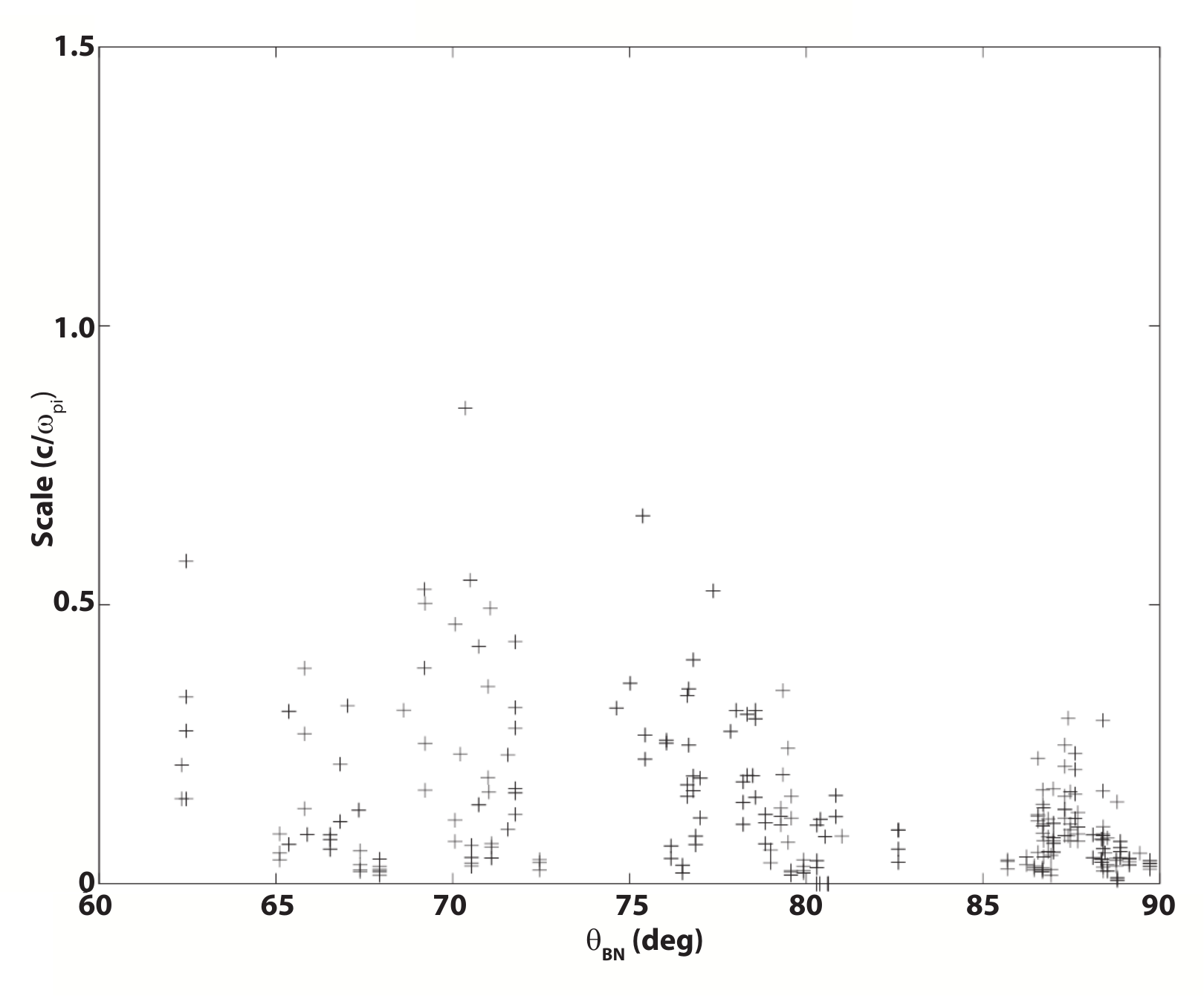}
\caption{Dependence of scale size on ${\theta}_{Bn}$. \citep[From][]{walker04:_elect}}
\label{fig:f_tbn}       
\end{figure}

    Figure~\ref{fig:f_tbn} shows a scatter plot of the relationship between ${\theta}_{Bn}$ and the scale size of the electric field enhancements. In general there appears to be a broad range of scales. However, as ${\theta}_{Bn}$ approaches $90^{\circ}$- the scale length decreases. For the shocks analysed with ${\theta}_{Bn}$ close to $90^{\circ}$ - the scale lengths are of the order of $2c/{\omega}_{pe}$. This compares favourably with theoretical estimates that for shocks close to perpendicular the scale width is estimated to be of the order of the electron inertial length as proposed by \citet{karpman64:_struc, galeev88:_fine}. This tendency corresponds exactly to dispersion dependence upon the angle ${\theta}_{Bn}$. 
    
\subsubsection{Amplitude}

\begin{figure}[t]
\centering
\includegraphics[width=0.7\textwidth]{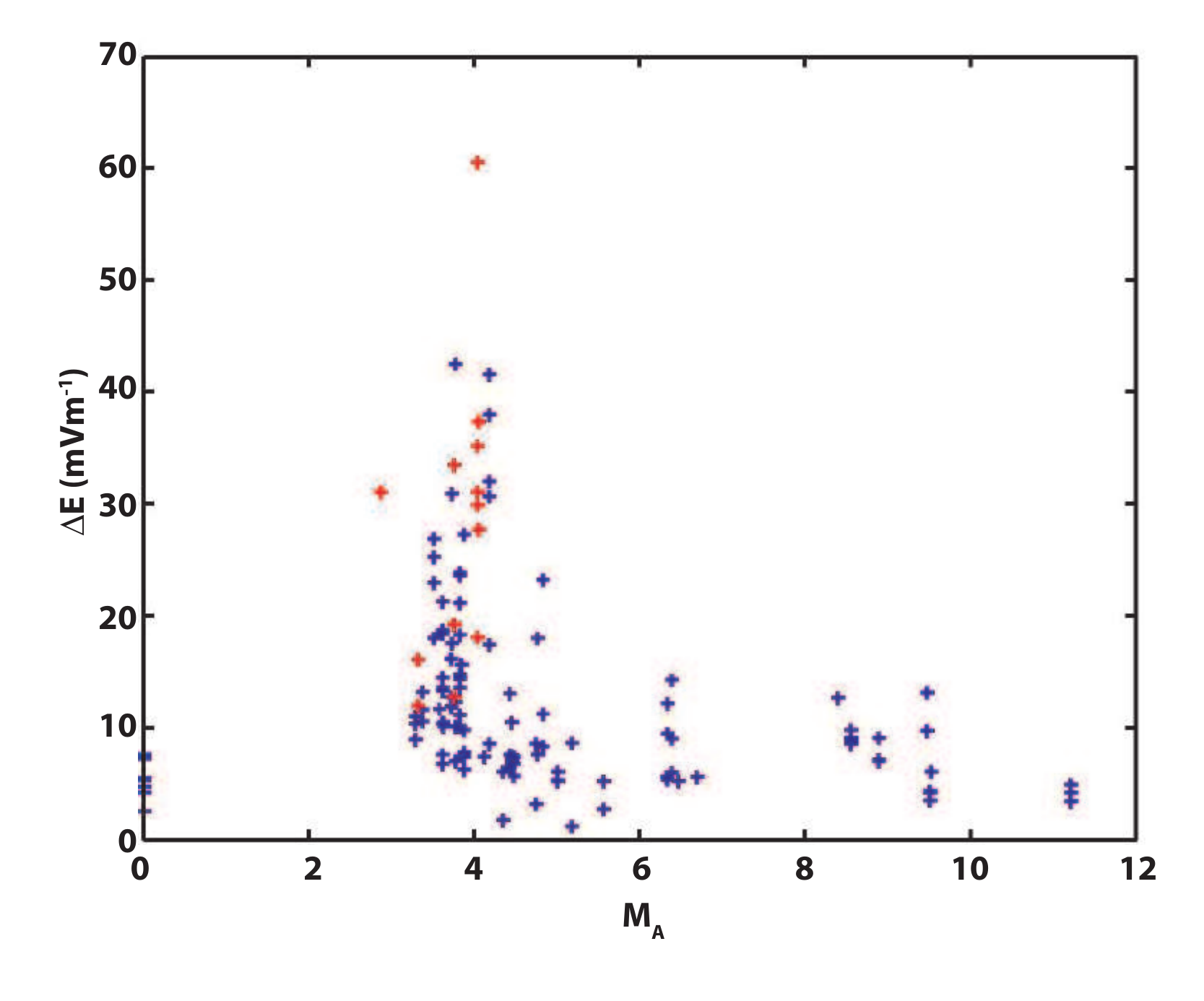}
\caption{Scatter plot showing the relationship between the amplitude of the electric field spikes as a function of Mach number. The red crosses are used to highlight the data for the shocks that occurred on March $31^{\rm st}$, 2001. \citep[Adapted from][]{walker04:_elect}}
\label{fig:ma_de}       
\end{figure}

    The examples presented above show that the increase in the electric field (${\Delta E}= E_{spike}-E_{upstream}$) observed during encounters with these spike-like structures varies between 4 and 70\mvm above the average field that is measured in the solar wind just upstream of the shock. In this section the relationship between this change ($\Delta E$) and the shock Mach number, and the angle ${\theta}_{Bn}$ is presented.
    Figure~\ref{fig:ma_de} shows a scatter plot of the peak amplitude observed in the electric field spike event ($\Delta E$) as a function of the shock Mach number $M_{A}$. For shocks whose Mach number $M_{A}>5$ there is a fairly constant trend in which ${\Delta }E < 15$\mvm. In the Mach number range $3 <M_{A}< 5$ the range of observed amplitudes varies between 5 and 60\mvm. It appears that in this Mach number range the electric field potential becomes more important than for low and high Mach number shocks that corresponds to dependence of the electrostatic potential upon the Mach number \citep{dimmock12:_mach_clust}. The red crosses highlight the shocks observed on March $31^{\rm st}$, 2001. All of these shocks fall into this range of Mach numbers. This set of shocks seems to possess Mach numbers corresponding to supercritical range and having large number density, at about 7\%, of alpha particles (Maksimovic, private communication). Their structure seems to resemble that of electrostatic sub-shocks similar to those observed in laboratory plasmas \citep{eselevich71:_isomag}. A characteristic signature of sub-shocks is the occurrence of small scale electrostatic fluctuations such as those observed on this particular day. Ion sound sub-shocks have been observed in laboratory plasmas with scales of the order of 100 Debye lengths. For the shocks observed in March 31$^{st}$ 2001, the scale is closer to characteristic scale of the fast magnetosonic mode \citep{balikhin02:_obser}.

\begin{figure}[t]
\centering
\includegraphics[width=0.6\textwidth]{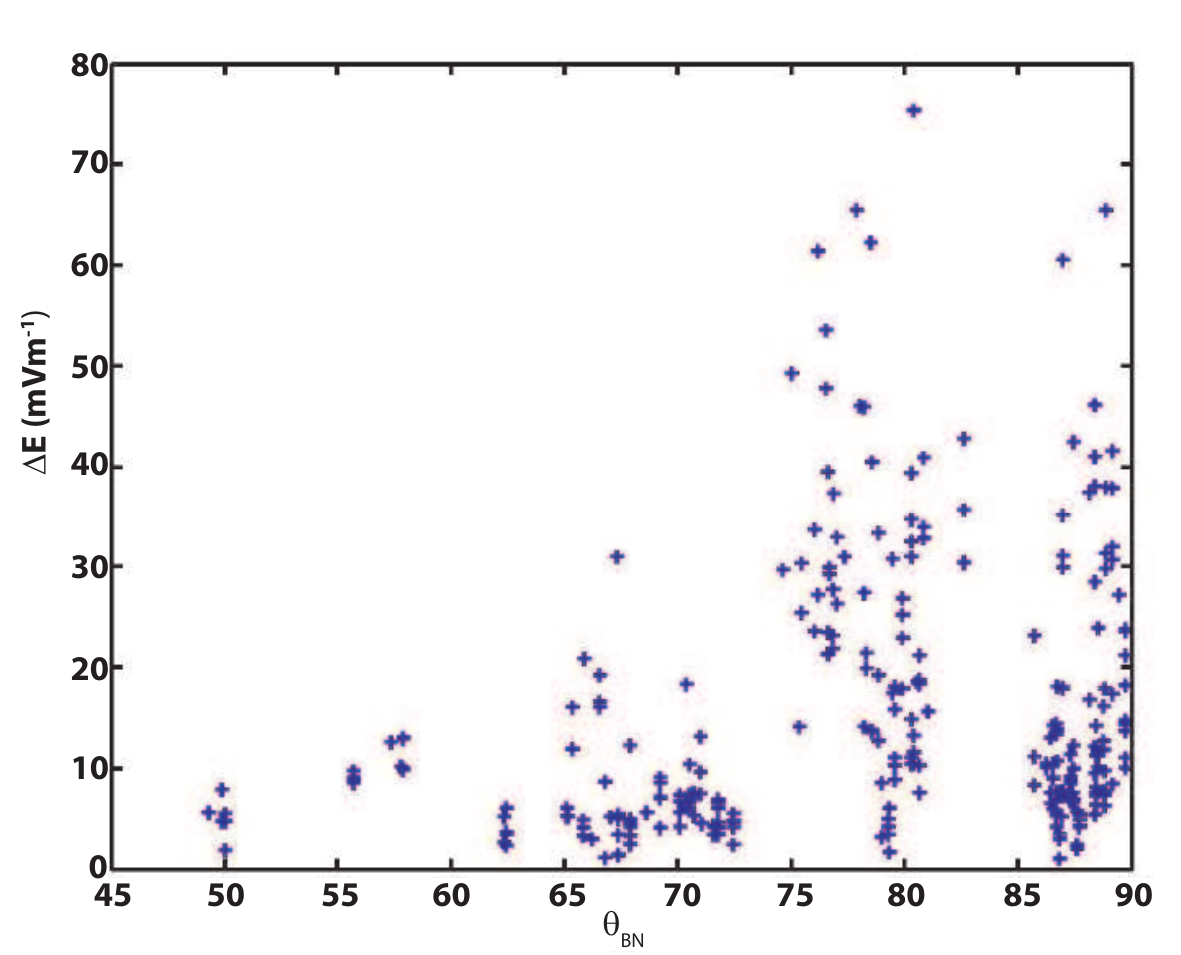}
\caption{The relationship between $\Delta E$ and ${\theta}_{Bn}$. \citep[Adapted from][]{walker04:_elect}}
\label{fig:f_beta_de}       
\end{figure}

    The relationship between ${\Delta}E$ and ${\theta}_{Bn}$ is shown in Figure~\ref{fig:f_beta_de}. It clearly shows that as ${\theta}_{Bn}$ approaches $90^{\circ}$- the range of the observed amplitudes of the electric field spikes increases.

\subsection{Conclusions}
\label{subsec_conc} 
In this Section we presented the changes observed in the electric field during the crossing of a number of quasi-perpendicular bow shock. It is shown that the electric field is enhanced during the crossing of the shock and that the scale size over which this enhancement is observed is larger than that of the macroscopic magnetic ramp region. Within the whole shock region, short lived electrostatic structures are observed that are intensified in the ramp region. The scale size of these structures is of the order of a few $c/{\omega}_{pe}$: and was shown to decrease as ${\theta}_{Bn}$ approaches $90^{\circ}$- which corresponds to the dependencies following from theoretical model based on consideration of the shock as mainly dispersive nonlinear structure  \citep{galeev88:_fine, krasnoselskikh02:_nonst}. The amplitudes of these structures is typically of the order of 5-20\mvm: but under special circumstances may reach as high as 70\mvm. The highest amplitudes appear to be observed for shocks whose Mach number is in the range 3 to 5. This may be an indication that such shocks have quasi-electrostatic sub-shocks inside the main ramp transition. It was also demonstrated that these small scale structures make a substantial contribution to the overall change in potential observed across the shock and that the potential change is not linear.

\section{Dispersive nature of High Mach number shocks: Poynting flux of oblique whistler waves \label{sec:upstr_waves}}

It is well known that a subcritical shock has a nonlinear whistler wave train upstream of its front \citep{sagdeev66:_cooper_phenom_shock_waves_collis_plasm, mellott85:_subcr}. The major transition of such a dispersive shock, the ramp, behaves as the largest peak of the whistler precursor wave package \citep{karpman73:_elect, kennel85:_quart_centur, galeev89:_mach, krasnoselskikh02:_nonst}. The presence of whistler/fast magnetosonic precursor wave trains in supercritical shocks was experimentally established in \citet{balikhin97:_nonst_low_frequen_turbul_quasi_shock_front, krasnoselskikh91, oka06:_whist_mach}. These whistler waves have rather large amplitudes and their role in energy transformation and redistribution between different particle populations and in the formation of the shock front structure is still an open question. The energy source responsible for the generation of these waves is the subject of active debate in shock physics  \citep[see][]{galeev88:_fine, galeev89:_mach, krasnoselskikh02:_nonst, matsukiyo06:_mach, comisel11:_non}. Often the precursor waves are almost phase-standing in the shock frame. However, if they are generated by the ramp region as the dispersive precursor their group velocity can still be greater than zero in the shock reference frame, which would allow energy flow in the form of Poynting flux to be emitted towards the upstream of the shock transition.
On the other hand, if the waves are generated by instabilities related to reflected ions their energy flux will be directed from the upstream region towards the shock ramp. The goal of this Section is to address this problem, to present the direct measurement of the Poynting flux of the upstream whistler waves aiming to establish the direction of the Poynting flux.

Below we establish the energy source of the waves by calculating the Poynting flux of the waves in the Normal Incidence Frame (NIF) of the shock, using multi-satellite Cluster data from crossings of the Earth's bowshock  \citep{escoubet97:_clust_scien, bale05:_quasi_shock_struc_proces}.
Two events with supercritical Alfv\'enic Mach numbers are analyzed.
In both cases it is found that the shocks show dispersive behaviour with the Poynting flux directed in upstream direction.

Poynting flux is not a Lorentz invariant and therefore depends on the frame of reference. To evaluate the value and direction of the Poynting flux with respect to the the shock we transform the electric field to the Normal Incidence Frame (NIF). 
The normal $\hat{\mathbf{n}}=+\hat{\mathbf{x}}$ which also serves as the $x$-coordinate direction in the NIF system is obtained by four-spacecraft timing, $\hat{\mathbf{z}}$ is the direction of maximum varying magnetic field obtained from a minimum variance analysis, and $\hat{\mathbf{y}}$ is the direction of the convection electric field which completes the right-handed system.

The transformation from the spacecraft frame to the NIF is given by 
$
\mathbf{E}_{NIF} = \mathbf{E}_{sc} + \mathbf{v}\times\mathbf{B}.
$
The total velocity required for this transformation is defined by $\mathbf{v}=\mathbf{v}_{sh}+\mathbf{v}_{NIF}$ where $\mathbf{v}_{sh}=v_{sh}\hat{\mathbf{n}}$ is the shock velocity, $\mathbf{v}_{NIF}=\hat{\mathbf{n}}\times(\mathbf{v}_u \times \hat{\mathbf{n}})$ is the NIF velocity and $\mathbf{v}_u$ is the solar wind velocity.

A general shift of reference frame, coordinate transformation, and evaluation of the complete Poynting vector requires knowledge of the full six-dimensional electromagnetic field (three electric and three magnetic components). The Cluster spacecraft, however, only measures the two components of the elecric field in the spin-plane of the spacecraft, while the third component normal to the spin-plane is not measured. 
To reconstruct the third component we use the assumption that for the wave electric and magnetic fields the condition $\mathbf{E}\cdot\mathbf{B}=0$ holds. 
While this is a true condition for the cross-shock (DC) electric field, it holds well for whistler wave electric fields at lower frequencies.

We study two quasi-perpendicular high Mach number shocks encountered by the Cluster  multi-spacecraft mission \citep{escoubet97:_clust_scien}. The first shock was observed around 04:53:40~Universal Time (UT) on 20-Jan-2003, and the second around 07:07:00~UT on 24-Jan-2001. 
We use data from the EFW (electric field) \citep{gustafsson97:_elect_field_wave_exper_clust}, FGM (DC magnetic field) \citep{balogh97:_clust_magnet_field_inves} and STAFF (wave magnetic field) instruments \citep{decreau97:_whisp_reson_sound_wave_analy} from spacecraft 2 (for the 2003 shock) and spacecraft 3 (for the 2001 shock). 
The shock normal $\hat{\mathbf{n}}$ is established by assuming a planar shock and using the time of crossing of the four spacecraft and their relative positions \citep{paschmann98:_analy_method_multi_spacec_data}. 

\begin{figure}
\begin{minipage}{0.45\linewidth}
\centering
\includegraphics[width=2.4in]{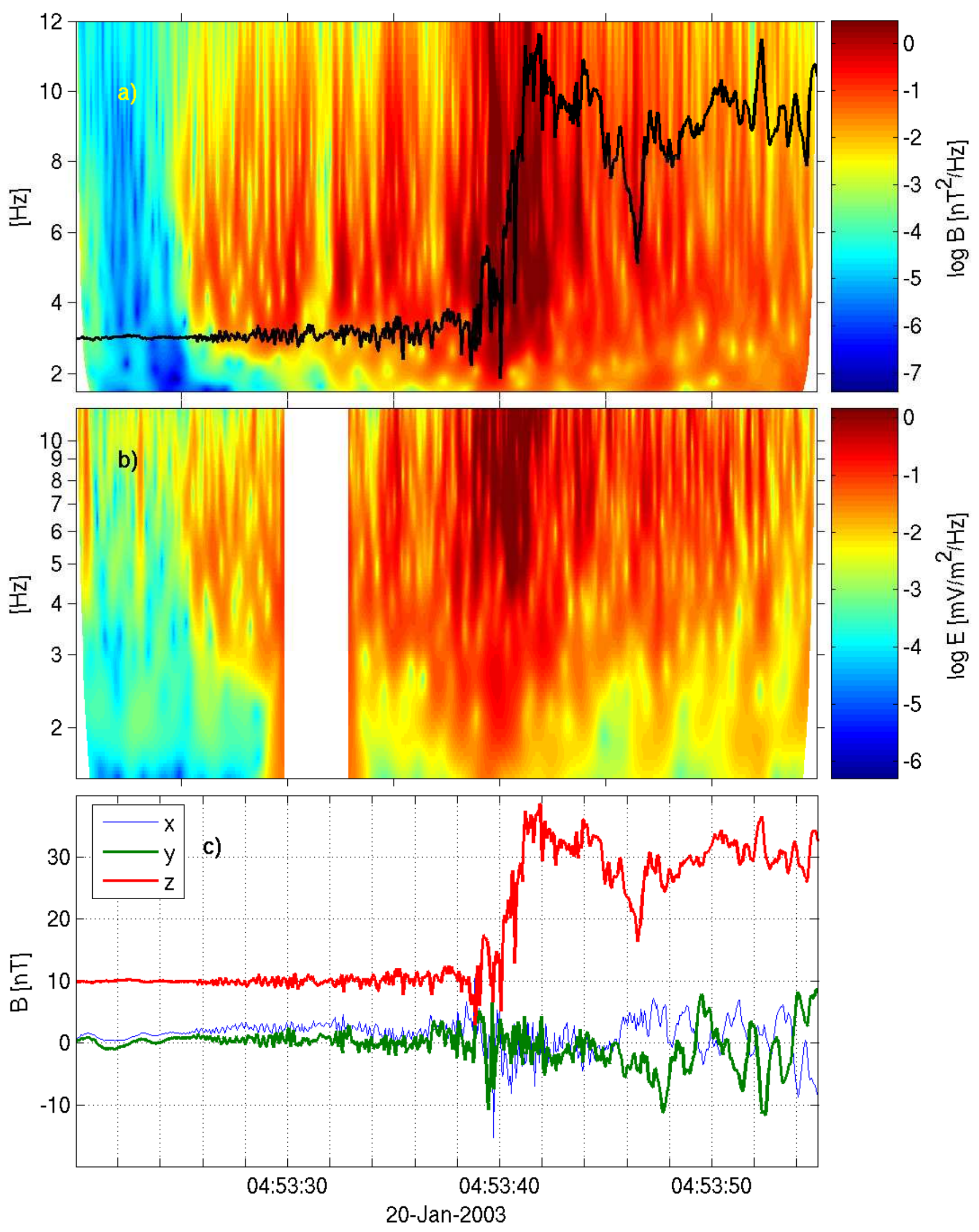}
\caption{\label{fig:shock2003fields} Magnetic and electric fields 
in the Normal Incidence Frame (NIF) of a high Mach number shock. 
a) Power spectra of the magnetic field (STAFF). The black line is the DC total magnetic field, included to show the waves in relation to the shock ramp structure. b) Power spectra of the electric field (EFW). The data gap is due to instrumental interference. c) The magnetic field in NIF coordinates $B_{\textrm{NIF}}$. 
\citep[Adapted from][]{sundkvist12:_disper_natur_high_mach_number}}\end{minipage}%
\qquad
\begin{minipage}{0.45\linewidth}
\centering
\includegraphics[width=2.4in]{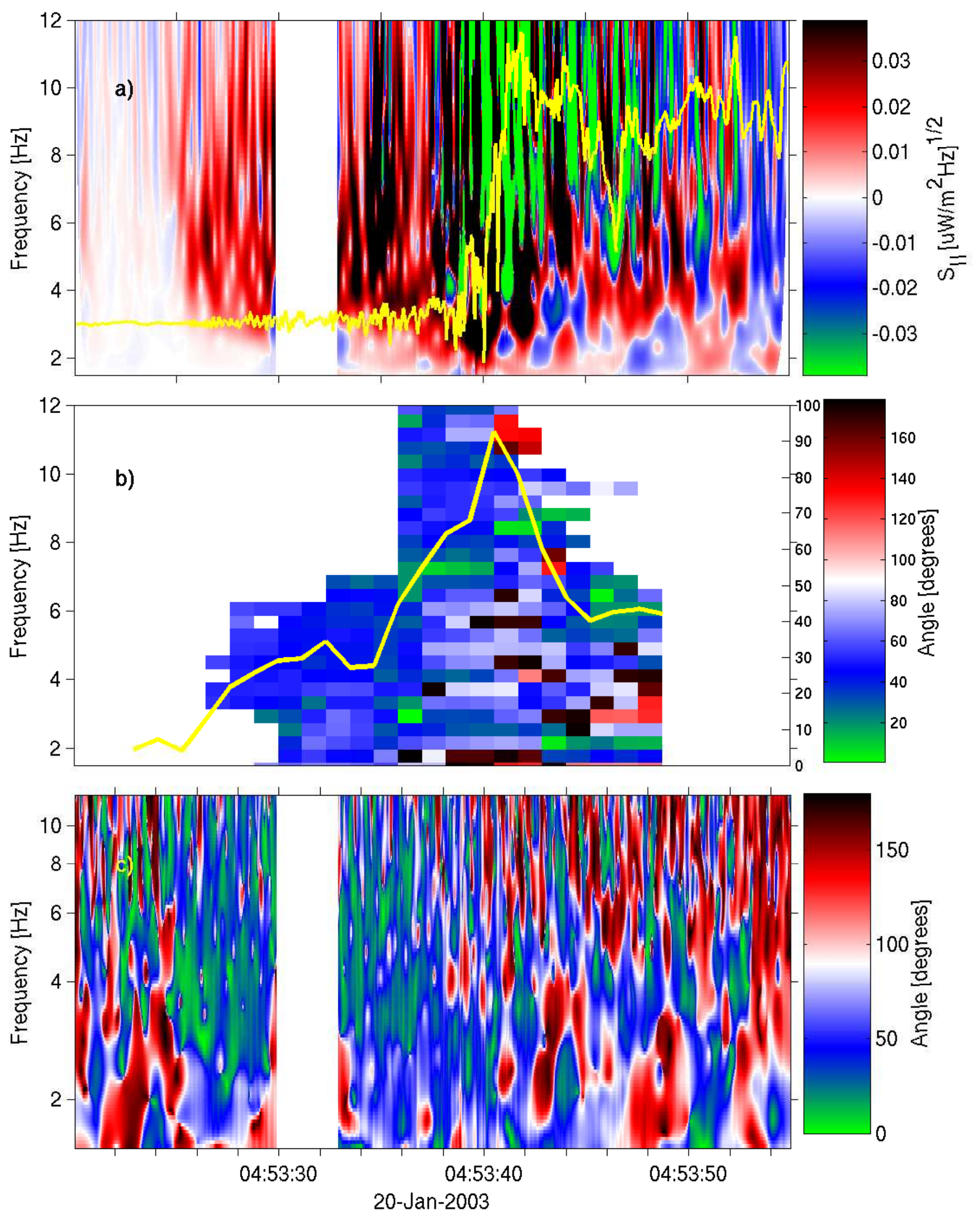}
\caption{\label{fig:shock2003poynt} 
Poynting flux in the Normal Incidence Frame (NIF) of the same shock as in Figure~(\ref{fig:shock2003fields}). 
a) Poynting flux $S_{||}$ projected on the local $\mathbf{B}_0$ in the NIF, where red corresponds to the upstream flux away from the shock. b) Angle $\theta_{k,B}$ between $\hat{\mathbf{k}}$ and magnetic field $\mathbf{B}_0$. The yellow line represents the average over all frequencies (right scale). c) Angle between Poynting flux $\mathbf{S}$ and $\mathbf{B}_0$. \citep[Adapted from][]{sundkvist12:_disper_natur_high_mach_number}}
\end{minipage}
\end{figure}

The first shock analyzed had an upstream $\theta_{Bn}\sim 85 ^\circ$ and an Alv\'enic Mach number $M_A\sim 5.5$.
The electric and magnetic fields in the shock front region are characterized by waves, with stronger amplitudes closer to the ramp, see Figure~(\ref{fig:shock2003fields}).
The waves have frequencies $f_{cp} < f$, where $f_{cp}\sim 0.1$~Hz is the proton gyrofrequency, and right-handed polarization looking along the magnetic field vector and thus belong to the magnetosonic/whistler mode. 
The direction of the wave-vector $\hat{\mathbf{k}}$ was determined by the Means method \citep{means72:_use}, which uses the imaginary part of the three-dimensional magnetic field spectral matrix. 
The angle $\theta_{kB}$ between the wave vector and the local ambient magnetic field is shown as a function of frequency in Figure~(\ref{fig:shock2003poynt})b. The average value $\left<\theta_{kB}\right>$ in the shock front region is $\sim 10-50^\circ$ (right-hand scale). The whistler waves are thus 
oblique with respect to the local magnetic field, as well as to the shock normal. 
The angle increases continously as the shock front is approached and 
$\theta_{kB}\rightarrow 90^\circ$ at the ramp, reflecting the quasi-perpendicular nature of the shock. This smooth transition stresses the nature of the shock as a dispersive nonlinear whistler wave.

\begin{figure}
\centering
\includegraphics[width=0.5\columnwidth]{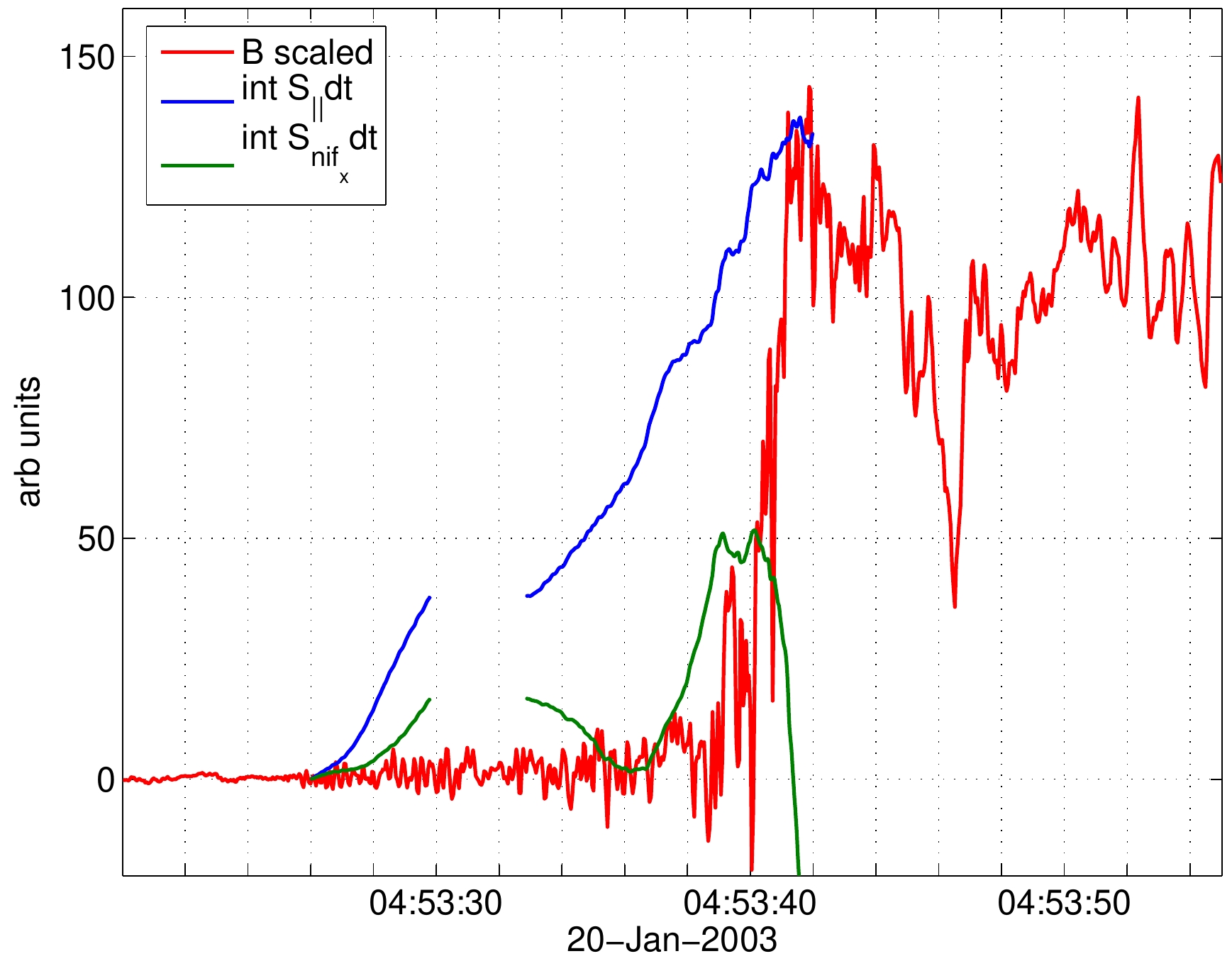}
\caption{\label{fig:integrated_pf} Poynting flux integrated along the spacecraft trajectory. The blue line is a projection along the ambient magnetic field $\int S_{||} dt$ 
and the green line is a projection on the shock normal  $\int \mathbf{S} \cdot \hat{\mathbf{n}} dt$. The red line shows the scaled magnetic field $B_o$ for reference. \citep[Adapted from][]{sundkvist12:_disper_natur_high_mach_number}}
\end{figure}

Since Poynting flux is a second-order quantity the electric and magnetic fields in the NIF were wavelet transformed (Morlet width 5.36) and the cross-product $\mathbf{S}_f = 1/\mu_0 \mathbf{E}_f \times \mathbf{B}_f$ formed in frequency space. 
The calculated Poynting flux is therefore distributed in both time and frequency.
The projection of the Poynting flux distribution along the magnetic field $S_{||}=\mathbf{S}_f \cdot \mathbf{B}_0 / |\mathbf{B}_0|$ using an instantaneous value of $\mathbf{B}_0$ is plotted in Figure~(\ref{fig:shock2003poynt})a, where the colors red (upstream) and blue (downstream) show the direction of the flux. 
We note that in the front region of the shock the Poynting flux is everywhere directed upstream (red), away from the shock. 
In the downstream area there is a mixture of blue, green and red, where there is more turbulence and the waves are no longer coherent.
The upstream and slightly oblique direction of the Poynting flux is further quantified in the instantaneous angle $\theta_{S,B}$ between the Poynting flux and the ambient magnetic field, plotted in Figure~(\ref{fig:shock2003poynt})c. 
Figure~(\ref{fig:integrated_pf}) shows the Poynting flux along the spacecraft trajectory, with integrated power over frequencies corresponding to the waves in Figure~(\ref{fig:shock2003poynt})a, $2 < f < 10$~Hz. In this figure the slope is the important characteristic. Positive slope means Poynting flux carried upstream, and negative slope downstream. From the figure it is evident that the source of the Poynting flux is associated with the shock ramp. The data gap and associated plateau are due to instrumental interference.

Another important characteristic established is that the Poynting flux direction is oblique with respect to the shock normal as well as the background magnetic field. 
This can be explained by analyzing how the phase velocity for whistler waves depends on this angle. 
The phase velocity of a wave propagating in the plane of the shock normal $\hat{\mathbf{n}}$ and background magnetic field $\mathbf{B}_0$,
having an angle $\alpha$ with respect to the shock normal is
$
V_{ph}=\frac{1}{2}\sqrt{\frac{m_{i}}{m_{e}}}\cos (\theta _{Bn}-\alpha ).
$
Its projection on the direction of the shock normal is 
$
V_{ph,\hat{\mathbf{n}}}=
V_{ph}\cos \alpha =\frac{1}{2}\sqrt{\frac{m_{i}}{m_{e}}}\cos (\theta
_{Bn}-\alpha )\cos \alpha .
$
Its maximum value can be found to be equal to 
$
\max({V_{ph,\hat{\mathbf{n}}}})=
\frac{1}{4}\sqrt{\frac{m_{i}}{m_{e}}}(1+\cos
\theta _{Bn}),
$
thus the projected phase speed can be larger than the whistler critical velocity given above. The above analysis also explains the observation of oblique whistler wave trains found in computer simulations of purely perpendicular shocks \citep{hellinger07:_emiss}.
So even in the case of shocks having Mach numbers larger than the whistler critical Mach number, whistler waves oblique with respect to the shock normal can remain quasi-standing.

\begin{figure}
\centering
\includegraphics[width=.6\columnwidth]{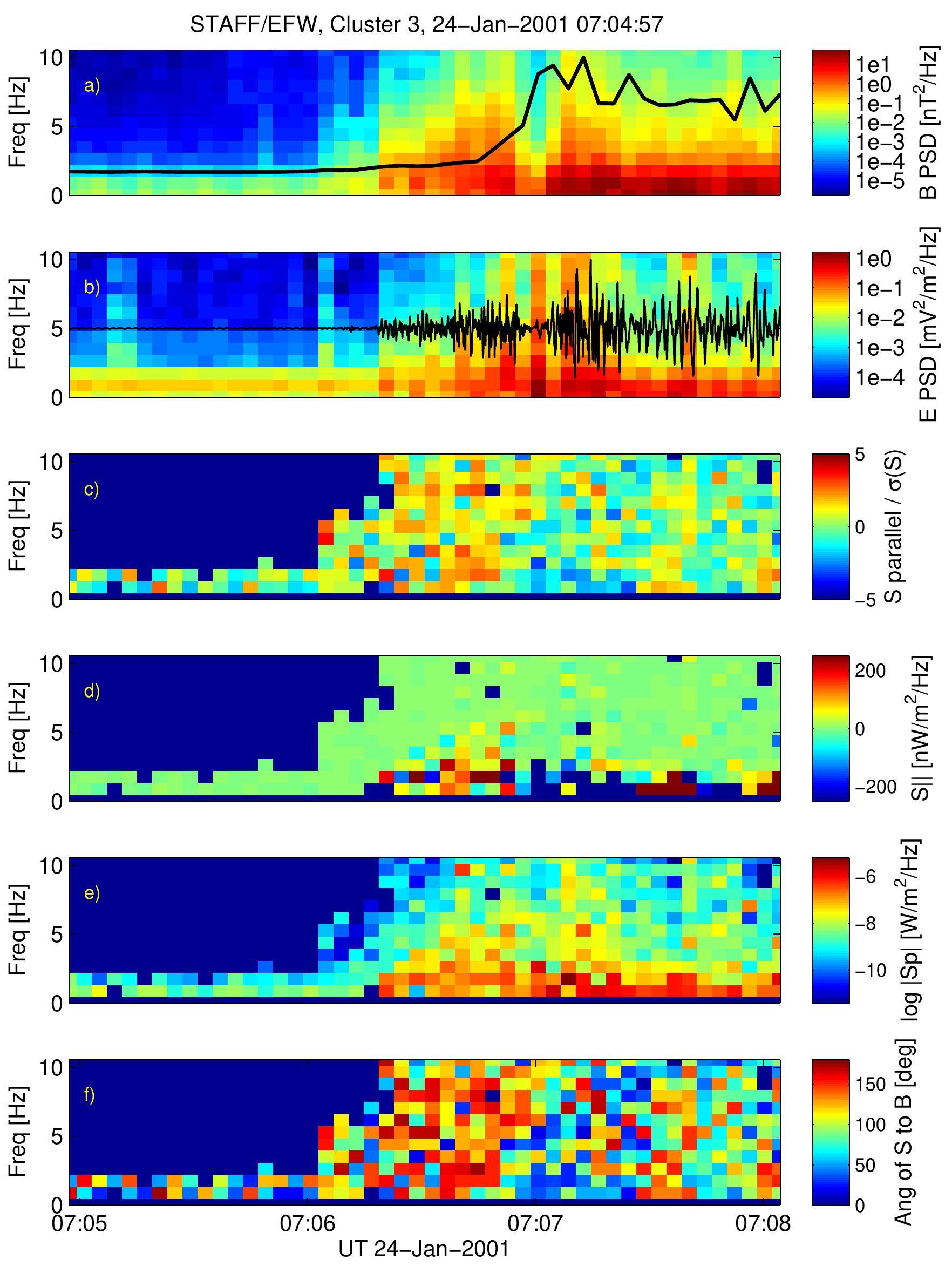}
\caption{\label{fig:reforming_shock} Poynting flux 
derived from electric and magnetic fields for a high Mach number shock. a) Wave magnetic field and averaged $B_0$. b) Wave electric field. c) $S_{||}$ normalized by its standard deviation (yellow and red corresponds to upstream flux). d) $S_{||}$. e) $\log_{10} S_{||}$. f) Angle of S to $\mathbf{B}_0$ (red meaning upstream). \citep[Adapted from][]{sundkvist12:_disper_natur_high_mach_number}}
\end{figure}

The second analyzed shock crossing on 24-Jan-2001 is shown in Figure~(\ref{fig:reforming_shock}). 
This is a reforming high Mach number shock ($M_A\sim11$) and has been analyzed in detail in \cite{lobzin07:_nonst_mach}. 
Both of the shocks discussed by  were analyzed using wavelet as well as Fast Fourier Transform (FFT) dynamic spectra techniques. 
We present the second shock using the FFT analysis, to show that the conclusions are not technique dependent.
The upstream whistler waves, Figure~(\ref{fig:reforming_shock})a,b, again have an overall Poynting flux upstream, away from the shock in the normal incidence frame, evident from the red and yellow (upstream) colors of $S_{||}$ (Figure~(\ref{fig:reforming_shock}), panels c through f). 
For this shock the ambient magnetic field was directed in the opposite direction, so that  $180^{\circ}$ (red) means upstream in Figure~(\ref{fig:reforming_shock}). 

The power flux given by the Poynting vector shows unambiguously that they carry energy over a broad frequency range from the shock ramp towards the upstream solar wind, starting from the position of the shock front. This leads to conclusion that the results of the analysis are consistent with a theoretical model \citep{galeev88:_fine, krasnoselskikh02:_nonst}  that considers the shock steepening to be balanced by the effect of dispersion in addition to dissipation. As the shock steepens, nonlinearities transfer energy to shorter wavelengths of the spectrum, and is ultimately carried away from the shock as dispersive whistler wave trains. 
This analysis demonstrates that for high Mach number shocks, dispersive effects are dominant for the formation and stability of the shock front. 
Since the whistler waves are strongly damped upstream of the shock, we infer that they can play the role of an intermediate step in the energy re-partition problem, with the energy ultimately being dissipated through wave-particle interaction.

\section{Electron Heating Scale at High Mach number Quasiperpendicular Shocks}
\label{sec:elec_heat}

From the discussion in the previous sections the energy repartition amongst particle populations in quasiperpendicular shocks is a multi-scale process related to the spatial and temporal structure of the electromagnetic fields within the shock layer. While the major features of the large scale
ion heating are known, the electron heating and smaller scale fields
remain poorly understood and controversial. In this section we will discuss the scale of the electron temperature gradient based on the possibility of obtaining unprecedented high time resolution electron distributions measured in situ by the Cluster spacecraft recently discussed by \citet{schwartz11:_elect_temper_gradien_scale_collis_shock}. The authors discovered that approximately half of the electron heating coincides with a
narrow dispersive layer several electron inertial lengths
($c/\omega_{pe}$) thick. Consequently, it gives one more argument that the nonlinear steepening is limited by wave dispersion. The DC electric field associated with the electron pressure gradient must also vary over these small
scales, strongly influencing the efficiency of shocks as cosmic ray
accelerators.

\begin{figure}
\includegraphics[width=\columnwidth]{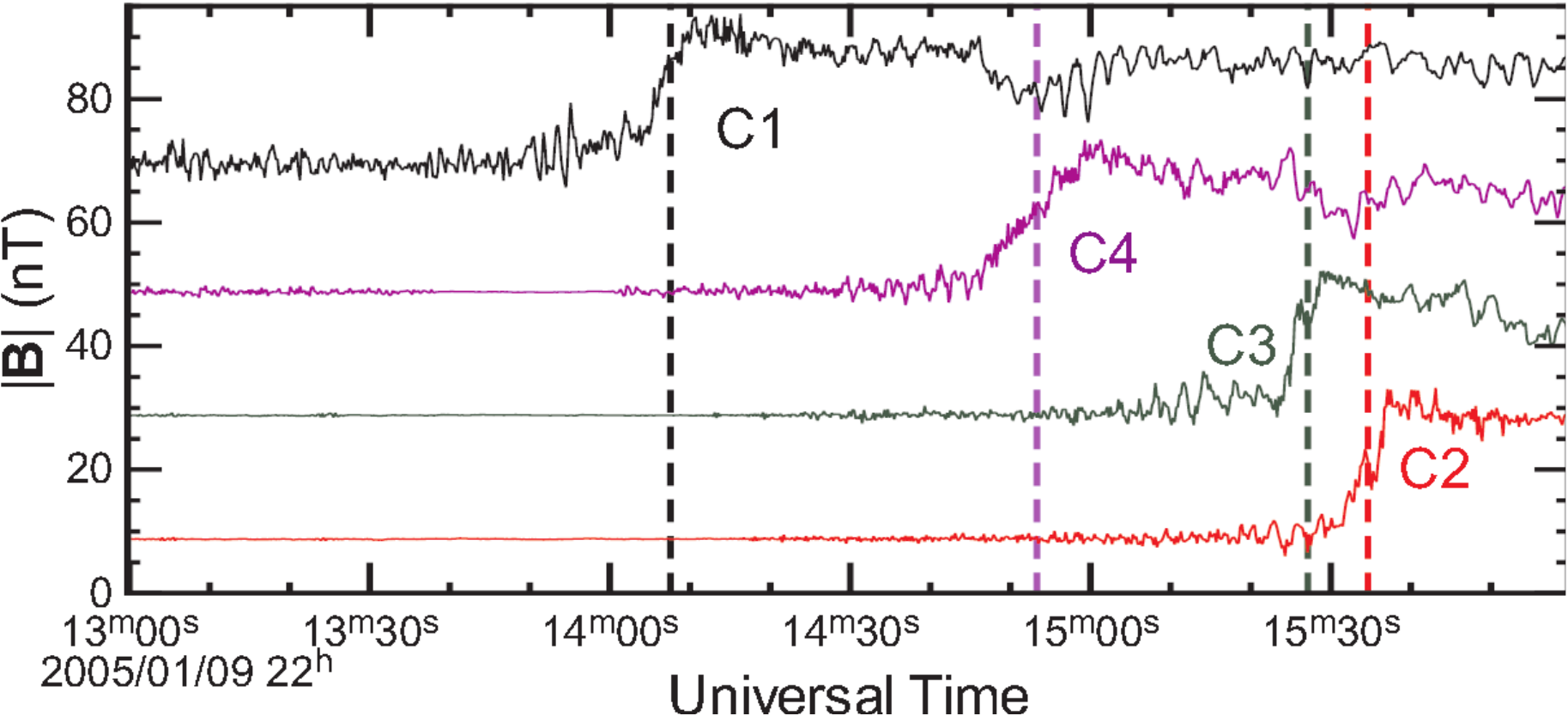}
\caption{Magnetic field data at a crossing of the Earth's bow shock by the 4 Cluster spacecraft on 9 Jan.\ 2005. Traces have been shifted by 20~nT for clarity. The dashed lines show the times of the steep ramp. \citep[Adapted from][]{schwartz11:_elect_temper_gradien_scale_collis_shock}. }
\label{fig:timing}
\end{figure}

The 4 Cluster spacecraft \citep{escoubet97:_clust_scien} are unique in their ability to remove the time-space ambiguity in time series data taken by in situ space plasma instrumentation. By timing the passage of an event at each corner of the tetrahedron formed by the 4 spacecraft, the planar orientation and speed of the event can be determined. We employ this technique to convert the time series of data to distance along the shock normal \cite{schwartz98:_shock_discon_normal_mach_number_relat_param}. Figure~\ref{fig:timing} illustrates the identification of the steep shock ramp that we use as event times.

The electron instrument on Cluster measures fluxes at several energies in a half-plane containing the spacecraft spin axis. These measurements form an azimuthal wedge divided into 12 polar directions from aligned to anti-aligned with the spin axis, and are repeated at 125--250\,ms intervals. A full 3D distribution covering all azimuths is thus built up over 1  spin ($\sim 4$\,s). However, when the magnetic field is roughly aligned with the  spin axis, each  wedge contains a full set of pitch angles from $0^\circ$ to $180^\circ$. Under these circumstances, and assuming gyrotropy, the full pitch angle distribution function is available at $\leq 250$\,ms resolution.

We rebin the raw electron data into pitch angles $\alpha$ relative to the instantaneous magnetic field. We calculate pseudo-densities and temperatures  for each pitch angle bin as if the distribution were isotropic, e.g., $n(90^\circ) = 4\pi \int f(v,\alpha=90^\circ)\,v^2 dv$. These pseudo-moments better characterise the phase space distributions in the $\parallel, \perp$ directions than the full $T_{\parallel,\perp}$ moments \citep[cf.\ Fig~9 of][]{mitchell12:_elect}.

\subsection{Results and Conclusions}

\begin{figure}
\includegraphics[width=\columnwidth]{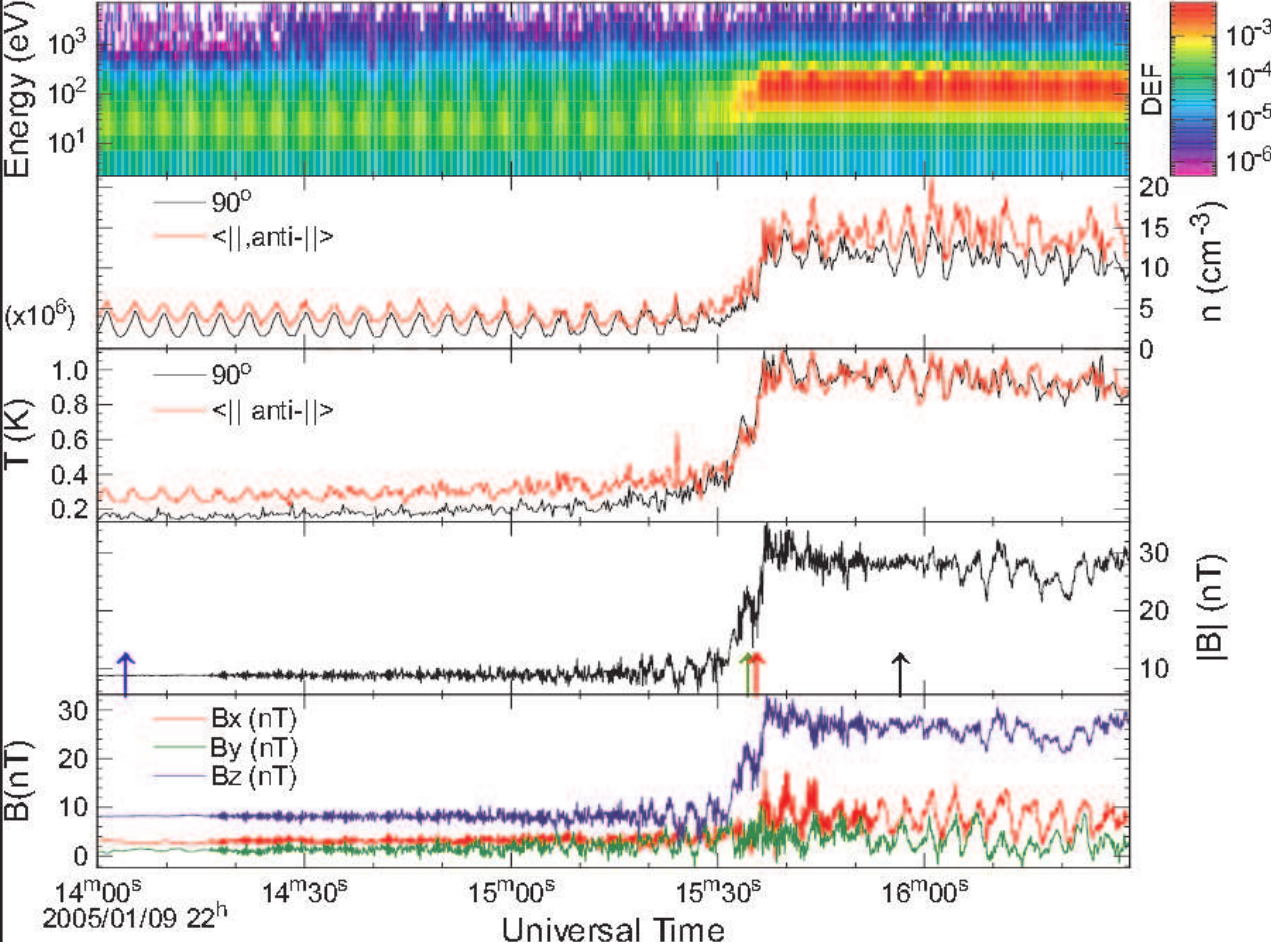}
\caption{Overview of data from Cluster 2 on 2005 Jan 9. From top to bottom: Omni-directional electron energy-time spectrogram @ 250\,ms resolution, electron pseudo-density, electron pseudo-temperatures (see Method), magnetic field magnitude, and field components. Arrows in the fourth panel show locations of the cuts presented in Figure~\ref{fig:cuts}. \citep[Adapted from][]{schwartz11:_elect_temper_gradien_scale_collis_shock}. }
\label{fig:overview}
\end{figure}

An overview of the data for 2005 Jan 9 is shown in Figure~\ref{fig:overview}. The transition from unshocked solar wind plasma to the shocked magnetosheath occurs around 22:15:30.  Although the solar wind flow is a factor of 10 slower than the electron thermal speed, some residual modulation at the spin period is evident in the data. We have averaged the parallel  and anti-parallel ($\alpha=0, 180^\circ$) moments so that the second and third panels of Figure~\ref{fig:overview} reveal the pseudo-parallel and perpendicular moments. Note that the pseudo-densities $n(\alpha)$ are not, and from their definition above need not be, equal. The bottom two panels show increasing oscillations and a gradual ``foot'' ahead of a steeper magnetic ``ramp'' region. The dominant $\hat{\vec{z}}$ magnetic field component is nearly aligned with the spin axis, enabling the parallel and perpendicular moments to be available in every $0.25$\,s wedge as described above. Figure~\ref{fig:overview} already suggests the main result namely that the rise in electron temperature closely follows even the steepest ramp of the magnetic field.

\begin{figure}
\includegraphics[width=\columnwidth]{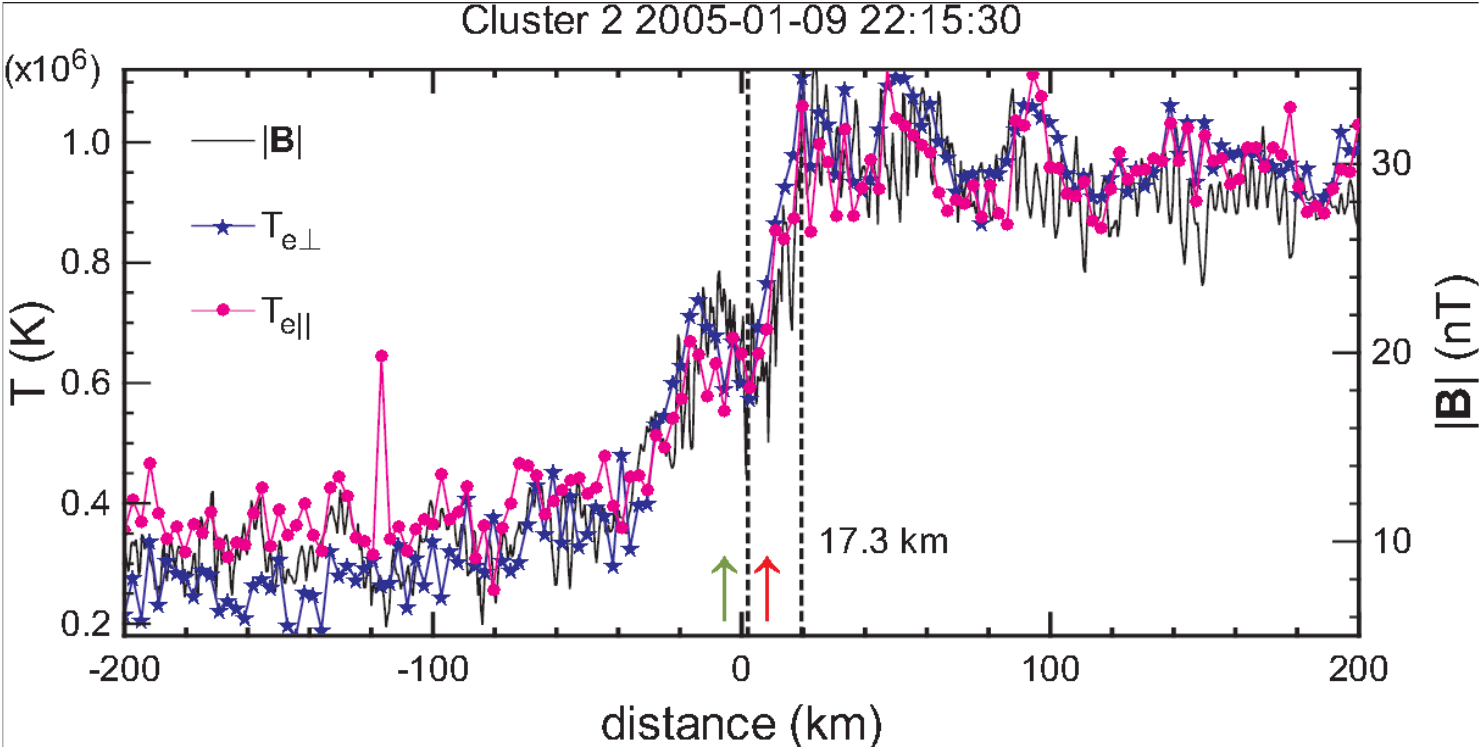}
\caption{Magnetic field (solid) and electron temperature (symbols) as a function of distance from the shock ramp. Roughly half the temperature rise occurs within the region 17.3\,km wide between the dashed vertical lines corresponding to 6.4 electron inertial lengths ($c/\omega_{pe}$). \citep[Adapted from][]{schwartz11:_elect_temper_gradien_scale_collis_shock}.}
\label{fig:distance}
\end{figure}

Figure~\ref{fig:distance} shows that both the parallel and perpendicular electron temperatures closely track the steep rise in magnetic field, with half the electron heating taking place on a scale of 17.3\,km, corresponding to 6.4 electron inertial lengths and a small fraction (0.15) of an ion inertial length. Although much of the electron dynamics is linked to the DC electric and magnetic fields within the ramp \citep{feldman83:_elect_ebs, goodrich84, scudder95:_ae_review_physic_elect_heatin_collis,lefebvre07:_elect} and is therefore reversible (the distribution function in this limiting case might be dependent upon energy and adiabatic invariant in de Hoffmann-Teller reference frame), the fact that both $T_{e\parallel}$ and $T_{e\perp}$ rise together suggests an inflation of the particle phase space distribution that is not reversible, due primarily to the filling in and/or entrapment of electrons in regions of phase space that would otherwise be inaccessible. 

\begin{figure}
\includegraphics[width=\columnwidth]{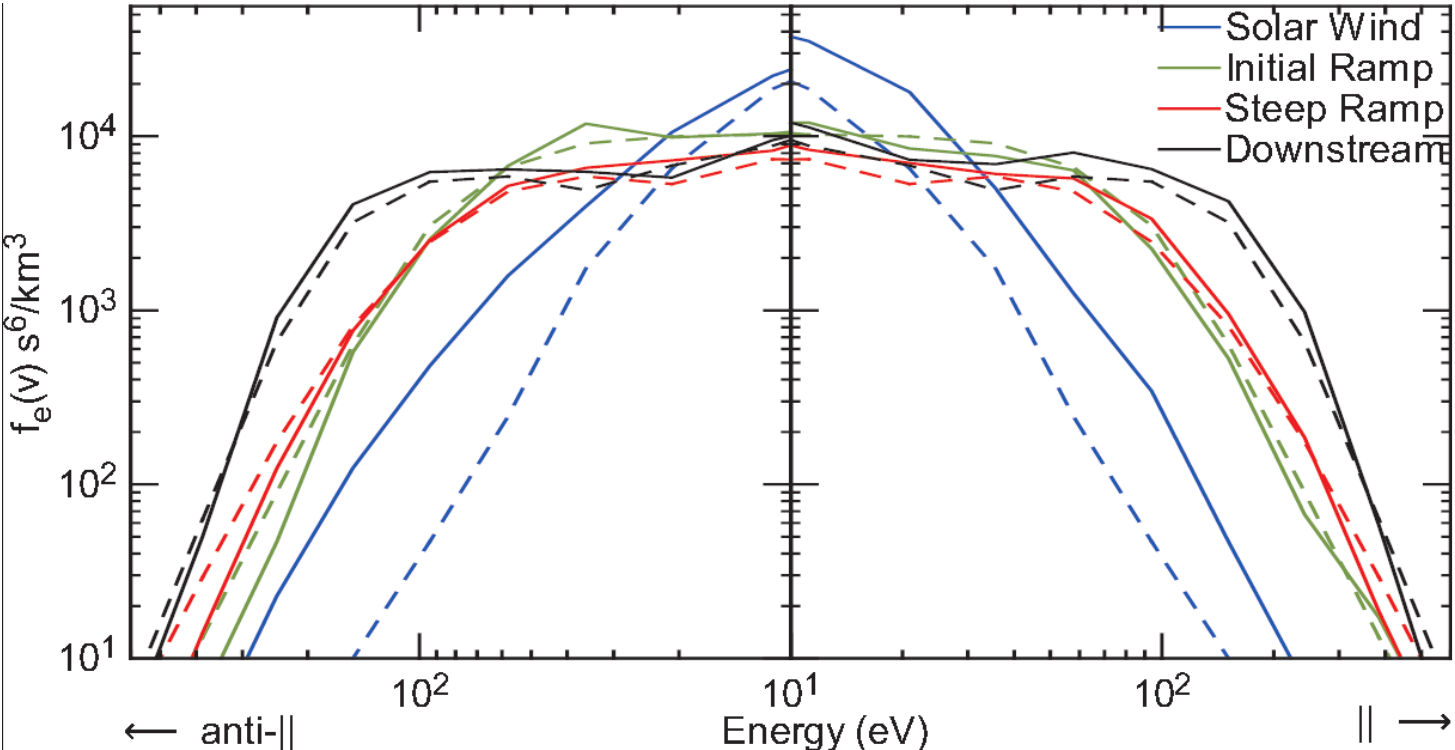}
\caption{Cuts of the electron distribution functions in the solar wind, initial ramp, steep ramp, and downstream along (solid) and perpendicular (dashed) to the magnetic field. The locations of the cuts are indicated along the axes in Figs.~\ref{fig:overview} and \ref{fig:distance}. Note the solar wind halo drift evident in the anti-aligned direction and the absence of features within the steep ramp. \citep[Adapted from][]{schwartz11:_elect_temper_gradien_scale_collis_shock}. }
\label{fig:cuts}
\end{figure}

This infilling can be seen in the cuts of the distributions shown in Figure~\ref{fig:cuts}. Within the steep ramp, the inflated distribution is evident, with the flat-topped infilled region already at its downstream level. This supports the notion that the temperature profiles shown in Figure~\ref{fig:distance} really do represent irreversible heating. Interestingly, Figure~\ref{fig:cuts} shows that features previously reported with the ramp, e.g., the beam vestige of the solar wind peak \citep{feldman83:_elect_ebs}, are present only in the more gradual initial rise that precedes the steep ramp. That beam has been totally eroded by the time this electron scale ramp is encountered.

Thus the electron heating occurs over scales that are significantly smaller than the convected proton gyro-scale $V_n/\Omega_{ci}$ invoked in \citet{bale03:_densit_trans_scale_quasip_collis_shock} and also smaller than the ion inertial length that might be anticipated due to micro-instabilities within the shock current layer \citep{papadopoulos85:_microin, matsukiyo06:_mach}. 

Recent statistical studies \citep{hobara10:_statis} argued that previous fits to a proxy of the plasma density profile \citep{bale03:_densit_trans_scale_quasip_collis_shock}  mixed contributions from the more extended foot region governed by reflected gyrating ions. Restricting the measurements to just the steep ramp, they report  widths in the range 3--55\,$c/\omega_{pe}$ with a decreasing trend as the Mach number increases. They interpreted their work in terms of shock steepening limited by the dispersion of electron whistler waves, with dispersion relation $\omega = \Omega_{ce} \cos \theta_{Bn} \left(k^2 c^2/\omega_{pe}^2\right)$. The limiting case of a wave capable of phase standing in the incident flow has a wavelength that can be written 
\[ 
\frac{\lambda}{c/\omega_{pe}} = 2\pi \frac{\cos\theta_{Bn}}{M_A} \sqrt{\frac{m_i}{m_e}} 
\]
The results from Table~\ref{tab:params} yield a value of 9.2 for this ratio, comparable to the 6.4 electron inertial lengths given above. The fact that supercritical shocks steepen to this whistler limit suggests that dissipation processes are insufficient to broaden the transition further.

\begin{table}
\caption{Shock Parameters 2005 Jan 9 @ 22:15}
\label{tab:params}
\begin{tabular}{lrl}
\hline
Parameter & \textnormal{Value} & \\\hline
$V_{shock}$  & +10.8&\kms \\
Unshocked magnetic field $\vec{B}_u{}^\dagger$ & (3.07, 1.35, 8.14)&nT\\
Unshocked electron density & 4.0 &cm$^{-3}$\\
Location (Earth radii) & (12.3, 13.3, -6.7)& $R_e$\\
$\hat{\vec{n}}$ shock normal (timing) & (0.855, 0.418, -0.307)&\\
$\hat{\vec{n}}$ (model) \cite{schwartz98:_shock_discon_normal_mach_number_relat_param} §& (0.904, 0.383, -0.189)& \\
$V_n \equiv \vec{V}\cdot\hat{\vec{n}} $ (shock rest frame) & 373 & \kms \\
Alfv\'en Mach no. $M_A$ & 3.8 &\\
Magnetosonic Mach no. $M_{ms}$ & 3.0 & \\
$\theta_{Bnu} \equiv \angle\vec{B},\hat{\vec{n}}$ & 83 &${}^\circ$\\
Plasma ion $\beta_i$ & 0.4 &\\
Plasma electron $\beta_e$ & 0.34 & \\
Electron inertial length $c/\omega_{pe}$ & 2.7  & km\\
Ion inertial length $c/\omega_{pi}$ & 117 & km\\
$V_n/\Omega_{ciu}{}^{\dagger\dagger}$ & 443 & km\\
$V_n/\Omega_{cis}$ & 139 &km\\
Whistler wavelength $\lambda$ & 24.8 & km\\
Electron Larmor radius $r_{Le\,u}$ & 1.01 & km\\
\hline
\multicolumn{3}{p{0.95\columnwidth}}{${}^\dagger$All vectors are in the GSE frame of reference. Subscripts ``$s$'' (``$u$'') denote quantities in the (un)shocked region.}\\
\multicolumn{3}{p{0.95\columnwidth}}{${}^{\dagger\dagger}\Omega_{ci}\equiv eB/m_p$ is the proton gyrofrequency}\\
\end{tabular}
\end{table}

It should come as no surprise that the steepening of a fast mode (right-handed) wave results in a right-handed whistler signature. Indeed, the non-coplanar component of the magnetic field \citep{thomsen87}, responsible for the difference in the shock electrostatic potential when viewed in different shock rest frames \citep{goodrich84}, is right-handed.  There is new evidence \citep{sundkvist12:_disper_natur_high_mach_number}  that the wave Poynting flux is directed away from the ramp region upstream as expected for dispersion-limited steepening. 

The present study  measures directly the actual temperature profile of the electrons. The result confirms that nonlinear steepening proceeds down to scales limited by whistler dispersion. We have argued that this represents irreversible heating, implying that dissipation is operative on this, or probably smaller, scales.

We have attempted a similar analysis on other shock crossings observed by Cluster, with consistent findings. Suitable events are rare, since they require the combination of a slowly moving shock and favorable magnetic field orientations. Future space missions need to be proposed to target electron physics and hence should provide numerous examples for statistical studies.

\begin{figure}
\includegraphics[width=\columnwidth]{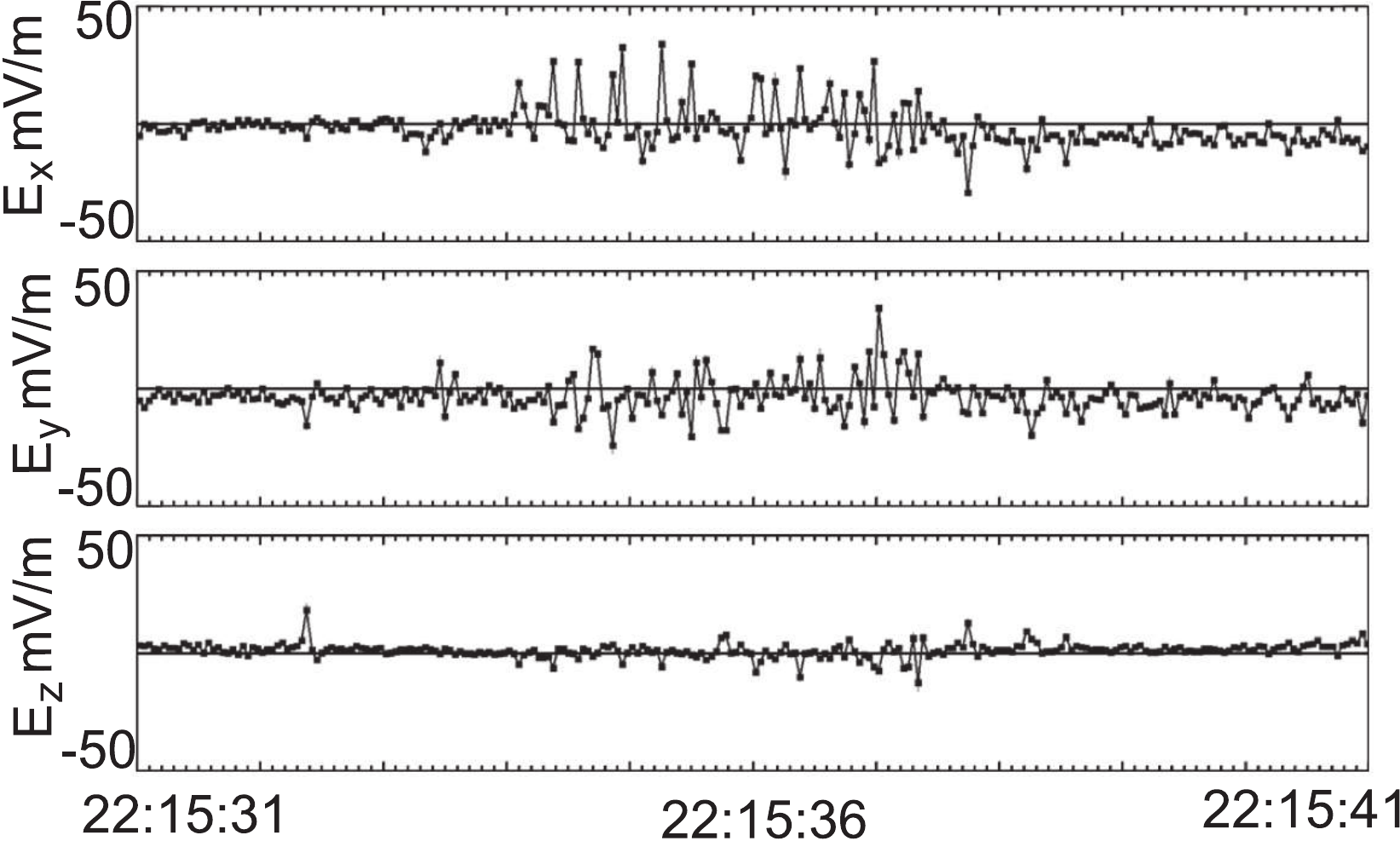}
\caption{Electric field measured onboard Cluster 2 satellite from 22:15:31 to 22:15:41. Electric field bursty spikes having amplitudes of 20-40\mvm having duration of the order of 0.1 sec are clearly seen during ramp crossing. Figure is provided by F. Mozer.}
\label{fig:spikes}
\end{figure}

What process(es) are actually responsible for \hbox{(sub-)}whistler-scale dissipation? The overall inflation in phase space is linked to the action of the cross-shock electrostatic potential in concert with the magnetic mirror forces. Some or all of the potential may be concentrated in intense spikes \citep{bale07:_measur} that may break the adiabaticity of electron phase space trajectories despite a ramp thickness which, in our example, is 20 times the local electron gyroradius. It is worth noting that the localized spikes of the electric field are present inside the ramp region. Figure~\ref{fig:spikes} represents 10 second interval of electric field measurements inside the ramp by Cluster 2, from 22:15:30 to 22:15:40 that is relatively short time with respect to time of the shock crossing but corresponds to ramp region (courtesy of F. Mozer). One can clearly see quite intense bursts of the electric field having amplitudes as large as 20-30\mvm. These bursts are very similar to those reported in Section~\ref{sec:magneticscales}. Such electric field bursts can be one of the possible sources of electron heating and scattering.    

Another candidate processes \citep[e.g.][]{balikhin94:_kinem} responsible for in-filling regions of phase space, in some of which electrons are trapped, include wave scattering \citep{scudder86:_3, veltri93:_elect_2} and demagnetization \citep{balikhin94:_kinem}; these will require further analysis and simulations. 

Our discovery of short scale electron heating has an important consequence for  electron and ion acceleration. Gradient drift and surfing mechanisms are sensitive to the scale of the field transitions \citep{zank96:_inter}, becoming very efficient at scales comparable to those reported here.

\section{Can anomalous resistivity account for energy dissipation and electron heating.}
\label{sec:anom_res}

The major results reported in this Section were first published in \citet{balikhin05:_ion, walker08:_lower}.

\subsection{Ion Sound Wave Packets at the Quasiperpendicular Shock Front.}

\begin{figure}
\centering
\mbox{\subfigure{\includegraphics[width=1.8in]{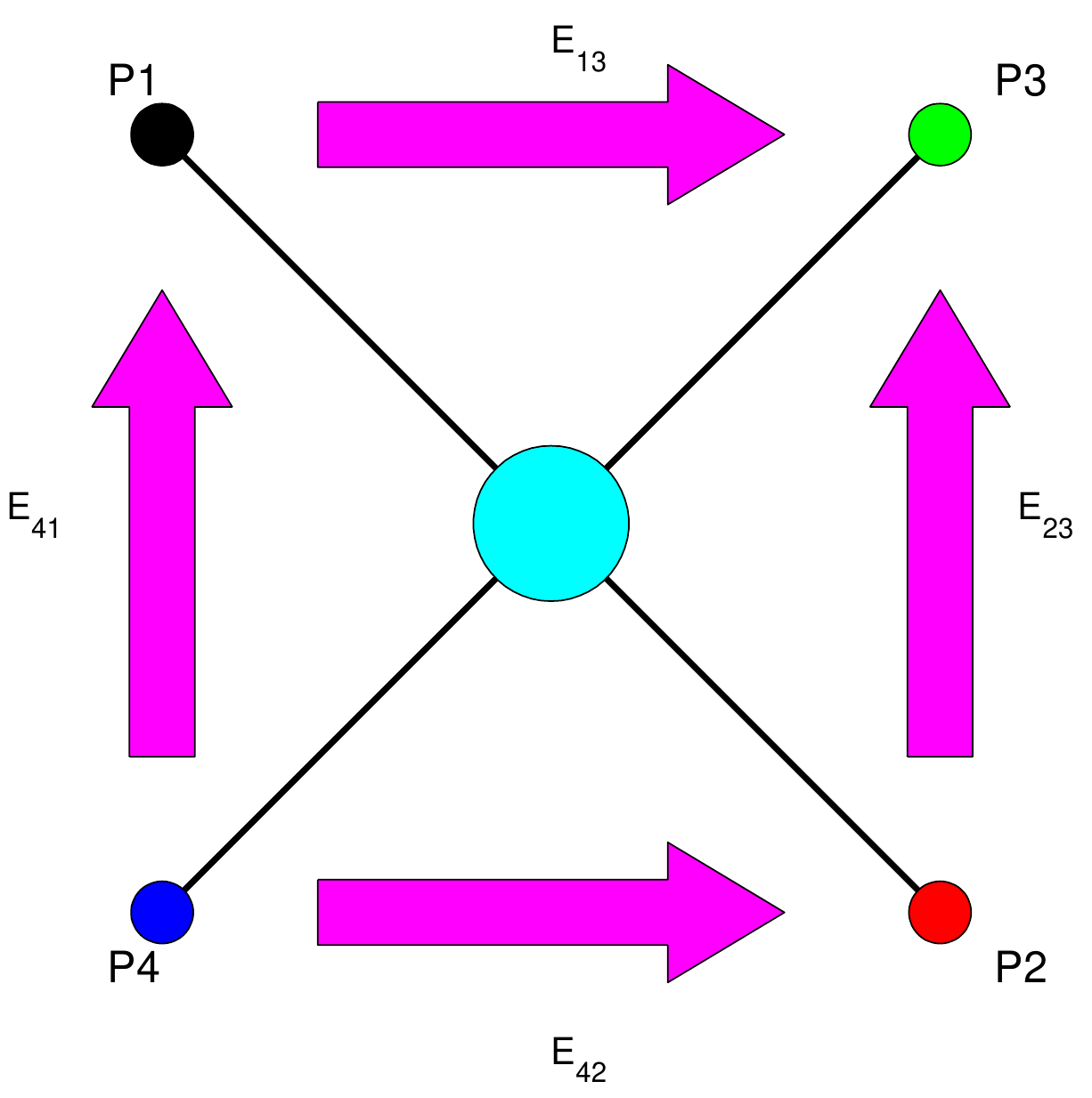}}\quad
\subfigure{\includegraphics[width=2.6in]{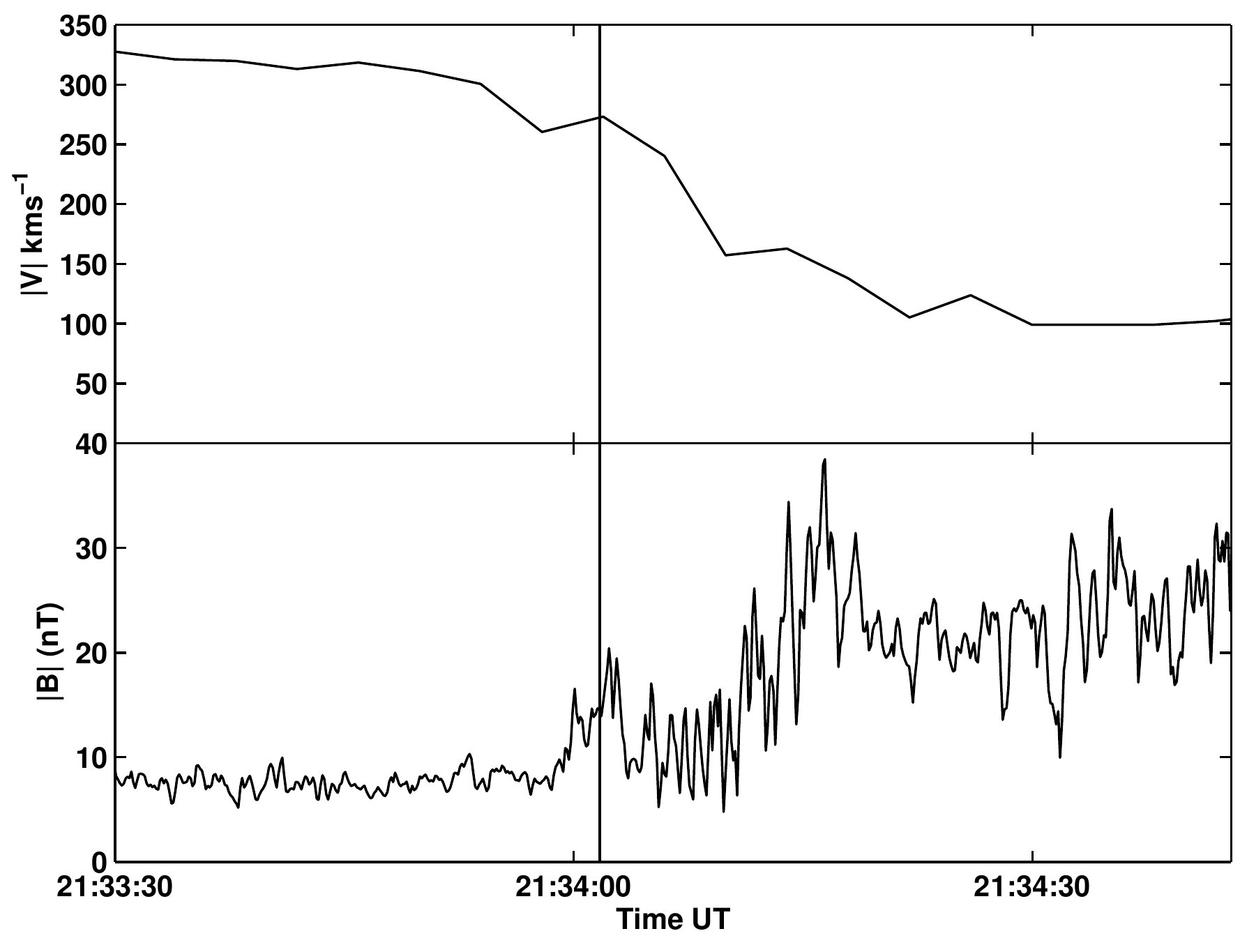} }}
\caption{The left panel shows the configuration of the EFW electric field probes and illustrates the electric fields calculated from them. The right hand panel shows the magnitude of the magnetic field (bottom) and ion bulk
flow (top) measured during the shock crossing that occurred on
February $26^{th}$, 2002 at 2134 UT. \citep[Adapted from][]{balikhin05:_ion}. \label{fig5.1} }
\end{figure}

The data used in both articles presenting observations of short scale waves were collected by the EFW instrument on board Cluster satellites. EFW uses two pairs of spherical probes in the satellite spin plane situated on the ends of wire booms whose length is 44 m as shown in the left hand panel of Figure~\ref{fig5.1}. Thus, the distance between probes adjacent/opposite to one another is $\sim 62/88 {\rm m}$ respectfully. Normally, the EFW instruments return the electric field calculated as the difference in probe potentials between probes 1 and 2 ($E_{12}$) and 3 and 4 ($E_{34}$) with a sampling rate of either 25Hz (normal science mode) or 450Hz (burst science mode). The individual probe potentials are also available with a time resolution of 5Hz. In addition to these standard 
modes, there is a triggered internal burst mode. Using this mode, data for a 
short time period may be captured with a much higher sampling rate. The EFW data 
that has been analysed in this study consists of internal burst mode data 
comprising the four individual probe potentials sampled at 9kHz for 
periods of around 10 seconds. Since the internal burst data is captured and 
stored depending upon some criteria, it may be that although the shock region 
was targeted for data collection, the waveforms returned may not have been 
captured in the shock front itself. To this end, a search was made to find 
possible candidate events by cross referencing the list of Cluster shock 
crossings for 2002 with the list of periods for which internal burst data are 
available. This resulted in a list of 10 possible events. Of these events, a 
comparison between the FGM magnetic field measurements and the time periods for 
which EFW internal burst data were available showed that there were only two 
shocks for which the period of internal burst data lay solely in the foot region 
of the shock. Of these, one shock possessed a magnetic profile that was highly 
turbulent and difficult to interpret and was also eliminated from further analysis. This left just one clean shock on which to perform the analysis.

The internal burst data sets are the only ones generated by 
EFW that contain the individual probe potentials at a high enough sampling rate 
to investigate waves and turbulence at frequencies around the lower-hybrid 
frequency in the vicinity of the terrestrial bow shock (10-30Hz). By 
using the individual probe potentials it is possible to compute two parallel 
electric field components one on either side of the satellite. For example, the 
probe pairs 1,3 and 4,2 maybe used to compute electric field components $E_{13}$ 
and $E_{42}$ whose directions are parallel and are spatially separated by a 
distance of $\sim $62.2m. This technique has previously been used by 
\citet{balikhin05:_ion} and \citet{tjulin03:_lower} to study small scale 
electric field structures and waves and is similar to the short baseline interferometry techniques employed in the analysis of data from sounding rockets \citep{pincon97:_obser}.
Since the probe potentials can be used to calculate two parallel electric field 
vectors it is possible to use the phase differencing technique to determine the wave vector ${\bf k}$. This method may also be used to examine the polarisation 
characteristics of the wave in question. In this case, the phase differencing 
algorithm is applied to a pair of perpendicular components of the electric field 
(as opposed to the parallel field components mentioned above). The resulting 
histogram of the phase difference as a function of frequency yields a vertical 
line of constant phase difference with respect to frequency at a phase 
difference of zero for a linearly polarised wave and $\pm \pi /2$ for a 
circularly polarised wave. Thus, this technique may be used to help distinguish 
between a linearly polarised lower hybrid wave and a circularly polarised 
whistler mode wave, both of which have been observed at these frequencies. This method is used in preference to an examination of the coherency \citep[see for example][]{krasnoselskikh91} due to the short duration of the wave packets.

The magnetometer data, used to put the electric field measurements into context 
within the shock front and compute the lower hybrid resonance frequency, come 
from the FGM instruments \citep{balogh97:_clust_magnet_field_inves} and made publically available 
through the Cluster Active Archive. These measurements typically have a sampling 
rate of 22Hz.

\begin{figure}
\centering
\includegraphics[width=3in]{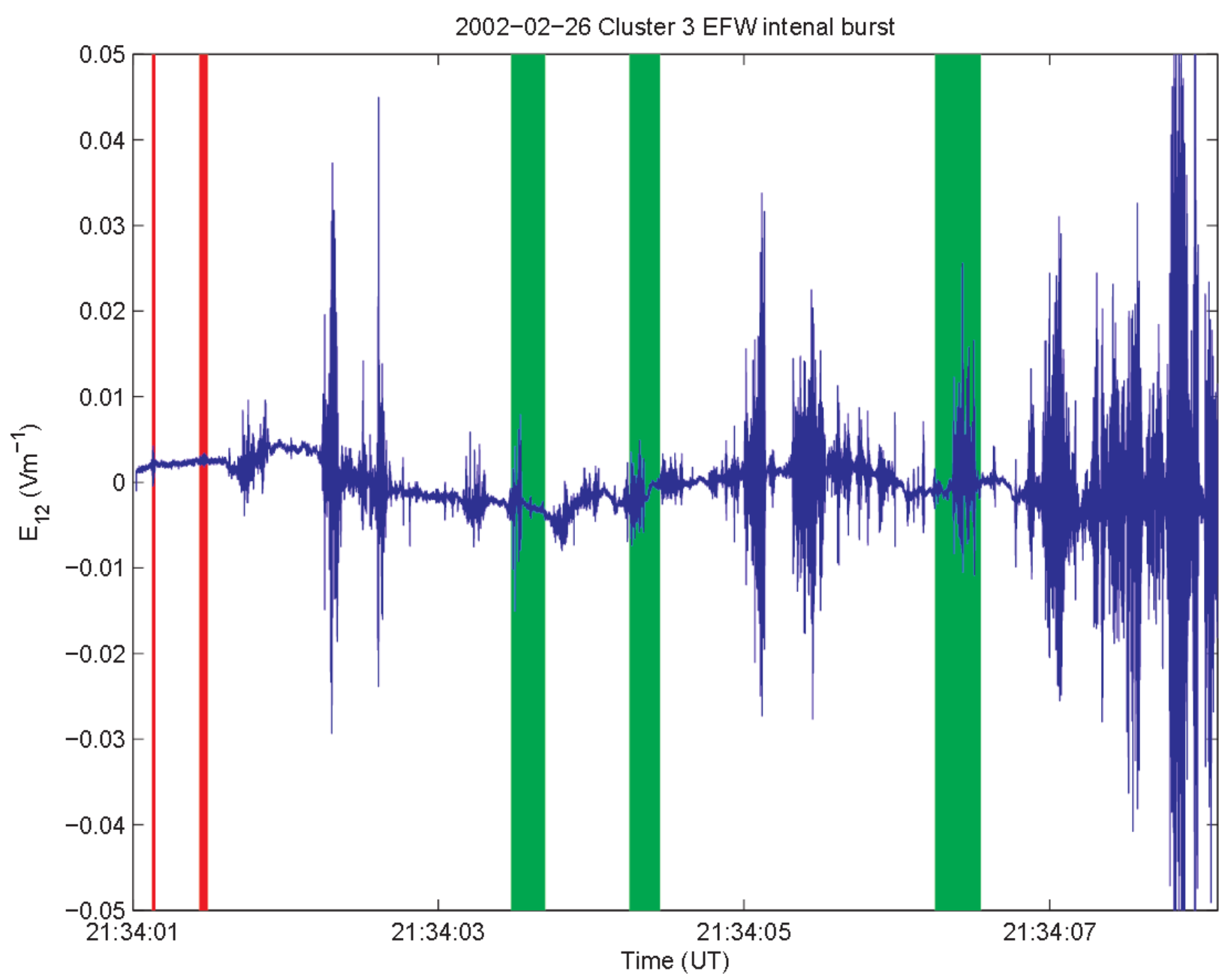}
\caption{Waveforms of electric field measurements during shock front crossing on 
February $26^{th}$, 2002 at 2134 UT. The vertical lines mark periods where the waves were registered. As it will be shown later red vertical lines mark periods where the waves were identified as ion-sound, two first green columns as lower-hybrid electrostatic waves, third green column as whistler waves in lower hybrid frequency range.
\label{fig:times_is_lh}}
\end{figure}

All the data presented in this Section were recorded during one shock crossing on  February $26^{th},$2002 at around 2134 UT during the time intervals marked on Figure~\ref{fig:times_is_lh} by vertical lines. Red lines mark the periods of registration of ion sound waves, green lines the periods of registration of lower hybrid and whistler waves. 

During this period the Cluster satellites were situated in the foot region of a
quasiperpendicular shock ($\theta_{Bn}\sim 55^{\circ}$, $M_{A} \sim 4.3$). The EFW instrument onboard Cluster 3 was triggered to operate in internal burst mode for a few seconds. 

Two parallel electric field vectors of these electric field measurements lie
in the same direction and have a perpendicular separation of $\sim 62.2$m in the direction $P_{2}$ to $P_{3}$. The availability of two closely spaced, simultaneous measurements enables the use of phase differencing techniques
\citep{balikhin97:_exper} for the identification of propagation modes for waves with coherence lengths down to a few Debye lengths based upon single satellite
measurements. Since there is no component measured normal to the spin
plane, the separation between temporal and spatial variations is possible
only in the spacecraft spin plane. As a consequence, phase differencing
methods are limited to the determination of the projection of the
${\bf k}$-vector in the spin plane. In most cases, however, this can provide
enough information to identify the plasma wave mode. This approach was 
implemented in these studies. Plasma measurements were obtained from the CIS HIA (ions) and PEACE (electron) instruments. Magnetic field data were obtained from FGM. It should be noted that the spin vector of the Cluster satellites is almost coincident (to within 5$^{\circ}$) with the $z$ GSE axis.

The ion bulk velocity (top panel) and the magnitude of the magnetic field
(lower panel) as measured by Cluster 3 spacecraft are plotted in the right hand panel of 
Figure~\ref{fig5.1}. Initially, the spacecraft was in the solar
wind. The foot region was encountered just before 2134UT and the shock ramp was crossed around 2134:12.5UT. The plasma bulk velocity began to decrease around 2133:50. Shortly before 2134 low frequency oscillations were observed in the magnetic field, a feature commonly observed in the foot region of supercritical shocks. The beginning of the foot region is characterised by a large amplitude, nonlinear structure similar to those previously reported by \citet{walker99:_ramp}. A comparison of magnetic field and plasma data show that this structure is not a partial penetration of the ramp. The present study is limited to the short interval at the beginning of the internal burst mode indicated by the vertical line and coincides with the foot region. 

\begin{figure}
\centering
\mbox{\subfigure{\includegraphics[width=2in]{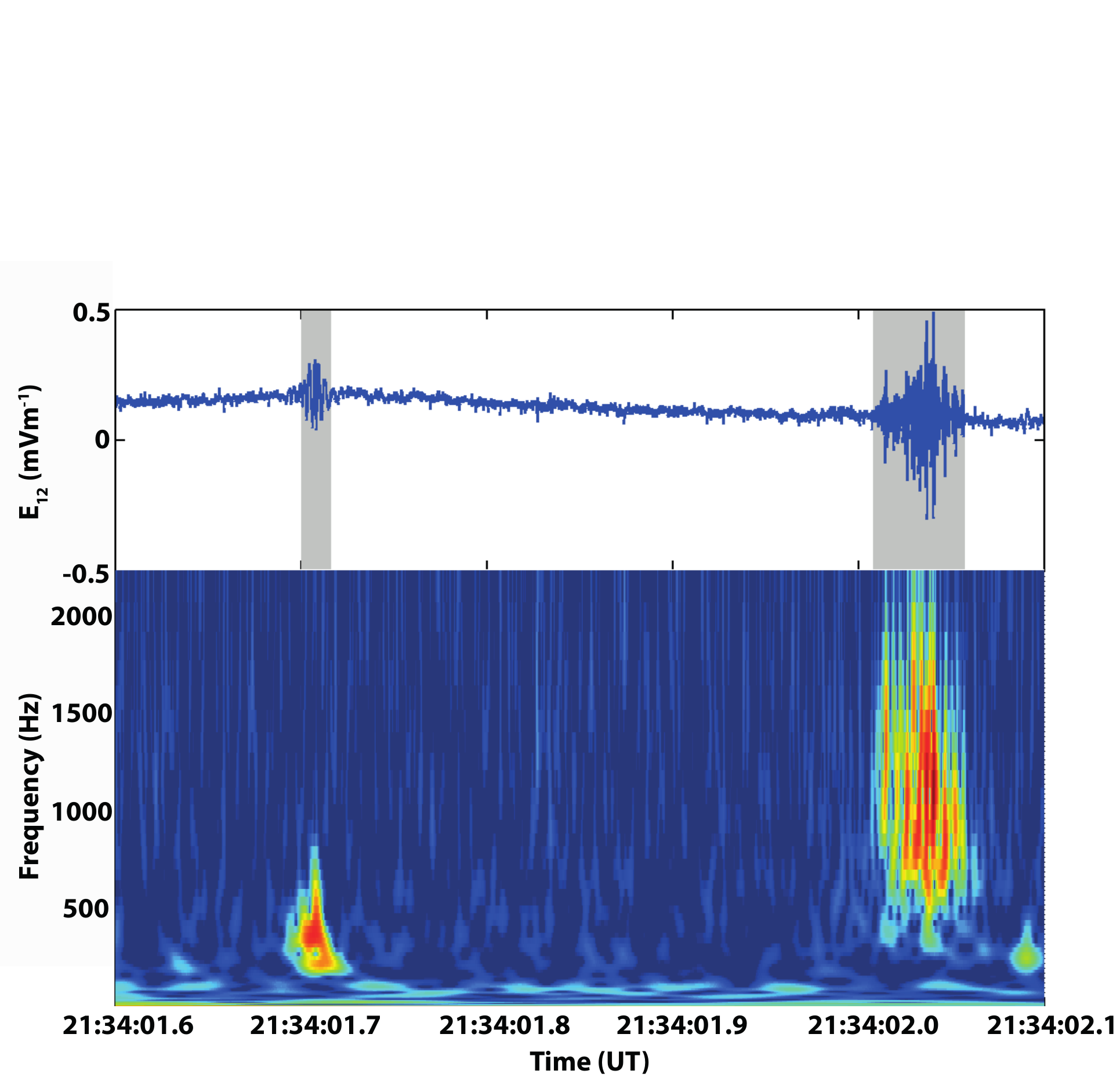}}\quad
\subfigure{\includegraphics[width=2.5in,viewport=0 190 440 430,clip=true]{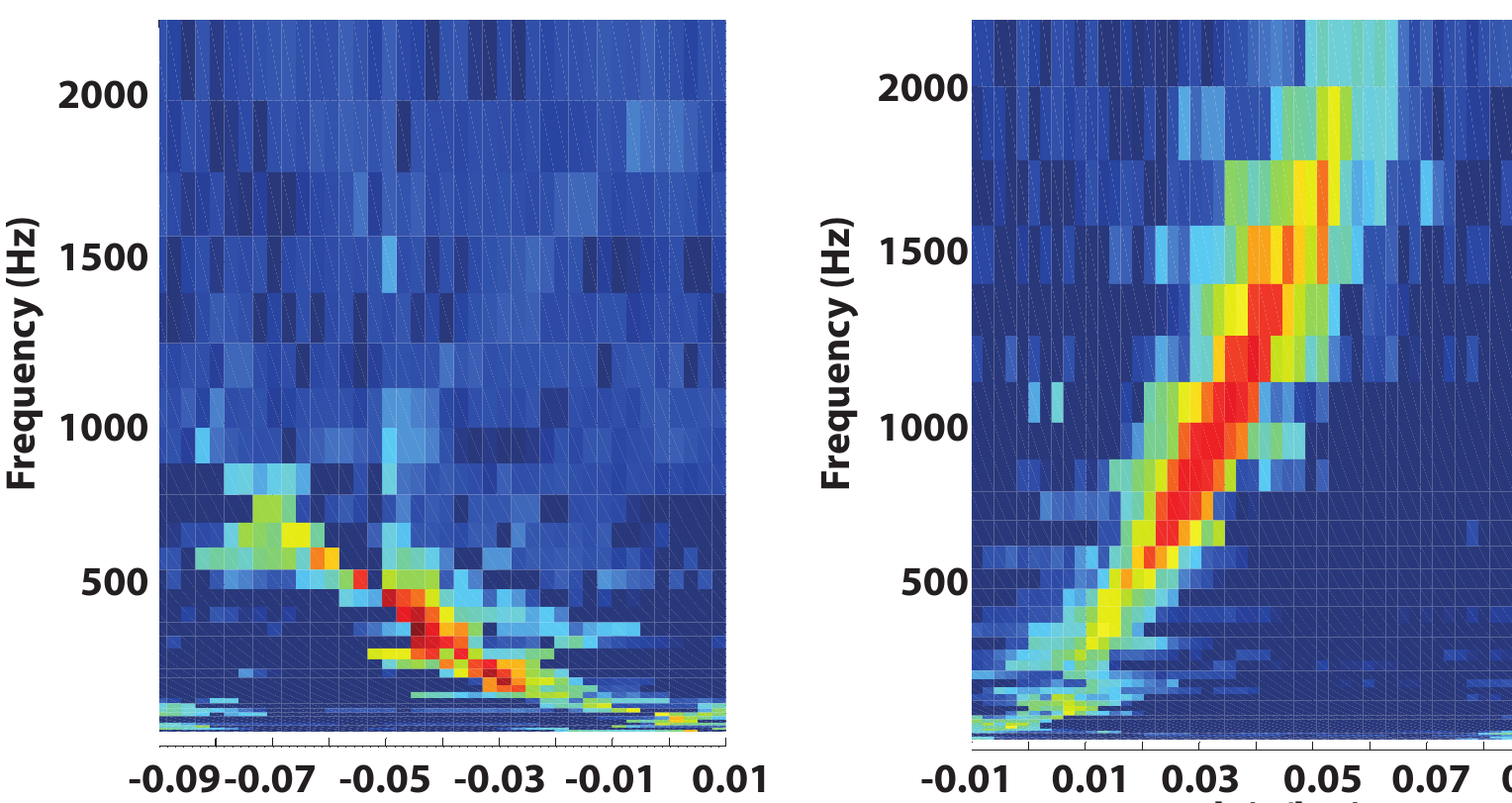} }}
\caption{Left: The waveform (top panel) and wavelet spectrogram (bottom panel)of the electric field computed from the difference in potential between
probes 3 and 1. Centre and right: Examples of the $f-k$ spectrograms for the first (left) and second (right) wave packets. \citep[Adapted from][]{balikhin05:_ion} \label{fig5.2}}
\end{figure}

The electric field component $E_{31}$ as measured during the initial part
of the internal burst mode interval is shown in the upper left panel of
Figure~\ref{fig5.2} and its Morlet wavelet spectra is shown in the lower left panel. The electric field fluctuations show a pair of well defined wave packets centered around 2134:01.6 and 2134:02.05UT. Their frequency ranges are 100-800Hz and 250-2000Hz respectively. We present here the results of the identification of these wave packets to illustrate the use of the technique and its results. 

The $f-k_{23}$ spectrum, as shown in the center and right hand panels of Figure~\ref{fig5.2}, is a histogram representation of the distribution of wave energy in frequency-$k$ space for the first wave packet \citep{balikhin97:_exper}. The $f-k$ spectrum shows a well developed ridge like maxima, the shape of which indicates the wave dispersion relation projected along the $k_{23}$ direction.
This result may be combined with a similar dispersion along the $k_{13}$ to yield the wave vector projection in the satellite spin plane.

\begin{figure}
\centering
\includegraphics[width=2in]{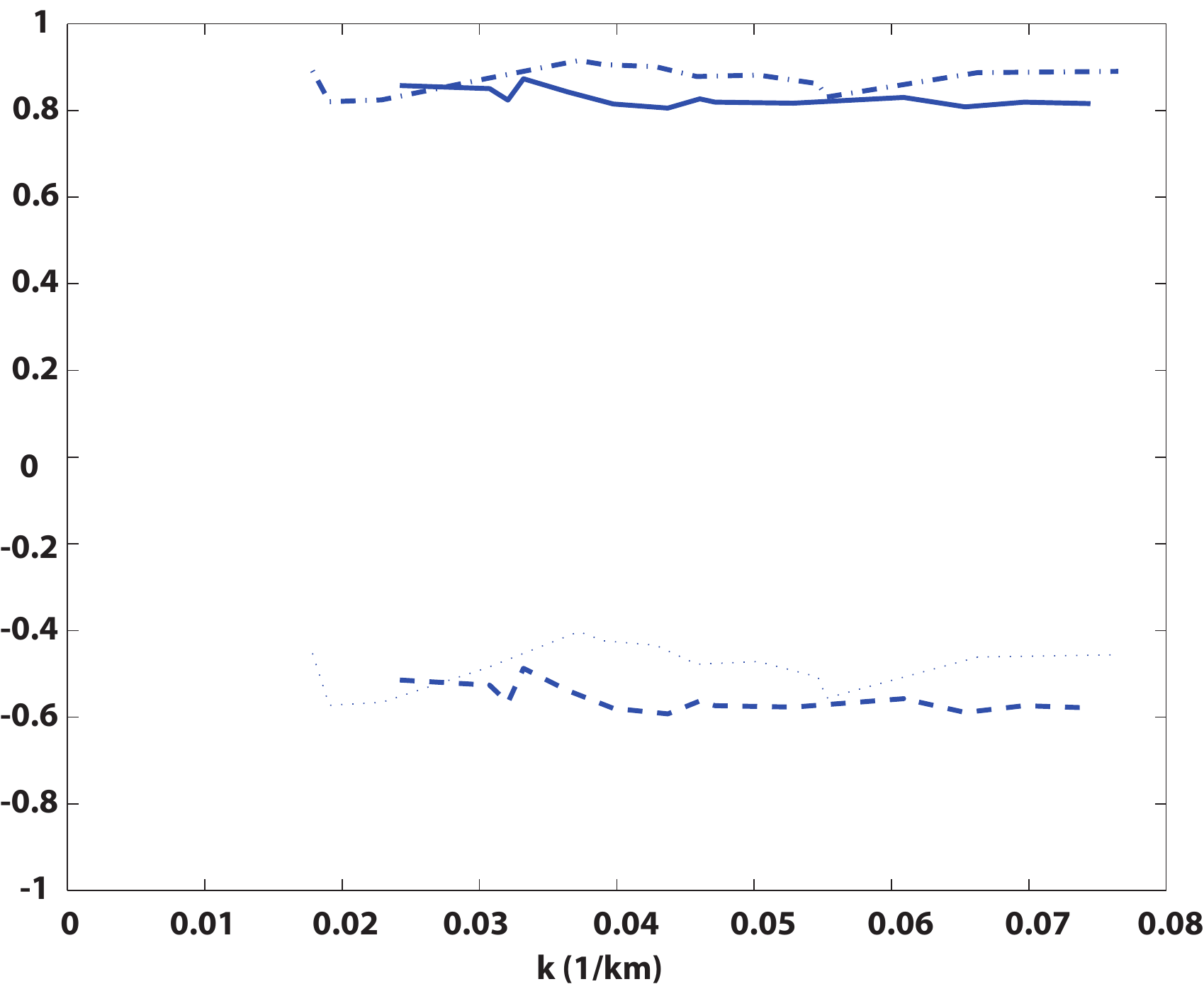}
\caption{A comparison of the wave vector directions for the two wave packets. The dotted and dashed lines represent the  X component of wave vector for events 1 and 2 respectively. The corresponding Y components are shown by the dash-dotted and solid lines. \citep[Adapted from][]{balikhin05:_ion}\label{fig5.3}}
\end{figure}

Since the angle between the spacecraft spin plane and the GSE XY plane
is small, we will consider that the projection into the spin plane is the same as that into the GSE XY plane. The projection of the dispersion relation into the GSE XY plane is shown as the solid line in Figure~\ref{fig5.3} for the first (centre anel in Figure~\ref{fig5.2}) and second (right on previous Figure) wave packets. The observed frequency range of the first wave packet ($100-800$Hz) corresponds to approximately $ 0.25-1.9
\Omega_{ce}$,  and
the magnitude of wave vector projection is in the range 0.015 $< k_1<$ 0.075m$^{-1}$. For this interval the electron temperature is $T_e \sim 17$eV and plasma density $n_i=9.7$cm$^{-3}$. This leads to an estimate for the Debye length of $\lambda_d\approx 10$m. Thus the observed values of for the projection of $k$ correspond to $\approx 8-40 \lambda_d$.

The satellite frame dispersion relation in the satellite spin plane is shown by the solid line in Figure~\ref{fig5.3}. It's phase velocity is in the range 40-70kms$^{-1}$. The Doppler shift can be estimated as the scalar product of the solar wind velocity and spin plane wave vector component. This estimation of the Doppler shift term is shown as a dashed line. It has the same sign as the phase velocity and is always greater than the observed wave dispersion indicating that in the plasma frame the waves propagate in the direction opposite to that of the
solar wind, but are convected Earthwards by the plasma flow. This convection reverses the direction of propagation in the satellite frame.  The average
angle between the spin plane projections of wave vector and the magnetic
field is about $20^{\circ}$. 

The second wave packet analysed was observed $\approx 0.3$ seconds after the first. The electric field waveforms (not shown) again indicate a good correlation between the corresponding electric field components measured by different probe pairs. The $f-k_{23}$ spectrum calculated for this wave packet is shown in the right panel of Figure~\ref{fig5.2}. The ridge like maxima in these spectra correspond to the projections of the wave dispersion relation in the direction $k_{23}$. The resulting dispersion relation is shown as the solid line in Figure~\ref{fig5.3} . Its frequency range is $250-2000$Hz ($\approx 0.6-4.9 \Omega_{ce}$), and the magnitude of wave vector projections is in the range $\approx 0.018-0.075 $m$^{-1}$. For this wave packet, the satellite frame phase velocity is in the range 150-160kms$^{-1}$. The range of wave vectors and angle of propagation with respect to the magnetic field for the second wave packet coincide with those determined for the first. Even more surprising is the fact that the angle between the two wave vector projections is less than $5^{\circ}$. The dashed line in Figure~\ref{fig5.3} shows the estimation of the Doppler shift. It can be seen that the Doppler shift term for the second wave packet is less than that of the observed frequency and so the second wave packet propagates in the same direction in both the satellite and plasma frames. Therefore the first and second wave packets propagate in opposite directions in the plasma frame. While for the second wave packet the satellite frame phase velocity is the sum of of its plasma frame velocity and the solar wind convection speed for the first wave packet it is their difference. That explains why in the satellite frame the second wave packet propagates faster than the first one.
The use of multipoint measurements enables one to separate temporal
and spatial variations. In the current study it is possible to distinguish which of these two wave modes was observed. Thus we have a method that is independent of using the observed frequency criterion formulated by \citet{gurnett85:_plasm}. For this interval $|B|\sim$14.8nT and hence the local
 electron cyclotron frequency $f_{e}={\Omega_{ce}/{2 pi}}\sim$415Hz. As can be seen from the $f-k$ spectra shown in Figure~\ref{fig5.2} that
the maximum wave energy of the first wave packet occurs at a frequency 
lower than $f_{e}$. According to the classification used by \citet{gurnett85:_plasm}
this should be a whistler wave packet whose dispersion relation may be written as (neglecting thermal corrections)
$\omega^2=\Omega^2_{ce}\cos^2\theta_{Bk} k^2c^2/(k^2c^2+\omega^2_{pe})$,
where $\theta_{Bk}$, $\Omega_{e}$, $\omega_{pe}$ are the angle between the wave
vector and the magnetic field, the electron cyclotron and electron plasma
frequencies respectively. The wave vectors for the first wave packet lie in the range $(kc/\omega)^2\sim 30-150$ and therefore correspond to
the electrostatic limit of the mode for which the plasma frame frequency 
should be $\sim\omega_{e}\cos\theta_{Bk}$. If we estimate the 
angle $\theta_{Bk}$ using the angle between the projections of
wave vector and the magnetic field in the spin plane, the plasma frame
frequency can be estimated as $f_{e}\cos\theta_{Bk}\sim$280Hz. For the wavevectors found, $0.015 <k<0.0075$m$^{-1}$ the electrostatic whistler phase velocity varies in the range $24<v_{ph}< 112$kms$^{-1}$ in the plasma rest frame. In the spacecraft frame the slowest waves would reverse their direction of propagation, so that waves propagating in both directions would be observed. However, it has been shown earlier that all waves are propagating in the same direction. Moreover, for the strongly dispersive electrostatic whistler the phase velocity should vary by a factor of two or more over the observed range of wavevectors, while the actual variation is within 20\% only. These arguments exclude the possibility that the observed mode is the whistler in the electrostatic regime.

The other possibility is the ion-sound mode. Since we are limited to spin plane measurements of wave vectors only order of magnitude estimations of the wave parameters can be made. For such crude calculations we will disregard the factor $\theta_{Bk}\sim 18^{\circ}$ in dispersion of ion-sound waves and use the simplified form $\omega=kv_{is}/\sqrt{1+k^2\lambda^2_D}$ where
$v_{is}=\sqrt{k_{b}T_e/m_i}$ is the ion-sound velocity and $k_{b}$ is Boltzmann's constant. During the time interval in which both waves packets were observed $v_{is}\approx$ 40kms$^{-1}$. Thus, in the plasma rest frame the wave phase velocity should be in the range $0.80v_{is}<v_{ph.pf}<0.99v_{is}$. This velocity dispersion is very close to the observations. If observed waves are indeed ion-sound waves their plasma frame frequency should be in the range $\approx 75-100$ Hz, much lower than the observed frequency. This disagreement can be attributed to the Doppler shifts estimated as $\frac{|k|}{2\pi}V_{sw}\sim 600-3000$Hz. In reality the Doppler shift is smaller due to the angle between the wave vector and the solar wind velocity. 

The above arguments indicate that the first wave packet consists of ion-sound waves. As previously mentioned, the wave vectors for the second packet have exactly the same range as the first. Therefore, all arguments used above to deduce the wave mode of the first wave packet are valid for the second. The main
difference between these two wave packets is in the sign of the Doppler
shift. For the first wave packet, the observed frequency is the
difference between the Doppler shift and the plasma frame frequency whilst for
the second it is their sum. It can be seen that they almost coincide for the whole range of observed waves. The angle between the averaged propagation directions of these wave packets is $<5^{\circ}$. This coincidence in the parameters for these two wave packets, observed at clearly distinct periods of time can only be explained by their simultaneous generation at the same location. The generation of ion-sound waves at the shock front are usually attributed either to electric currents or the strong electron temperature gradients in the ramp. Both waves packets were observed upstream of the ramp and carried by the solar wind flow towards it. Since there appear to be no strong gradients in the electron temperature in the foot these waves are probably generated by electric currents. The very short duration of these waves indicates that the current layer might be localized in space and time. Such small scale current layers have been predicted by a nonstationary model of the shock front \citep{krasnoselskikh85:_nonlin, galeev88:_fine, galeev89:_mach,balikhin97:_nonst_low_frequen_turbul_quasi_shock_front,walker99:_ramp}. In this model quasiperiodic steepening of and overturning of the shock ramp takes place leading to the ejection of a nonlinear whistler wave into the upstream region. The amplitude of these nonlinear structures can be comparable to the $|B|$ changes in the ramp itself \citep{walker99:_ramp} and will be associated with localised currents responsible for the ion-sound waves.

\subsection{Observations of lower-hybrid waves}\label{subsec:low_hybrid}

The data set used in this study was collected by the EFW instrument in the same burst mode regime as in previous case onboard the Cluster satellites using onboard timing provided by the DWP instrument \citep{woolliscroft97:_digit_wave_proces_exper_clust}.

\begin{figure}
\begin{minipage}{0.35\linewidth}
\centering
\includegraphics[width=1.8in]{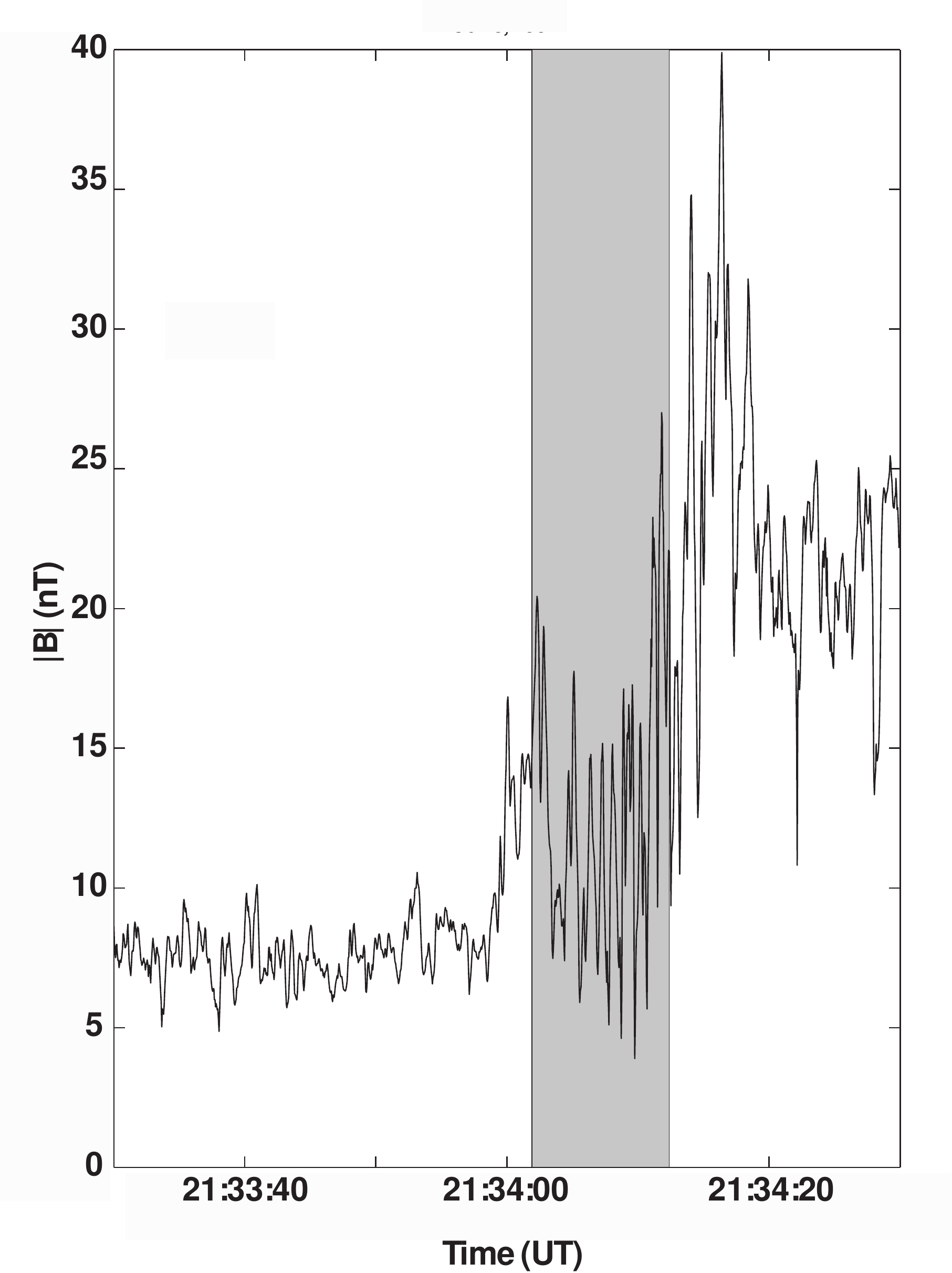}
\end{minipage}%
\qquad
\begin{minipage}{0.6\linewidth}
\centering
\includegraphics[width=2.6in,angle=270]{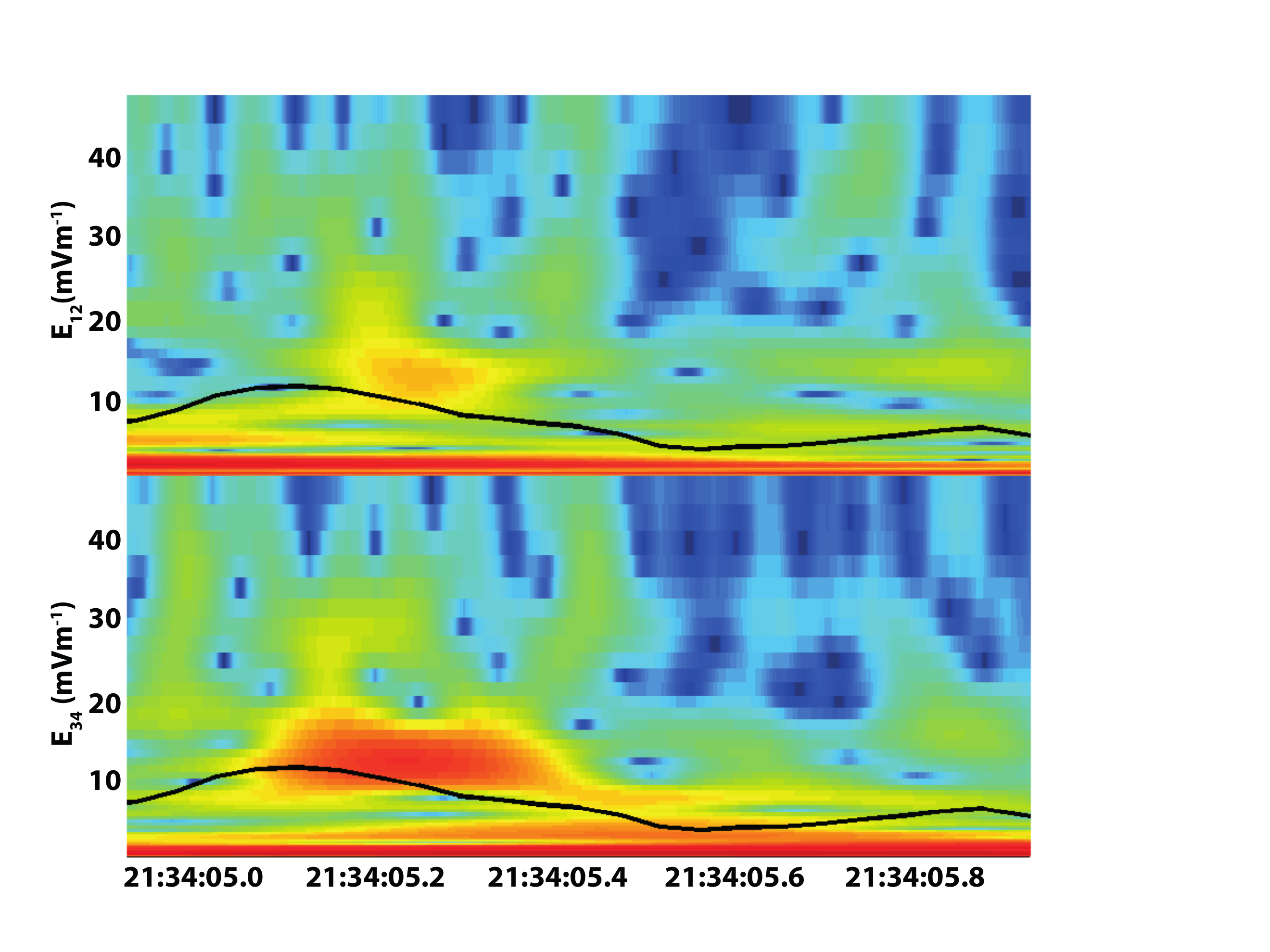}
\end{minipage}
\caption{Panel(1): The magnetic profile of the bow shock crossing observed by Cluster 3 on February 26$^{th}$, 2002 just after 2134UT. The period for which EFW internal burst mode data is available is indicated by the shaded region. Panel(2): The wavelet dynamic spectrogram of electric fields E$_{12}$ (top) and 
E$_{34}$ showing a the occurrence of oscillations just above the lower-hybrid 
resonance frequency (black line) for event 1. \citep[Adapted from][]{walker08:_lower}. \label{fig5.4}}
\end{figure}

Figure~\ref{fig5.4}(1) shows an overview of the magnetic profile of the shock 
encountered on February 26$^{th}$ at 21:34UT. From Figure~\ref{fig5.4}(1) it can 
be seen that Cluster 3 first encountered the foot region of the shock just 
before 21:34UT, finally crossing the ramp and entering the downstream region at 
approximately 2134:15UT. Here the EFW internal burst data 
selection was triggered at 21:34:01.922UT and lasted for a period of 10.47 
seconds as indicated by the shaded region in the figure.

The analysis presented here was performed on data recorded on February 26$^{th}$, 2002 just after 21:34UT on spacecraft 3. This quasi-perpendicular shock crossing took place on an inbound pass at a position (12.0, -1.60, 8.07)R$_{E}$. As can be seen from Figure~\ref{fig5.4}(1) the whole period of internal burst data was collected in the foot region of the shock. The initial increase in the magnetic field profile at around 21:34UT as has been shown above and published in \citet{balikhin05:_ion} to be part of the foot region rather than a partial ramp crossing.

During the 10.5 second period for which EFW internal burst data is available 
there were several short periods when the electric field measurements indicated 
that there were oscillations occurring at or just above the local lower-hybrid 
resonance frequency. In the following subsections the properties of the waves observed are discussed.

\begin{figure}
\begin{minipage}{0.35\linewidth}
\centering
\includegraphics[width=2.2in]{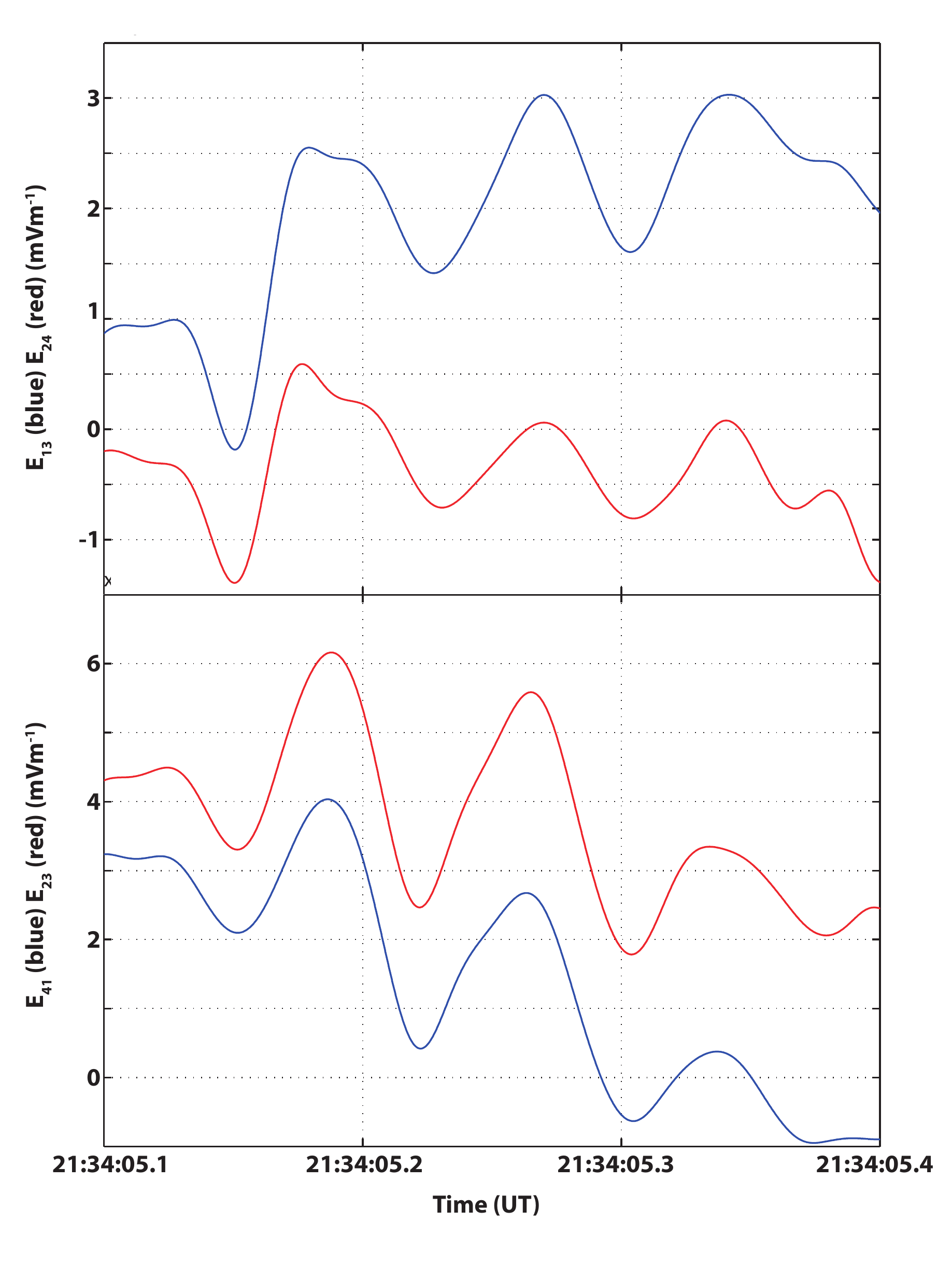}
\end{minipage}%
\qquad
\begin{minipage}{0.6\linewidth}
\centering
\includegraphics[height=2.8in]{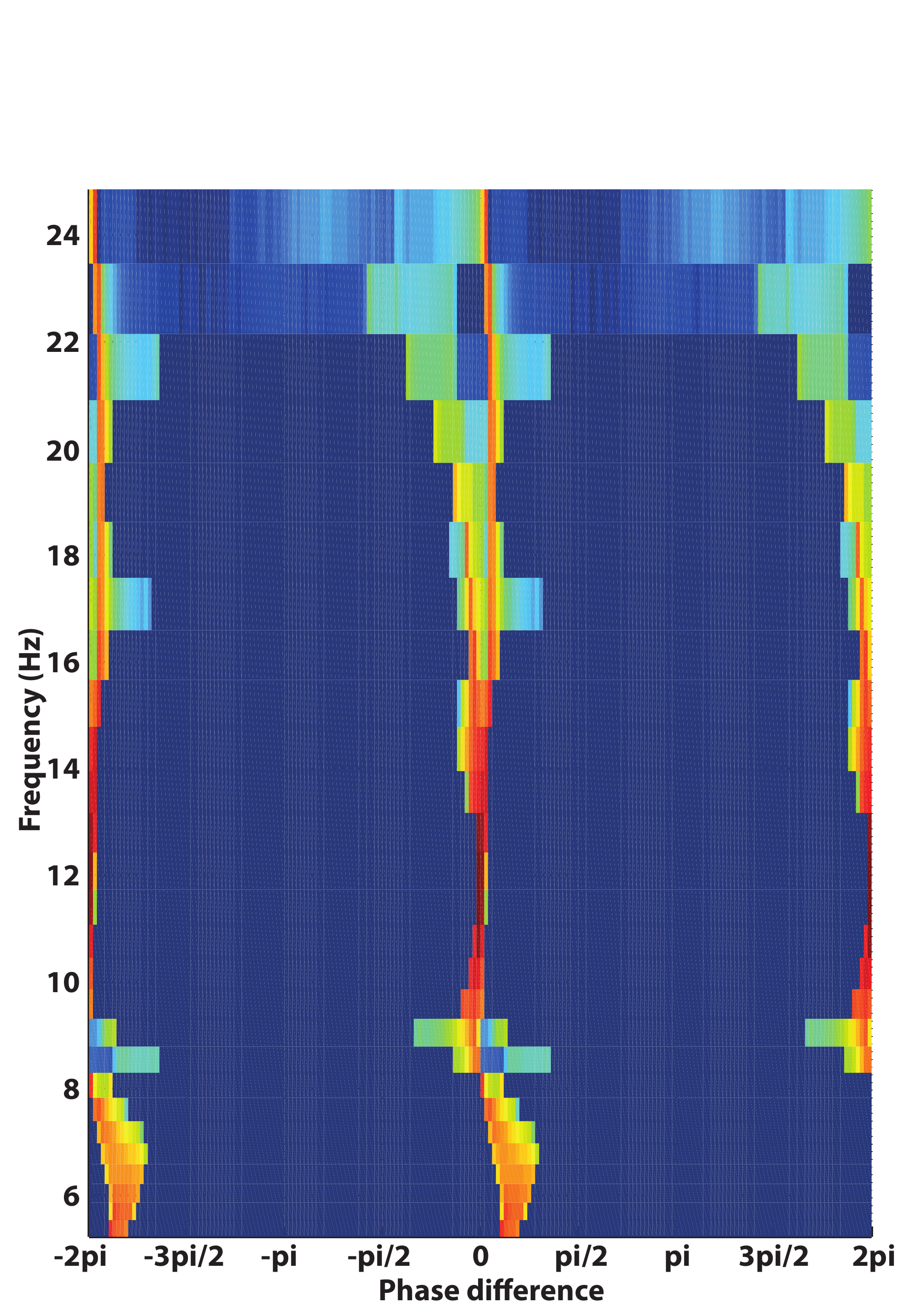}
\end{minipage}
\caption{Panel(1): The $\omega-k$ joint spectrum calculated from the phase differences 
measured between the two parallel electric field components $E_{13}$ and 
$E_{42}$ for event 1. Panel (2): The electric field waveforms E$_{13}$ and E$_{42}$ (top panel) and 
E$_{41}$ and E$_{23}$ (lower panel) for the first event. \citep[Adapted from][]{walker08:_lower}.\label{fig5.5}}
\end{figure}

The first event occurred just after 2134:05UT. 
Figure~\ref{fig5.4}(2) shows a dynamic spectrogram of the electric fields 
E$_{12}$ (top) and E$_{34}$ measured between probes P$_{1}$ and P$_{2}$ and 
probes P$_{3}$ and P$_{4}$ respectively calculated using a Morlet wavelet 
transform. The black line represents the lower-hybrid resonance frequency. It is 
clearly seen that at around 2134:05.2 and there is a packet of waves at 
a frequency between 10-20Hz, whose lower edge is just above the 
lower-hybrid resonance frequency. The duration of this wave packet is around 
3ms corresponding to a few wave periods. Having identified a possible 
occurrence of lower hybrid waves, the phase differencing technique was applied 
to parallel electric field vectors in an attempt to compute the dispersion 
relation of the waves and hence provide an unambiguous identification of the 
wave mode. However in this case using the spin plane electric field components $E_{13}$ and $E_{42}$ in the frequency range of interest (10-20Hz) no measurable dispersion of the waves was observed on scales of the separation distance of the probe pairs (62.2m), see Figure~\ref{fig5.5}(1). 

This implies that the wave travels over the spacecraft at rather high speed so 
that there is virtually no difference in the phase of the wave measured at the 
two points on either side of the satellite. This was also evident in the waveform of the electric field signals. A comparison of the waveforms (Figure~\ref{fig5.5}(2)) shows that the two measurements which are observed to vary in phase which indicates that whatever passed over the satellite has a scale much larger than the individual probe separation distances. 

So, one should conclude that the phase differencing method appears to be unable to show any dispersion in the parallel electric field vectors that means that this method cannot be used to reliably identify the wave packet as being lower hybrid. As a result, one needs to investigate some other wave properties of the wave packet to determine if they are compatible with the lower hybrid mode. As it was mentioned above that by applying the phase differencing method to 
perpendicular components of the electric field it should be possible to 
determine whether the wave packet is linearly or circularly polarised. To this end one should calculate the phase difference between two pairs of probes. The result of this calculation gives the estimate that in the frequency range 
10-20Hz the value of the phase difference is around zero. This result 
indicates that the wave possesses linear polarisation. This leads to the conclusion that the wave mode that is observed in this case corresponds to a lower-hybrid mode.

Similar analysis of the second event occurred around 2134:04.5 on February 
26$^{th}$, 2002 shown the very same result, namely, it shows the phase difference of zero which again indicates that the wave is propagating with a large phase speed over the satellite and the wave packet possesses a linear polarisation and it exhibits properties that are consistent with propagation in in the lower hybrid mode.

The third event highlighted by the authors of the paper \citet{walker08:_lower} occurred between 2134:07.3 and 2134:07.45 UT on February 26$^{th}$, 2002. The wavelet dynamic spectrogram analysis showed a wave packet in the frequency range 10-15Hz that lies just above the lower hybrid resonance frequency. This wave packet was observed to drift in frequency as time increases. This change in frequency mirrors the change in the lower hybrid resonance frequency as calculated from the magnetic field. Analysis of parallel electric field components using the same phase differencing method again indicated zero phase difference between the components. However, a comparison of perpendicular electric field components indicated that a phase difference between two signals is equal to $\pi /2$. 
This leads to conclusion that the wave packet possesses circular polarisation and is thus not propagating in the lower hybrid mode. The circular polarisation indicates that this particular wave packet is propagating in the whistler mode.

\subsection{Estimates of efficient collision frequency using direct measurements of ion-sound and lower-hybrid waves}

A definition of the problem of conductivity relies on exchange of momentum
between electrons and waves assuming the current is mainly carried by
electrons. The conventional formula for plasma conductivity reads%
\[
\sigma =\frac{ne^{2}}{m_{e}\nu } 
\]%
where $n$ is the plasma number density and $\nu$ is the collision frequency of electrons with scattering centers, usually ions or neutrals with respect to momentum loss. When electrons excite some oscillations or waves as a result of instability development they also loose the momentum and this loss is referred as the anomalous momentim loss. In order to find effective collision frequency $\nu _{eff}$ one has to use the momentum conservation law in the system consisting of electrons and waves. In the case of instability this momentum exchange can be written as follows
\[
\nu _{eff}m_{e}n_{0}\overrightarrow{u}_{ed}=\frac{2}{(2\pi )^{3}}\int
d^{3}k\gamma _{k}W_{k}(\frac{\overrightarrow{k}}{\omega _{k}}) 
\]
where $\overrightarrow{u}_{ed}$ is the relative velocity of electrons
carrying current, $\gamma _{k}$ is the instability increment, $W_{k}$ is the
wave energy density that is defined as $W_{k}=\frac{\varepsilon _{0}\mid
E\mid ^{2}}{2}$, ${\mid E \mid}$ is the turbulent electric field amplitude. We have in the left hand side the rate of the electron momentum loss per unit time, and in the right hand side we have the momentum gain by waves due to instability. It follows then that 
\[
\nu _{eff}=\frac{2}{(2\pi )^{3}m_{e}n\mid \overrightarrow{u}_{ed}\mid }%
\left| \int d^{3}k\gamma _{ke}W_{k}(\frac{\overrightarrow{k}}{\omega _{k}}%
)\right| 
\]

If one would like to evaluate the efficient collision frequency from direct
measurements it is necessary to have an estimate of the wave energy. Using
this estimate one can evaluate the shock thickness that shock might have if
it would be determined by the anomalous collisions using the characteristic
length of the momentum loss:

\[
L_{an}=\frac{V_{sw}}{\nu _{eff}}. 
\]

The comparison of the thickness obtained from this estimate with the real
shock thickness can be used to evaluate the relative role of the efficient
collisions.

The standard estimate of the efficient collision frequency for ion sound
mode \citep{galeev84:_curren} reads%
\[
\nu _{eff}=\omega _{pe}\frac{W}{n_{0}k_{B}T_{e}} 
\]

Taking the estimate of the averaged electric field intensity $\langle E\rangle^{2}\approx 10^{-5}$ calculated making use the data of measurements (it varies from $\% 1\times 10^{-5}$ to $2$ $\times 10^{-5}$V/m) and density and temperature from observations $n=9.7$cm$^{-3},T_{e}=17$eV, $\omega _{p}=1.7\times
10^{5}\sec ^{-1}$, one can find $W=4.5\times 10^{-17}$,$W/(nk_{B}T_{e})=1.8\times 10^{-6}$,
\[
\nu _{eff}^{is}=0.3s^{-1},L_{an}=\frac{V_{sw}}{\nu _{eff}^{is}}.\approx 1200km
\]
, where $\nu _{eff}^{is}$ is anomalous collision frequency due to ion-sound wave activity, $L_{an}$ is the characteristic scale of anomalous energy exchange between electrons and ions. 
It is sufficiently larger than the electron inertial scale $c/\omega
_{pe}=1.76$km, and comparable with the thickness of the foot region.

Another group of waves, namely lower hybrid has maximum amplitudes of the
order of 10\mvm and average electric field energy density of the same order
of magnitude as ion sound waves (from $1\times 10^{-5}$ to $4$ $\times
10^{-5}$V/m). In order to evaluate the efficient collision frequency for
these waves one should take into account the properties of lower-hybrid
drift waves. To this end we shall rely on the study of lower hybrid drift
instability published by \citet{davidson77:_effec}. The maximum growth rate for these waves can be estimated as

\[
\gamma _{LH}\approx \alpha \Omega _{LH} 
\]%
where $\Omega _{LH}$ is lower hybrid frequency, coefficient $\alpha <1,$
typically $\alpha \sim 0.1$, and can reach values up to $0.3$. Taking maximum of the linear growth rate we can evaluate the upper limit of the effective collision frequency. The phase velocity of waves around the maximum of increment is of the order of ion thermal velocity of ions, and the drift velocity of electrons that carry the current can be estimated evaluating current velocity from macroscopic gradient of the magnetic field. This estimation gives the value comparable with ion thermal velocity. Thus the estimate of the efficient collision frequency in this case can be written as follows:

\[
\nu _{eff}\approx \alpha \Omega _{LH}\frac{m_i}{m_e}\frac{W}{nk_{B}T_{i}} 
\]

In the region of observations where $B=14$nT and lower hybrid frequency is
approximately equal to $56\sec ^{-1}$, thus the efficient collision
frequency for these waves is found to be of the of the order of

\[
\nu _{eff}\approx \alpha \Omega _{LH}\frac{m_i}{m_e}\frac{W}{nk_{B}T_{i}}\sim
0.1\times 2\times 10^{3}\times 56\times 10^{-6}\simeq 10^{-2}s^{-1} 
\]

that is sufficiently smaller than the efficient collision frequency for ion
sound waves. The characteristic dissipation scale

\[
L_{an}\approx \frac{350}{10^{-2}}km\approx 3.5\times 10^{4}km 
\]

and is sufficiently larger than the major characteristic scales of the shock
front.

This leads to the conclusion that the anomalous resistivity observed can not
account for the important dissipation rate. The characteristic scales of the
dissipation are too large compared to the shock transition features observed.

\section{Nonstationarity and reformation of high Mach number quasiperpendicular shocks: Cluster observations}
\label{sec:nonstationarity}

In this Section, using Cluster observations, we provide convincing evidence that high-Mach-number quasiperpendicular shocks are indeed nonstationary, and moreover, quasi-periodic shock front reformation takes place. Most of the  material of this Section was first published in \citet{lobzin07:_nonst_mach}.
\subsection{ An Example of a Typical Crossing of Nonstationary Quasiperpendicular Shock Wave}
\label{subsec2}
A number of magnetic field profiles of the quasiperpendicular terrestrial bow shock observed by Cluster triaxial flux gate magnetometers (FGM) \citep{balogh97:_clust_magnet_field_inves} in the period January-May 2001 were studied. It was found that nonstationarity seems to be typical for shocks with relatively high Mach numbers. Both from numerical simulations and experiments it follows that the details of this nonstationary behaviour of the shock front may depend strongly not only on the fast magnetosonic Mach number, $M_{f}$, but also on the upstream $\beta_{e,i}$ and the angle between the upstream magnetic field and the shock normal, ${\theta}_{Bn}$. For a detailed case study, a shock was chosen that could be considered as a typical quasiperpendicular, supercritical, high-Mach-number shock wave, namely the shock crossing that occurred on 24 January 2001 at 07:05:00-07:09:00. Indeed, from the available experimental data and with the use of the multi-spacecraft timing algorithm described by \citet{schwartz98:_shock_discon_normal_mach_number_relat_param} the following estimates were obtained: ${\beta}_e = 1.7$, ${\beta_i} = 2.0$, ${\theta}_{Bn} = 81^{\circ}$, $M_{A} = 10$, and $M_{f} = 5$. 
\begin{figure}
\includegraphics[width=4in]{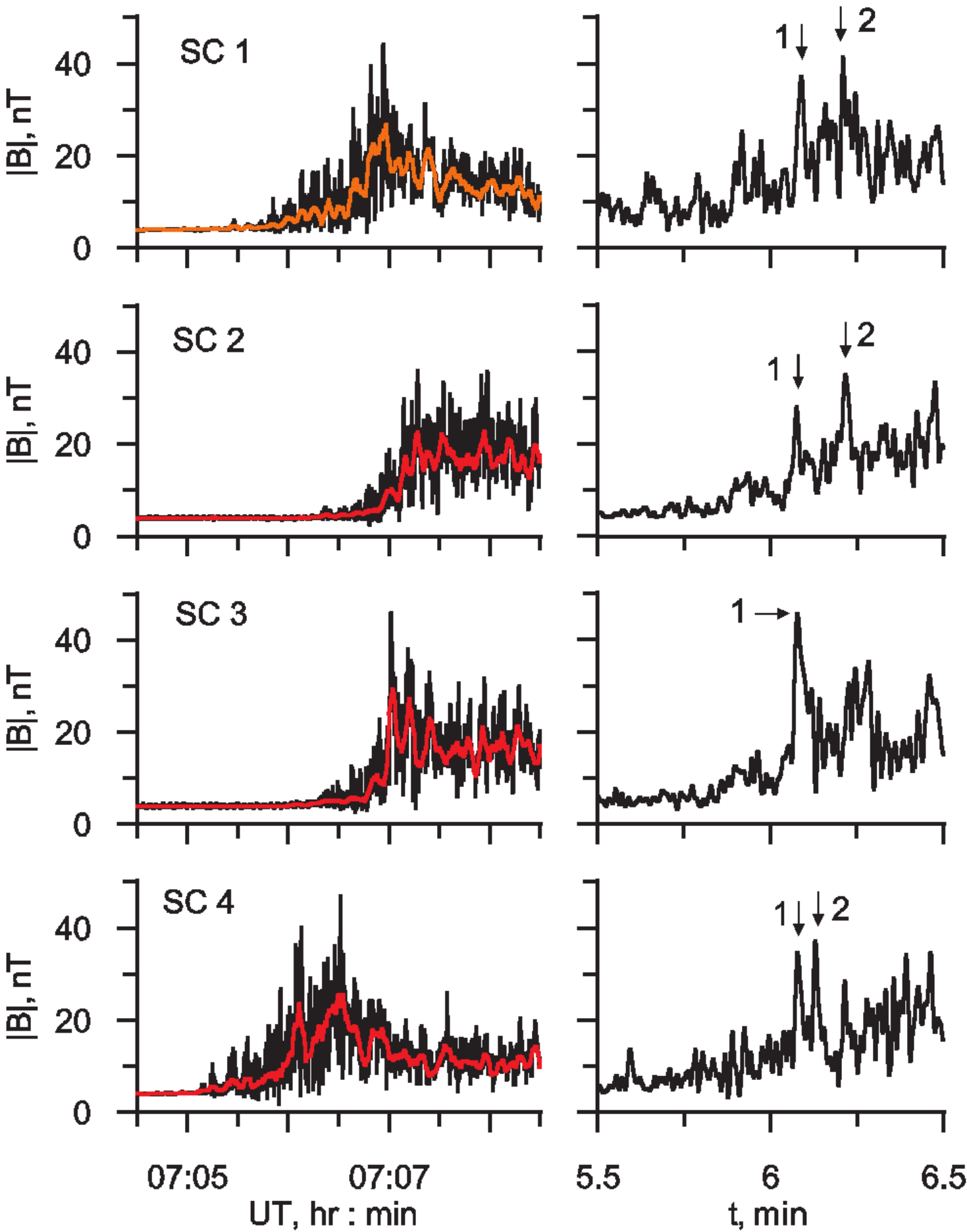}
\caption{The magnetic field profiles obtained by FGM experiments aboard four Cluster spacecraft during the Earth's bow shock crossing on 24 January 2001. (left) High-resolution magnetic field data (black line) and the data obtained by sliding averaging over 4 s time intervals (red line). (right) Vicinity of overshoots, with large peaks in the magnetic field magnitude. Oscillations with frequencies higher than 2 Hz were removed. To emphasize the similarity and differences of the profiles, the data for the first 3 spacecraft are shifted with respect to that for the 4th one. \citep[Adapted from][]{lobzin07:_nonst_mach}.}
\label{fig:nonstat1}       
\end{figure}

\begin{figure}
\includegraphics[width=4in]{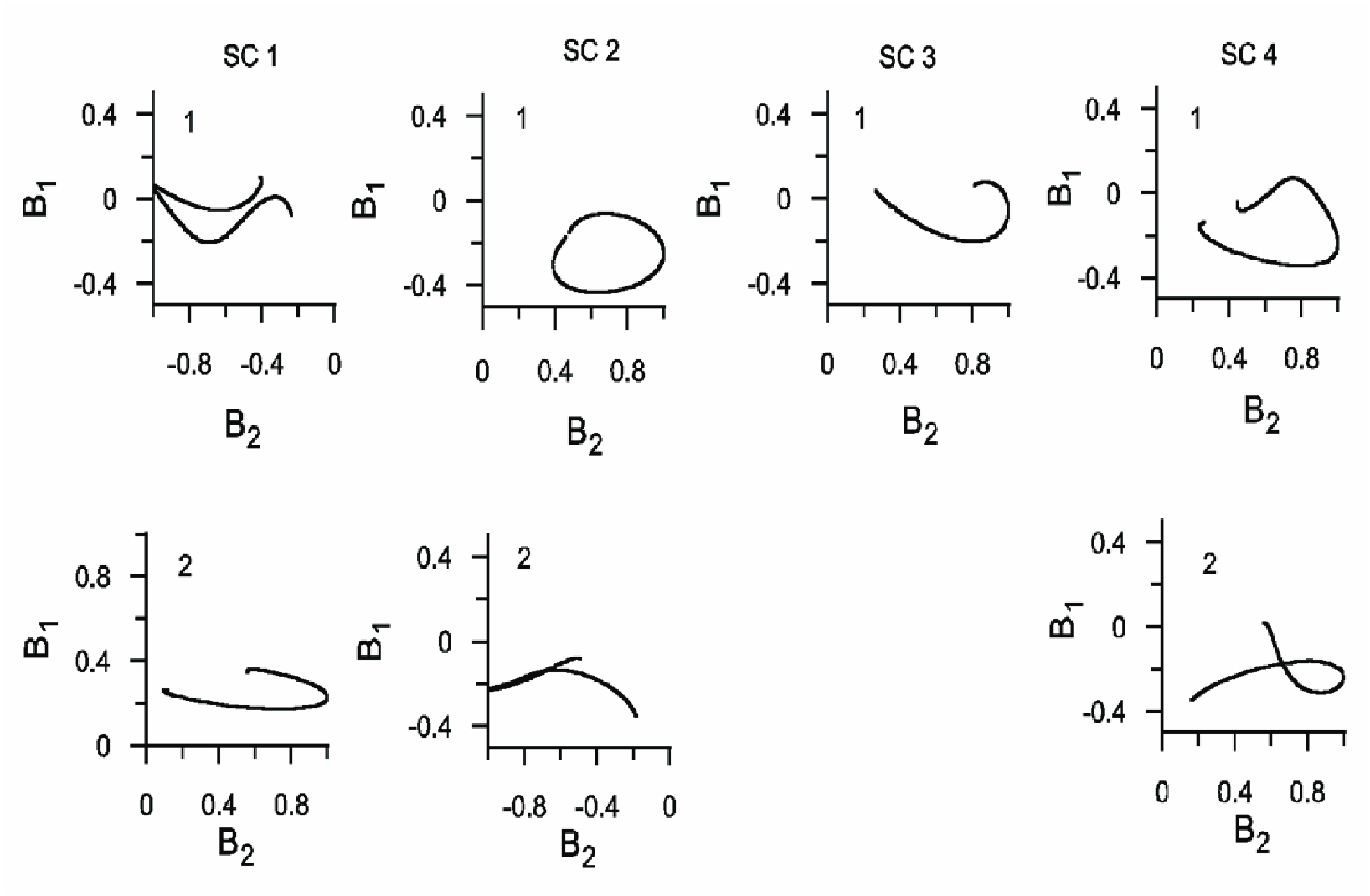}
\caption{Hodograms of magnetic field around their maxima for different satellites in the vicinity of supposed ramps.}
\label{fig:nonstat2}       
\end{figure}

Figure~\ref{fig:nonstat1} shows the magnetic field profiles measured by the Cluster FGM instruments on January 24 2001$^{th}$, 2001.The panels on the left show the full resolution data, sampled at 67 Hz (black line) and the result of averaging this data using a 4 second sliding window (red line). The panels on the right show the result of low pass filtering the data at 2Hz. This process enhances any large peaks in the magnetic field measurements. All the profiles can be considered as quite typical for high-Mach-number quasiperpendicular shock waves. From the averaged data shown by the red lines we observe that the shock front consists of a foot, a ramp, and at least one overshoot-undershoot cycle, i.e. large amplitude peak of the magnetic field at the end of the ramp region and following after it minimum. The small-scale oscillations of large amplitude are superimposed on this large-scale structure. To check whether these fluctuations are consistent with plane wave activity, the degree of polarization for the magnetic field waveforms obtained from STAFF experiment \citep{cornilleau-wehrlin97:_clust_spatio_tempor_analy_field}. By definition, the degree of polarization approaches a unity if and only if most of the energy is associated with a plane wave \citep{samson80:_some}. It was found that between the forward edge of the shock and the magnetic overshoot the oscillations in the frequency range 3-8 Hz have a high degree of polarization greater than 0.7 and that this polarization is elliptical. This wave activity can be considered as a whistler wave train nested within the shock \citep{galeev88:_fine, galeev89:_mach, krasnoselskikh02:_nonst}.
Obviously, the presence of whistler oscillations, due to their high amplitude, has a considerable impact on the large-scale shock structure. Indeed, averaging of magnetic field data reveals two regions resembling overshoots for SC4 whilst only one maximum is observed for SC1. The profiles for the other spacecraft appear to be more complicated. It follows from these considerations that the concepts of both overshoot and ramp, which must precede it, become ambiguous for such nonstationary shocks. Instead, we can speak about short scale large-amplitude structures embedded into the shock transition, with the forward edge of one of these structures playing a role of the ramp.
Figure~\ref{fig:nonstat1} also shows that the magnetic field profiles measured onboard the different spacecraft differ considerably from each other. Obviously, the number of large-amplitude peaks, their amplitudes, as well the positions within the shock front, are different. The waves observed by different spacecraft in the foot region are also different. In particular, from Figure~\ref{fig:nonstat1} (left) it is easily seen that the time interval between the beginning of the wave activity at the forward edge of the shock and the ramp crossing may differ by 10-20 s. This difference is substantial compared to the duration of the crossing of typical elements of the shock structure.
The distinctions found between observations from the different spacecraft are related to temporal rather than spatial variations in the structure of the shock front because the spacecraft separation is comparable with shock front thickness. Indeed, the distances between spacecraft lie within the range $380–980$ km. The foot thickness estimated with the use of the theoretical formula derived by  \citet{schwartz83:_ions} is equal to 550 km, in reasonable agreement with the observations, while the total shock front thickness is considerably larger. On the other hand, the maximum time lag between the crossings is about $3 T_{Bi}$, where $T_{Bi}$ is the ion gyroperiod $T_{Bi} = 15.5$ s. This time lag is larger than the period of the shock reformation.
Relying on theoretical considerations and results of numerical simulations, \citet{krasnoselskikh02:_nonst} argue that this type of nonstationarity is closely related to nonlinear whistler wave trains embedded into the shock front and that this is a typical property of quasiperpendicular high-Mach-number shocks.
Further evidence for the existence of whistler waves embedded within  the shock front can be seen from the rotational features of the magnetic field observed in the vicinity of the peaks, as shown in Figure~\ref{fig:nonstat2}, that are typical of whistler mode waves. 

The large-amplitude structures seen in the magnetic field profiles within the overshoot region and its vicinity have a characteristic time of about 2 s. To examine both the similarities and differences of these profiles, oscillations with frequencies higher than 2 Hz were removed by low pass filtering the data. The filtered data were then used to calculate a set of optimal cross-correlation coefficients for profile fragments that last ~35 s and include a portion of foot and the entire overshoot region. The highest correlation was found between SC1 and SC2, while the lowest one was between SC3 and SC4, a result that is in accordance with visual observations of the shifted profiles shown in Figure~\ref{fig:nonstat1} (right). An additional analysis of the relative position of the spacecraft tetrahedron and the shock reveals that the similarity of the shock profiles seems to depend mainly on the time interval between the shock crossings and/or the spacecraft separation measured along the shock normal rather than on the distance along the shock surface which is in accordance with the interpretation that the observed variations are temporal rather than spatial.

\begin{figure}
\includegraphics[width=4in]{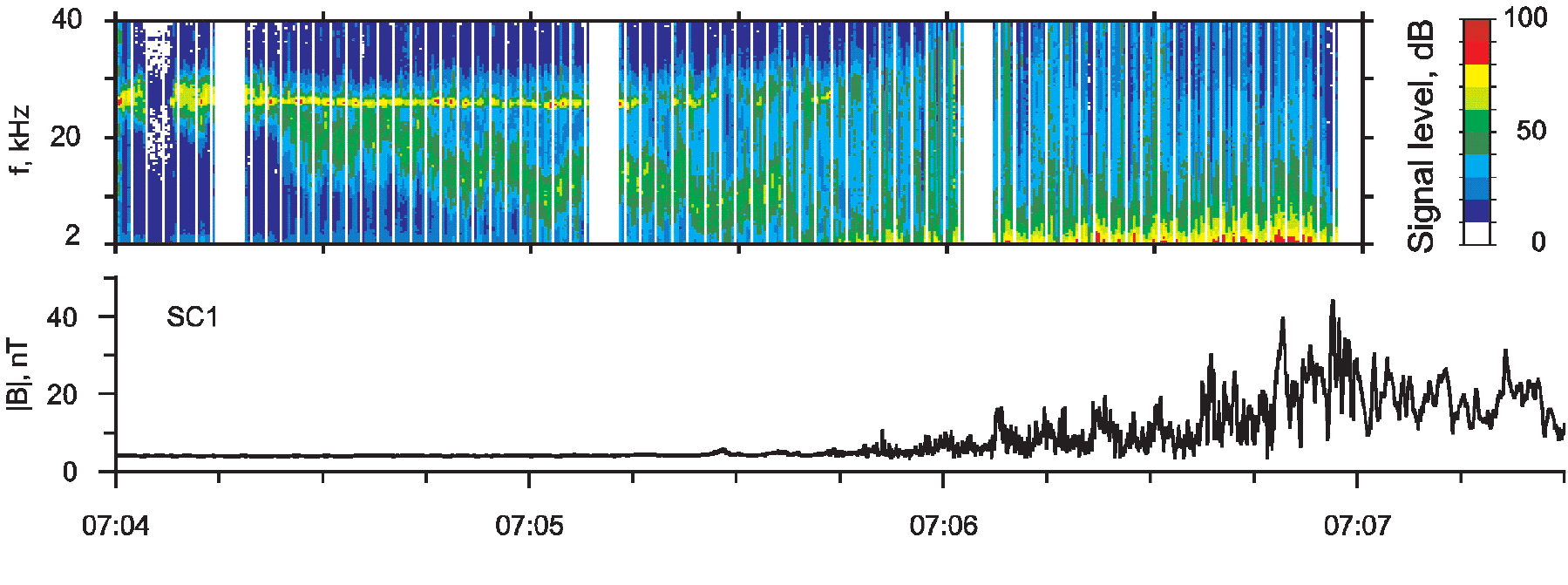}
\caption {(top) Electric field spectra and (bottom) magnetic field profile obtained during the Earth's bow shock crossing on 24 January 2001 aboard SC1. The frequency-time spectrogram is measured by the Whisper experiment. The vertical white bands correspond to the time intervals when no data were obtained in the natural wave mode. The wave intensity is colour coded with the reference level of $10^{-7}$ V$_{rms}$/ Hz$^{1/2}$, where rms is the root mean square to notify the averaged level of the fluctuating electric field variance. The magnetic field profile is obtained by FGM experiment. The time scales for the both panels are the same.\citep[Adapted from][]{lobzin07:_nonst_mach}.}

\label{fig:nonstat3}       
\end{figure}

Further evidence favoring the nonstationarity of this bow shock crossing comes from WHISPER measurements. In passive mode this experiment provides electric field spectra of natural emissions in the frequency range 2-80 kHz \citep{decreau97:_whisp_reson_sound_wave_analy}. The frequency-time spectrogram obtained by WHISPER experiment aboard SC1 is shown in Figure~\ref{fig:nonstat3}, together with the magnetic field profile with the same time scale. The bow shock crossing can be identified by a substantial enhancement in the electric field fluctuations within the frequency range 2–5 kHz. For SC1, maximum intensity for these oscillations is observed at 07:06:48 UT.
One of the most obvious features of these spectra is the presence of intense waves in the vicinity of the plasma frequency, $f_{pe} = 27$ kHz, together with downshifted oscillations. The most intense feature is a narrow-band Langmuir emission with a frequency in the vicinity of $f_{pe}$. As compared with Langmuir waves, the power density of downshifted oscillations is usually smaller, while the frequency band they occupy is considerably wider and can be as large as 15-20\% of the central frequency. Both the plasma waves and downshifted oscillations are considered to be typical of the electron foreshock region. It is commonly believed that Langmuir waves are generated by a plasma-beam instability, while for the downshifted oscillations two different mechanisms have been proposed, namely, the plasma-beam interaction, see \citep[see][]{lacombe85:_elect, fuselier85} and the loss-cone instability of electron cyclotron modes \citep{lobzin05:_gener}.
The mean frequency of the downshifted oscillations is not constant but varies within the range 0.2 - 1.0 $f_{pe}$. In addition, there exists a tendency for a large shift to occur in the vicinity of the shock front, while near the edge of the electron foreshock the shifts are considerably smaller. However, this tendency exists only on large time scales of about 1.0-1.5 min. For smaller scales, ~10-15 s, there are the large-amplitude variations of the mean frequency of the downshifted oscillations.
The peculiarities of the spectra described above can be explained as follows. The downshifted oscillations are produced by energetic electrons, which are reflected by the bow shock and move almost along the magnetic field lines. Because the solar wind is quiet during the time interval considered (indeed, Figure~\ref{fig:nonstat1} and Figure~\ref{fig:nonstat2} show that there are no significant variations of the magnetic field; the plasma bulk velocity is also approximately constant in the foreshock), the observed evolution of the wave spectra can only be attributed to variations of the suprathermal electron fluxes which are reflected from the bow shock and form the “rabbit ears” in the electron distributions upstream of the shock as was shown by \citet{lobzin05:_gener}. The reflection of electrons by a nearly perpendicular bow shock was studied by \cite{leroy84} and \cite{wu84:_fermi_energ}. They argued that the main characteristics of the distribution function of the reflected electrons depend first of all on the angle between the shock normal and upstream magnetic field, ${\theta_{B_n}}$, and to a lesser extent on the ratio of the maximum magnetic field to its upstream value and on the electrostatic potential jump in the de Hoffmann-Teller frame. Resulting from shock front nonstationarity, slow variations of the effective normal of the reflecting part of the shock will lead to considerable variations of number density, energy of reflected electrons, and/or loss-cone angle, thereby producing the observed variations of the downshifted wave spectra. Both theoretical considerations and numerical modeling show that a characteristic time of the shock front oscillations or reformation is comparable with the ion gyroperiod \citep[see][]{leroy82, krasnoselskikh02:_nonst, scholer03:_quasi}. The time scale of the spectra variations is also comparable with ion gyroperiod $T_{Bi}$, in accordance with our interpretation.

\subsection{Evidence for Shock Front Reformation} 
\label{subsec3}
As noted above, the magnetic field profiles for the shock under consideration have several nonstationary features. In this section, we consider large-amplitude structures, with a characteristic time of about 1–2 s and present the arguments in favor of front reformation for this particular bow shock crossing. Figure~\ref{fig:nonstat1}  (right) shows the magnetic field profiles obtained after low-pass filtering and shifting the data in time to clearly show the correspondence between the elements observed aboard different spacecraft. For three spacecraft there are two large narrow peaks in the overshoot region and its vicinity, while for SC3 there is only one peak in the corresponding region (see Figure ~\ref{fig:nonstat1} (right), where these peaks are shown by arrows and numbered). The amplitudes of these peaks, both absolute and relative, differ for different spacecraft. In addition, the distance between two adjacent peaks also varies, being the smallest for SC4 and the largest for SC2. Moreover, the single peak observed by SC3, which largest amplitude and relatively large width,  may be formed due to the coalescence of two separate peaks. The observed peaks in the overshoot region can be considered as a part of the nonstationary whistler wave packets since their rotational properties are clearly evident in Figure~\ref{fig:nonstat2}. These properties were argued to be an intrinsic element of the quasiperpendicular supercritical shock front structure \citep{krasnoselskikh02:_nonst}. In order to investigate these features further, an analysis of their polarization was performed using the minimum variance technique. The results provide additional evidence in favor of shock front nonstationarity. Indeed, the corresponding elements have different hodograms, which can be rather complicated. However, some of the elements have approximately circular polarization typical for large-amplitude whistlers as was stated in theoretical papers \citep{galeev88:_fine, galeev89:_mach, krasnoselskikh02:_nonst} and is evidenced on Figure~\ref{fig:nonstat2}.
A comparison of the magnetic field profiles, shown in Figure~\ref{fig:nonstat1} with the results of numerical simulations of high-Mach-number shock reformation \citep{krasnoselskikh02:_nonst} reveals a doubtless resemblance between them. Indeed, for large Mach numbers, quasiperiodic reformation of the shock front was observed in the simulations, with whistler wave packets playing a crucial role. In the first stage of the reformation cycle, a small-amplitude whistler perturbation upstream of the ramp is formed This perturbation grows and moves towards the ramp. When its amplitude exceeds that of the ramp, this disturbance begins to play the role of a new ramp, while the old one moves away downstream. The experimental results shown in Figures~\ref{fig:nonstat1} and~\ref{fig:nonstat2} resemble 4 different snapshots for the same shock undergoing the reformation.
\begin{figure}
\includegraphics[width=4in]{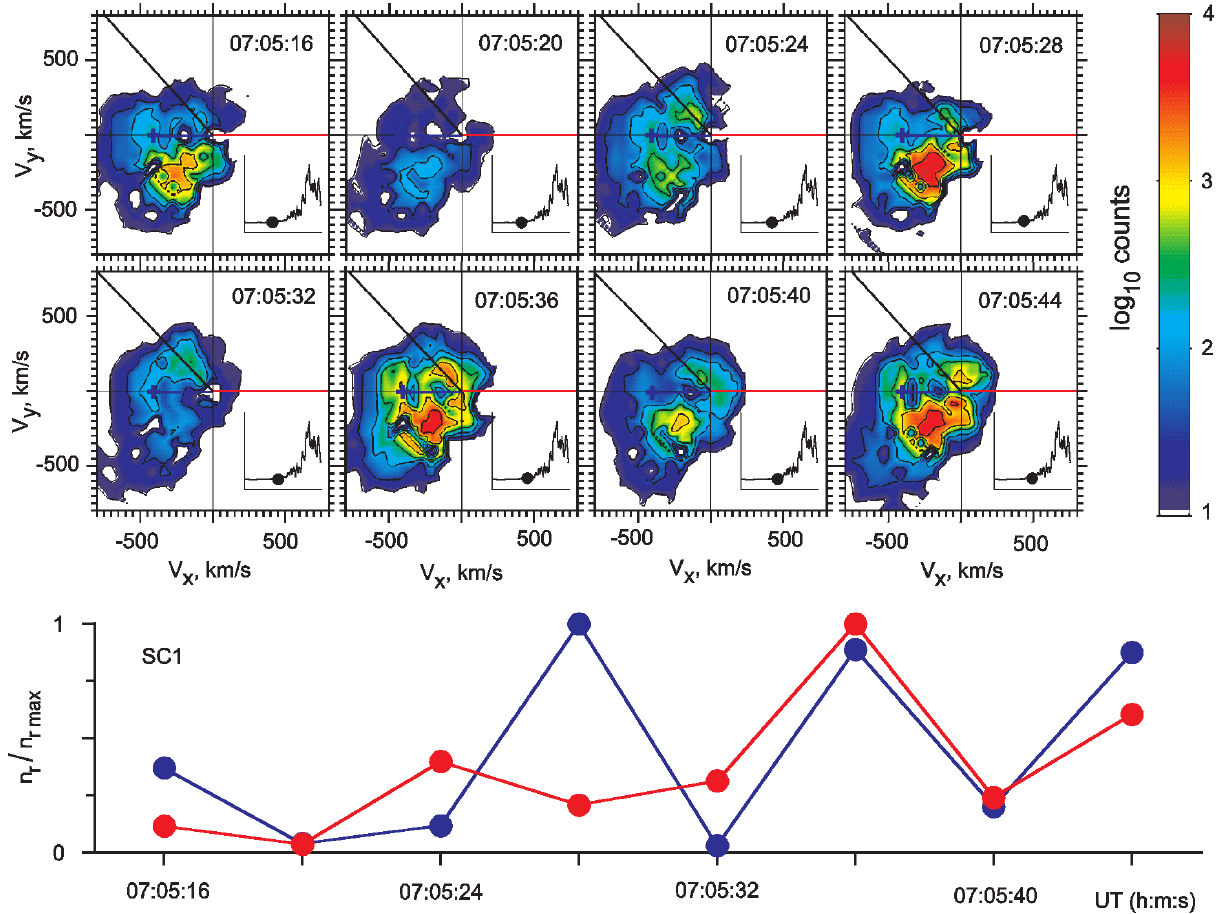}
\caption {(top) Ion velocity distributions obtained from CIS measurements within the forward part of the foot for the Earth's bow shock crossing on 24 January 2001 aboard SC1 and (bottom) temporal variations for the relative number of counts corresponding to reflected ions. The distributions were calculated in the GSE coordinates. In the bottom panel, a blue line corresponds to ions with $V_{y} < 0$, while a red line shows the data for $V_{y} > 0$. Strong variations of $n_r$, especially for $V_{y} < 0$, show that the reflection of ions is bursty. The relative positions where the measurements were made are indicated by the dots on the magnetic field profiles shown as inserts.\citep[Adapted from][]{lobzin07:_nonst_mach}.}
\label{fig:nonstat4}       
\end{figure}
The strongest evidence favouring the shock reformation comes from the CIS experiment, which measures the ion composition and full three-dimensional distributions for major ions with energies up to 40 keV/e \citep{reme97:_clust_ion_spect_cis_exper}. The time resolution of these measurements is about one spacecraft spin, 4 s. Figure~\ref{fig:nonstat4} shows 8 snapshots obtained at the upstream edge of the shock foot, where the disturbances of the solar wind magnetic field are still small. The Figure shows the number of counts vs a function of $V_x$ and $V_y$ in the GSE coordinate system; with the data being integratedin the $V_z$ direction. Reflected ions are observed for the first time at 07:05:16 (see the maximum of the number of counts in the quadrant corresponding to $V_{x} < 0$ and $V_{y} < 0$ in the first snapshot). In the time interval from 07:05:16 to 07:05:44, the position of this maximum in the velocity space does not change considerably. In addition, there exists a second population of reflected ions in the quadrant corresponding to $V_{x} < 0$ and $V_{y} > 0$. From the snapshots it is easily seen that the numbers of counts corresponding to the reflected ions show approximately periodic variations with a very large modulation depth and a period of about 8 s which corresponds to half of the proton gyroperiod $T_{Bi}$. To confirm this statement, we performed a summation of the number of counts corresponding to these populations, the results are approximately proportional to the corresponding number densities, $n_{r}$. The temporal evolution of these number densities normalized with respect to the corresponding maximum values for the time interval considered is shown in Figure~\ref{fig:nonstat4} (bottom). The quasiperiodic variations seem to be more pronounced for the first population (blue line), with the minimum-to-maximum ratio being as low as $\sim 3$\%. The number of counts for the second population also varies with approximately the same period, in phase with that for the first one. It is worth noting that the minimum number of counts corresponding to the reflected ions in this region is greater than the 'background noise' by a factor of 5, far beyond experimental errors, while for the maximum number of counts this factor is as large as ~30 if the 'noise' level is estimated in the unperturbed solar wind just before the shock crossing.
The observed peculiarities of the ion dynamics resemble the features found in the numerical simulations of \citet{krasnoselskikh02:_nonst}, where a quasiperiodic front reformation was observed for quasiperpendicular shocks with high Mach numbers. In particular, when the leading wave train before the ramp attained a large enough amplitude, a new population of reflected ions appeared upstream of the precursor. In other words, the reflection of ions is not stationary. It is quasiperiodically modified during the reformation process. In this case a spacecraft that moves slowly across the shock, will observe the quasiperiodic appearance/disappearance of reflected ions, in accordance with experimental results outlined above.

\subsection{Conclusions}
\label{sec:conclusions} 
In this Section we have presented a set of experimental results for a high-Mach-number ($M_{f} = 5$) quasiperpendicular ($\theta_{Bn} = 81^{\circ}$) bow shock crossing observed by Cluster spacecraft on 24 January 2001 at 07:05–07:09 UT. The structure of this shock gives a clear evidence of its nonstationary behavior. In particular, the magnetic field profiles measured by FGM experiments onboard different spacecraft differ considerably from each other. This difference is clearly seen for large-amplitude oscillations, which have relatively short scales of about 1-2 s and resemble nonlinear whistler soliton-like structures that is confirmed by analysis of their hodograms. WHISPER measurements reveal the presence downshifted oscillations within the electron foreshock, with nonmonotonic variations of their central frequency, the characteristic time for these variations is comparable with the proton gyroperiod, ${T_{Bi}} = 15.5 s$. From the analysis of data from CIS experiment it follows that the reflection of ions from the shock are also highly nonstationary. Moreover, it is shown that the reflection is bursty and the characteristic time for this process is also comparable with the ion gyroperiod. From numerous numerical simulations of quasiperpendicular shocks it is well-known that for high Mach numbers the shock becomes nonstationary. Moreover, front reformation can take place with a characteristic time comparable with the ion gyroperiod. The combination of the features outlined above for the bow shock crossing under consideration is the first convincing experimental evidence favoring the shock front reformation.

\section{Conclusions}
There exist several models of quasiperpendicular high Mach number shocks. Theoretical considerations and computer simulations on todays level are not capable to describe correctly all physical process that determine different aspects of shock physics. The only possibility to ensure that the theory or modelling correctly capture major physical effects is to rely on analysis of experimental data of direct in situ measurements onboard satellites. The best adapted for this goal are Cluster satellites since they allow to distinguish spatial and temporal variations and during the mission they had different intersatelllite distance that allows one Òto probeÓ the shock on different scales. The difficult task in such investigation program consists in formulation of the right questions to be addressed to data and to their analysis. Our aim was to determine major physical processes that define characteristics of the most important part of the shock front, its ramp and wave activity around it. The problems closely related to this major problem are electron heating mechanisms and transition of shock behaviour from stationary to nonstationary. We left beyond the scope of our Review many questions. One can mention ripples, remote sensing of the shock by field aligned beam, instabilities behind the shock front. We restricted ourselves by the analysis of scales of magnetic and electric fields, the scale of electron heating, determination of the source and generation mechanism of precursor whistler wave train and direct observation of the shock front reformation by Cluster satellites. Huge collection of data of statistical analysis and of the studies of individual events (case studies) leads to the conclusion that the ramp region of supercritical quasiperpendicular shock is nothing more than an intrinsic element of nonlinear dispersive wave structure slightly modified by reflected ions. This interpretation allows one to explain in a natural way the whole collection of data that we reported in this Review and to understand the transition from stationary to nonstationary shock behaviour when the Mach number exceeds nonlinear whistler critical Mach number. The role of anomalous resistivity is shown to be relatively weak with respect to effects of dispersion. Our study also points out several important opened questions. Presumably the most important is what is detailed mechanism of the electron heating and isotropization. Certainly, the evaluation of the role of anomalous resistivity can not be considered as the solved issue, the data set used is too poor to come to definite conclusions, thus this study still waits new measurements. We did not address the problem of particle acceleration and new results presented here will certainly have an impact on re-consideration of this important problem. 

\appendix

\section{Remark on comparison of computer simulation results with experimental data}
\label{sec:simulations}

Recently \citet{comisel11:_non} made an attempt to perform computer simulations that can properly reproduce the realistic physical conditions corresponding to observations. They modelled the shock dynamics using 1D PIC code with the realistic ion to electron mass ratio under conditions corresponding to shock conditions on 24th January 2001 that was observed by \citet{lobzin07:_nonst_mach}. The only difference between the model and real plasma parameters is an unrealistic ratio of $(\omega _{pe}/\Omega _{ce})$. 
The modelling results clearly showed that the shock indeed is nonstationary.
However, it was found that there are some important differences between the
results of the simulations and observations. The major differences can be
summarized as follows.
The electric fields observed in simulations in the close vicinity of the
shock front were much higher than the electric fields experimentally
registered.
The energy flux of waves observed in the foot region upstream of the shock
front was found to be directed toward downstream that clearly indicates that
the waves observed in simulations are generated by the beam of the reflected
ions and not by the ramp region as the dispersive mechanism predicts. 
This gives an indication that the properties of waves observed are much closer to short scale lower hybrid or lower hybrid drift waves described in Section~\ref{sec:anom_res} and not to those described in Sectons~\ref{sec:magneticscales} and~\ref{sec:electricscales}. The question arises where does this difference come from. To answer this question one should consider some scaling properties of equations describing dynamics of the shock. In order to do that let us re-write our equations in dimensionless form making use of natural variables
\[
\widetilde{v}=v/V_{A},\widetilde{t}=t\Omega _{ci},\widetilde{r}=r\Omega
_{ci}/V_{A},b=B/B_{0},e=qE/\Omega _{ci}M_A V_{A},\widetilde{n}=n/n_{0} 
\]%
The system have several dimensionless parameters that remain and should be taken into account. These are $\eta =\frac{m_{e}}{m_{i}}$ (the authors would
prefer the letter $\mu $ but it is already used for magnetic permeability)$,$
$\chi =\frac{\omega _{p}}{\Omega _{ce}}$, and certainly $\beta _{e,i}$.

To account for the principal difference between the real physical conditions and simulations let us consider where the parameter $\chi$ may play an important role. In dimensionless variables it appears in two Maxwell equations
\[
div\overrightarrow{E}=\frac{\eta }{\chi ^{2}}(\tilde{n}_{i}-\tilde{n}_{e}) 
\]%
\[
rot\overrightarrow{B}=(\tilde{n}_{i}\overrightarrow{\tilde{V_{i}}}-%
\tilde{n}_{e}\overrightarrow{\tilde{v}_{e}})+\frac{\eta }{\chi ^{2}}%
\frac{\partial \overrightarrow{E}}{\partial \tilde{t}}. 
\]

To clarify its role one can consider the properties of linear waves. One can note that the ratio of electric to magnetic field is determined by the refractive index of waves.
By definition it is 

\[
N=\frac{kc}{\omega } 
\]

and it is easy to see that it is exactly this ratio is used in determination of the electric to magnetic field ratio. In SI the system in dimensional variables reads 
\[
rot\overrightarrow{E}=-\frac{\partial \overrightarrow{B}}{\partial t} 
\]%
that leads to following estimate for linear waves:

\[
\lbrack \overrightarrow{k}\times \overrightarrow{E}]=\omega \overrightarrow{B%
}. 
\]

This can be re-written as follows: 
\[
\frac{cB}{E_{k\perp }}=\frac{kc}{\omega }=N.
\]

It can also be expressed in terms of phase velocity  
\[
N=\frac{c}{V_{ph}}.
\]

If we take the waves having velocities close to the shock front velocity
(approximately standing whistlers in a shock front reference frame) the velocity in the plasma reference frame is $%
V_{up}=V_{sw}=M_{A}V_{A}$, thus the refraction index is 
\[
N=\frac{c}{V_{sw}}=\frac{c}{M_{A}V_{A}}=\frac{\omega _{pi}}{M_{A}\Omega _{ci}%
}=\frac{\omega _{pe}}{M_{A}\Omega _{ce}}\frac{1}{\sqrt{\eta }}
\]

\[
N_{\exp }=\frac{Bc}{E_{k\perp }}=\frac{c}{V_{sw}}\left(\frac{\omega _{p}}{\Omega
_{ce}}=\frac{2.7\cdot 10^{4}}{1.2\cdot 10^{2}}\sim 230\right)
\]
where $E_{k\perp}$ is the electric field component perpendicular to the k-vector. 

On the 24th of January 2001 the solar wind velocity was $V_{swSW}=440$\kms.

\[
N_{\exp }\approx 700
\]

The maximum value of ratio $\frac{\omega _{p}}{\Omega _{c}}$ in simulations
is 8 thus 
\[
N_{sim}=23.
\]  
it is approximately 30 times smaller than in experiment, that means that for the
same level of fluctuations of the magnetic field the electric field fluctuations are 30 times stronger than in experiment.

According to our analysis of dimensionless parameters another important difference consists in similar overestimate of electric fields due to even small
deviations from quasi-neutrallity. One can suppose that this can lead to artificial increase of the role of quasi-electrostatic instabilities of short scale lower hybrid waves. As a result the dominant waves observed in simulations are similar to those reported in the Section "Anomalous resistivity", namely drift lower hybrid type waves. Presumably, the overestimation of the role of the electric field and consequently of short scale oscillations and consequently underestimation of the role of lower frequency standing precursor whistler waves results in the difference between observations and simulation results. If so, the simulated shock is really resistive while the observed one is certainly dispersive.
To evaluate the influence of this overestimate of the electric field let us
evaluate the electric field needed to reflect upstream ion flow assuming that
for efficient reflection the potential should be of the order of half energy of the incident ion.
Reflecting potential in nonlinear wave on the scale about 
\[
L=5\frac{c}{\omega _{pe}}=0.9\cdot10^{6}\rm{cm},\quad E_{ions}=\frac{m_iV^{2}}{2}=0.5\cdot10^{9}\rm{ev}%
\frac{V^{2}}{c^{2}}=1\rm{keV} 
\]

\begin{figure}
\centering
\includegraphics[width=4in]{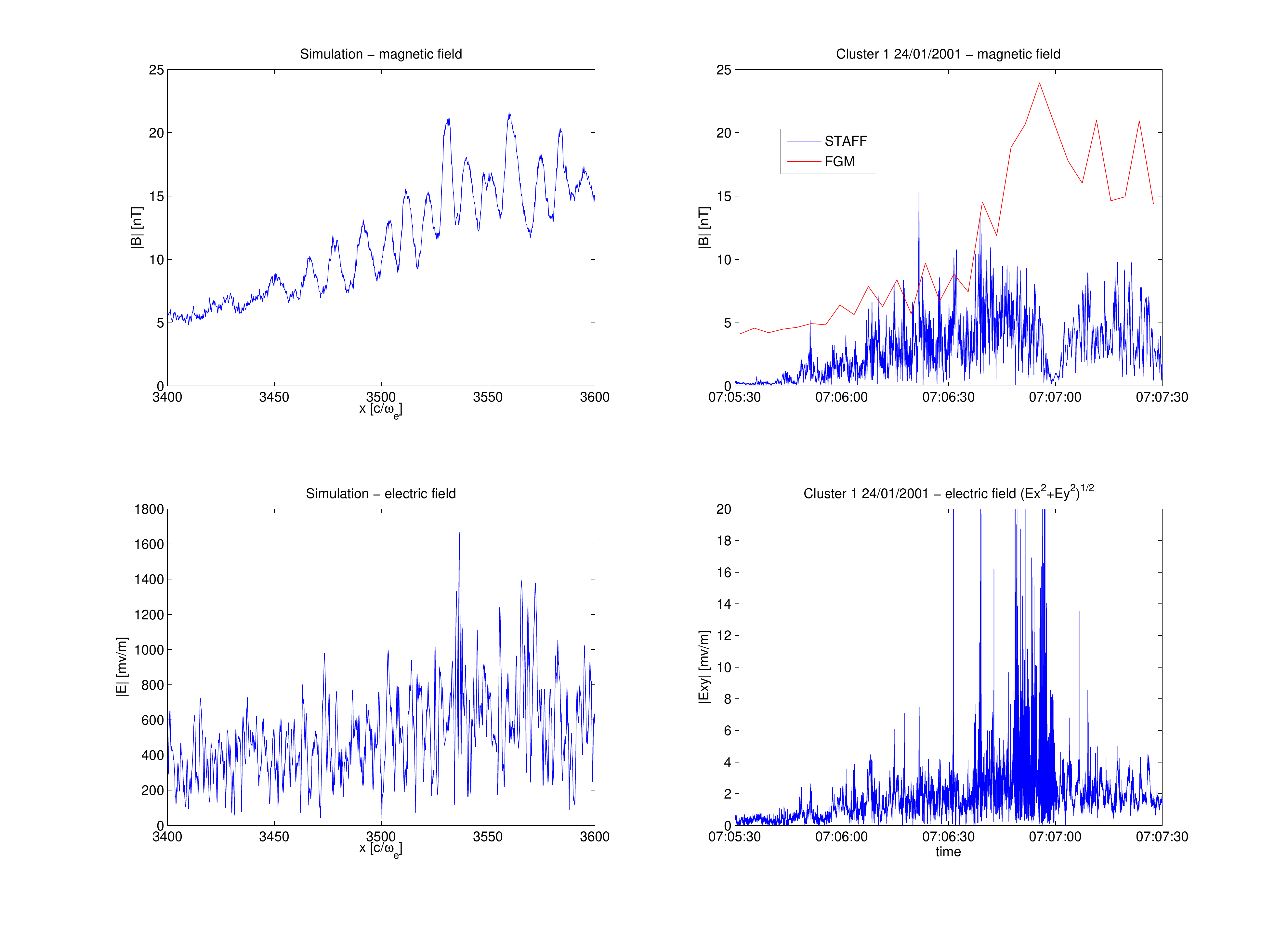}
\caption{Comparison of electric and magnetic fields observed on 24th of January 2001 by Cluster satellites and obtained in computer simulation by \cite{comisel11:_non}. Both panels show the fields in the vicinity of the foot/ramp regions. Top panels represent magnetic fields as measured by FGM instrument and STAFF search coil magnetometer (right), and obtained as a result of simulations(left). Bottom panels represent measured and simulated electric fields. One can see huge difference in amplitude of electric fields between measurements and simulations that results from artificial ratio of plasma to gyrofrequency and consequently unrealistic refractive index of waves. (Figure is provided by J. Soucek and H. Comisel) \label{fig:shk_comp}}
\end{figure}

This corresponds to the value of the electric field 
\[
E=\frac{0.5\cdot 10^{3}V}{2\cdot 0.9\cdot 10^{4}\rm{m}}=30 \rm{mVm}^{-1}
\]

\[
\delta B_{\exp }=\frac{N_{\exp }E}{c}=\frac{0.7\cdot 10^{3}\cdot 60\rm{V/m}\cdot
10^{-3}}{2\cdot 2\cdot 10^{8}}=1.5\cdot 10^{-4}\sim 15 {\rm nT}
\]
where $\delta B_{\exp }$ gives the idea of the magnetic field fluctuations really observed and obtained from the comparison with the electric field measurements. 
These effects are illustrated on the Figure\ref{fig:shk_comp} where left hand panels show electric and magnetic field fluctuations in units similar to those experimentally observed, and right hand panels show the data obtained by Cluster satellites for similar parameters (Mach number, angle and $\beta$). 
The magnetic field fluctuations in simulations that will be associated with
similar electric fields could be 30 times smaller. Thus electric fields
capable to trap and reflect ions are associated with the magnetic field fluctuations that are quite small, namely, less than $\sim 1$nT. The ion trapping and reflection can occur in small amplitude oscillations in the foot region that can not happen in real shock. Crucial change of ion dynamics certainly results in change of the characteristics of the shock front and wave activity around.

The goal of this remark is not to understate the role and importance of
computer simulations for the shock studies. We would like to point out that direct comparison of simulation results with the observations needs special attention and analysis of the simulation conditions to ensure that the process is properly described.

 \begin{acknowledgements}
V.K. acknowledges ISSI for supporting team activities of the project ÒHigh Mach Number ShocksÓ, CNES for financial support of this work through the series of grants Cluster Co-I DWP. V.K. is also grateful to A. Spitkovsky for providing a Figure and useful discussions of astrophysical shocks, A. Artemyev for assistance in preparation of the manuscript, and to R.Z. Sagdeev and A.A. Galeev, teachers who supported the interest to this topic for many years. The authors are grateful to A. Balogh for his kind support and encouragement during the compilation of this review. It is also our great pleasure to acknowledge the contribution of all Cluster PI's whose tireless efforts have provided the  
scientific community with a readily accessible, high quality, unique data collection without which this work could not have been completed. The authors are grateful to Referee who helped us to improve the manuscript. The final publication is available at link.springer.com.
\end{acknowledgements}

\section{Table of notations used in the article}
\label{sec:notation}
\begin{longtable}{| p{3cm} | p{8.3cm} |}

\hline \hline
\textbf{Parameter} & \textbf{Interpretation} \\
\hline \hline
\endfirsthead

\hline \hline
\textbf{Parameter} & \textbf{Interpretation} \\
\hline \hline
\endhead

\hline \hline
\endlastfoot

\hline \hline
\endfoot

$B$ & the magnitude of the magnetic field \\
$B_0$ & upstream magnetic field \\
$B_n$ & magnetic field component along the normal to the shock \\
$\beta =\frac{8\pi nT}{B^{2}}$ & the ratio of total particle thermal pressure to the magnetic field pressure \\
$\Delta B$ & change of the magnetic field through inhomogeneous layer \\
$\Delta t_{ij}$ & time difference of observation of shock front features such as electric field spikes by different satellites $i$ and $j$ \\
 ${\Delta E}$ & an amplitude of the electric field spike feature \\
$E_{x,y,z}$& electric field components along corresponding axes \\
$E_{spike}$ & maximum amplitude in electric field spike \\
$E_{ij}$ & electric field as measured by means of  probes $i$ and $j$ \\
$\vec{E}$ & electric field vector \\
$e_{ij}$ & electric potential difference between probes $i$ and $j$ onboard single satellite \\
$f = {\omega/{2 \pi}}$ & wave frequency \\
$f_{e}={\Omega_{ce}/{2 \pi}}$ & electron gyrofrequency (in Hz) \\
$\gamma _{LH}$ & growth rate of lower hybrid waves \\
$\vec k$ & wave-vector of a wave  \\
$k_{ij}$ & an estimate of the $\vec k$-vector component from electric field probes $i$ and $j$ measurements onboard one single satellite \\
$k_{\parallel}$ & parallel to the magnetic field component of the wave-vector \\
$l_{gr}$ & characteristic gradient scale inside the inhomogeneous layer \\
$L_d$ & dissipative scale \\
$L_{disp}$ & dispersive scale \\
$L_{i,e} = c/{\omega_p}$ & ion, electron inertial length \\
$L_{Br} $ & thickness of the ramp region of the shock as seen in magnetic field measurements \\
$L_{\phi}$ & characteristic scale of the electrostatic potential variation in the shock front \\
$L_{an}$ & characteristic scale of energy exchange due to anomalous resistivity\\
$L_{f}$ & characteristic spatial size of the magnetic foot\\
$L_{r}$ & width of the magnetic foot\\
$m_{i,e}$ & ion, electron mass \\
$M_A = {V_{up} } / {V_A}$ & Alfvenic Mach number, the ratio of the normal component of the upstream flow velocity to Alfven velocity\\
$M_{Ms} = V_{up}/V_{Ms}$ & magnetosonic Mach number\\
$M_{w} = V_{w,max}/V_A$ & nonlinear critical whistler Mach number\\
$n_{E}$ & shock front normal determined from timing of electric field spikes measured onboard four satellites\\
$n_{B}$ &shock front normal vector from magnetic field measurements\\
$n$ & plasma density\\
$\hat{\vec{n}}$ & shock front normal vector\\
$n_r$ & number density of reflected ions\\
$R_{Li} = V_{up}/{\Omega}_{ci} $ &convective ion gyroradius\\
$R_{Le}$ &electron Larmor radius\\
$S_{\parallel}$ & Poynting flux along the magnetic field\\
$T$ & total plasma temperature\\
$T_{i,e}$ & ion, electron temperature\\
$T_{e\parallel}$ & parallel to magnetic field electron temperature \\
$T_{e\perp}$ & electron temperature perpendicular to the magnetic field\\
$T_{Bi}$ & ion gyroperiod\\
$\overrightarrow{u}_{ed}$ & the relative velocity of electrons
carrying current \\
$V_A $ & Alfven velocity\\
$V_{up}$ & the normal component of the upstream velocity to the shock surface in its rest frame\\
$V_{Ms}$ & velocity of the magnetosonic wave propagating in the same direction as the shock to the background magnetic field\\
$V_{w,max}$  & highest possible velocity of nonlinear whistler wave that can stay in the upstream flow\\
$V_{sw}$ & solar wind velocity\\
$V_{ss}$ &relative shock spacecraft velocity\\
$v_{ph}$ & phase velocity of ion sound wave\\
${V_{ph}}$ & phase velocity of wave\\
$v_{is}$ &ion sound velocity\\
$V_{x,y,z}$ & velocity components along corresponding axes \\
$W_{k}$ & electric field energy density\\
$\lambda$ & wavelength of precursor whistler wave\\
$\lambda_D= \epsilon_0 k_B T_e/n e^2$ & Debye radius  \\
$\nu _{eff}$ & effective collision frequency due to wave particle interaction\\
$\theta _{Bn}$ & the angle between the magnetic field and shock front normal\\
$\theta_{kB}$  & angle between the magnetic field vector and the wave vector\\
$\omega_{lh}\sim\sqrt{\Omega_{ci}\Omega_{ce}}$ & lower hybrid frequency\\
${\omega}_{pi}$ & ion plasma frequency \\
${\Omega}_{ci,e} = {eB}/{m_{i,e}}$ & ion, electron gyrofrequency\\
${\omega}_p$ & electron plasma frequency \\
\hline
\end{longtable}

\bibliographystyle{aps-nameyear}      
\bibliography{ref}   

\end{document}